\documentclass[11pt,british]{article}
\usepackage[utf8]{inputenc}
\usepackage{mathtools}
\usepackage{comment}
\usepackage{amsmath}
\usepackage{pdflscape}
\usepackage{amsthm}
\usepackage{amssymb}
\usepackage{setspace}
\usepackage{microtype}
\usepackage{GGMath}
\usepackage{babel}
\usepackage{stmaryrd}
\usepackage{slashed}
\usepackage{cancel}
\usepackage{multirow}
\usepackage{bbm}
\usepackage{booktabs}
\usepackage{pgfplots}
\pgfplotsset{compat=newest}
\usepgfplotslibrary{fillbetween}

\usepackage{import}

\makeatletter
\usepackage{tikz}
\usepackage{tikz-cd}
\usetikzlibrary{3d} 
\usetikzlibrary{backgrounds}
\usepackage[framemethod=tikz]{mdframed}
\AtBeginEnvironment{mdframed}{%
\tikzset{every picture/.style={}}%
}
\mdfsetup{roundcorner=.5ex}

\theoremstyle{definition}
\newtheorem*{defn*}{Definition}

\usepackage{jheppub}                   
\usepackage[linecolor=blue,backgroundcolor=blue!25,bordercolor=blue,textsize=scriptsize]{todonotes}

\makeatletter
\gdef\@fpheader{\ }                    
\makeatother

\usepackage{slashed}	 	
\usepackage{amsfonts} 		
\SetTracking{encoding={*}, shape=sc}{0} 	
\usepackage{color} 		
\definecolor{darkblue}{rgb}{0.0,0.0,0.3} 	
\allowdisplaybreaks		
\date{\today} 		
\numberwithin{equation}{section}	

\makeatletter
\g@addto@macro\bfseries{\boldmath}
\makeatother

\let\originalleft\left
\let\originalright\right
\renewcommand{\left}{\mathopen{}\mathclose\bgroup\originalleft}
\renewcommand{\right}{\aftergroup\egroup\originalright}

\theoremstyle{definition}

\title{All-orders moduli for type II flux backgrounds}
\author[a]{George R. Smith,}
\emailAdd{g.smith19@imperial.ac.uk}
\author[b]{David Tennyson}
\emailAdd{dtennyson@tamu.edu}
\author[a]{and Daniel Waldram}
\emailAdd{d.waldram@imperial.ac.uk}

\affiliation[a]{Department of Physics,
	Imperial College London, \\
	Prince Consort Road, London, SW7 2AZ, UK}
	
\affiliation[b]{Mitchell Institute for Fundamental Physics and Astronomy, Texas A\&M University, College Station, TX, 77843, USA}
	
\abstract{We investigate the old problem of determining the exact bulk moduli of generic $\mathrm{SU}(3)$-structure flux backgrounds of type II string theory. Using techniques from generalised geometry, we show that the infinitesimal deformations are counted by a spectral sequence in which the vertical maps are either de Rham or Dolbeault differentials (depending on the type of the exceptional complex structure (ECS)) and the horizontal maps are linear maps constructed from the flux and intrinsic torsion. Our calculation is exact, covering all possible supergravity $\mathrm{SU}(3)$-structure flux backgrounds including those which are not conformally Calabi--Yau, and goes beyond the usual linear approximation in three important ways: (i) we allow for finite flux; (ii) we consider perturbative higher-derivative corrections to the supergravity action; and (iii) we consider obstructions arising from higher-order deformations. Despite these extensions we find that the spectral sequence reproduces the naïve expectations that come from considering the effective superpotential in the small-flux limit. In particular, by writing the moduli in a form that is independent of the Kähler potential on the space of ECSs, and arguing the superpotential does not receive higher-derivative corrections, we show that the spectral sequence is perturbatively exact. Further, preliminary results show that a Tian--Todorov-like lemma implies that all the obstructions vanish. This has implications for the tadpole conjecture, showing that such perturbative, higher-order effects do not provide a way of circumventing the bound.}

\subheader{\hfill\textrm{Imperial/TP/24/DW/1}\\
	\hspace*{\fill}\textrm{MI-HET-828}}

\begin{document}

\maketitle

\section{Introduction}

The study of four-dimensional superstring backgrounds preserving $\mathcal{N}=1$ supersymmetry is central to understanding the landscape of semi-realistic string models of particle physics. A key question is to identify the moduli of the background, that is deformations of the internal geometry that lead to massless scalar fields in the four-dimensional effective theory. In the absence of flux, the internal space is a manifold of special holonomy, for example, a Calabi--Yau space. The problem of identifying the massless scalars then translates into a well-studied problem of understanding the moduli space of special holonomy manifolds. 

With the inclusion of flux the problem of finding moduli becomes considerably more involved but also phenomenologically important since the addition of flux generically gives masses to some of the moduli of the special holonomy background. (Analysis of moduli in flux backgrounds is a large and very well-established field. For reviews see for example~\cite{Grana:2005jc,Blumenhagen:2006ci,McAllister:2023vgy}). This scenario is usually analysed in the limit where the fluxes are a linear perturbation of the special holonomy background, generating a potential that lifts some of the moduli, following~\cite{Taylor:1999ii,Dasgupta:1999ss,Giddings:2001yu}. Given the $\mathcal{N}=1$ supersymmmetry the potential takes the form 
\begin{equation}\label{eq:4d-EFT-potential}
     V = \; \stackrel{\text{F-terms}}{\ee^K\left(K^{I\bar{J}}D_I W D_{\bar{J}} \bar{W} - 3|W|^3\right)} - \frac{1}{2}\stackrel{\text{D-terms}}{\left(\re \tau^{-1}\right)^{\alpha\beta}D_\alpha D_\beta}.
\end{equation}
where $K$ is the Kähler potential for the moduli, $W$ is the flux-induced superpotential and $D_\alpha$ are flux-induced D-terms. Probably the most famous example of stabilisation is that of adding complex three-form flux $G_3$ to type IIB Calabi--Yau O3/O7-plane orientifolds discussed by Giddings--Kachru--Polchinski~\cite{Giddings:2001yu}, for which the superpotential has the Gukov--Vafa--Witten form~\cite{Gukov:1999ya,Taylor:1999ii} 
\begin{equation}
    W \sim \int_{\text{CY}_{3}}G_{3}\wedge\Omega , \label{eq:W-GKP} 
\end{equation}
where $\Omega$ is the holomorphic three-form on the Calabi--Yau space. Requiring the four-dimensional background to be supersymmetric and Minkowski, sets the F-terms to zero, $D_IW=0$, and $W=0$, giving the Graña--Polchinski solution~\cite{Grana:2000jj,Grana:2001xn,Gubser:2000vg}, and generically appears to fix all the complex structure (and also axion-dilaton) moduli. It has recently been  stressed that this final conclusion is maybe naive: the ``tadpole conjecture''~\cite{Bena:2020xrh} states that the stabilisation is actually severely constrained by the tadpole bound on the fluxes, so that giving masses to all the complex structure moduli is not possible when number of moduli is large. 

The standard analysis is valid in the following regime. First one is in the ``large-radius'' limit $\ell_s/R\ll1$ where $\ell_s$ is the string scale and $R$ is the compactification radius, in which  higher-derivative corrections to the leading supergravity approximation can be ignored. Second, one takes the flux perturbation to be small compared to the Kaluza--Klein scale, so that a massive Kaluza--Klein mode cannot become light under the perturbation. Using the flux quantisation condition, this requires $N_{\text{flux}}\ll(R/\ell_s)^{p-1}$ for a $p$-form flux, where $N_{\text{flux}}$ is the number of units of flux. In the large-radius limit, violating this constraint requires large $N_{\text{flux}}$. However, the number of flux quanta is bounded by tadpole conditions and so typically cannot be too big and so the standard scenario appears to apply. 

Nonetheless, one might wonder if one can go beyond this approximation. First one might consider solutions still in the large-radius limit, but violate the ``small-flux'' approximation. This could be either because comparatively large $N_{\text{flux}}$ is possible, or simply that locally there are regions where the flux is comparable to the Kaluza--Klein scale, even if the average, integrated flux satisfies the small-flux constraint. The geometry (or certain regions of the geometry) is then far from Calabi--Yau and so, in general, one needs new tools to identify the moduli.\footnote{For the particular case of three-form flux in IIB the finite-flux large-radius supersymmetric background famously is a warped Calabi--Yau space~\cite{Grana:2000jj,Grana:2001xn} and this special form means that progress can be made in identifying the moduli (see for example~\cite{DeWolfe:2002nn,Giddings:2005ff,Shiu:2008ry,Douglas:2008jx,Frey:2008xw} and subsequent work).} More ambitiously, one might ask whether one can also include higher-derivative supergravity corrections. Intriguingly, there is a non-renormalisation theorem that states, for the class of O3/O7-plane, IIB backgrounds, the (small-flux) superpotential~\eqref{eq:W-GKP} is correct to all orders in perturbation theory~\cite{Burgess:2005jx}. If the moduli space only depends on the form of the superpotential, such a non-renormalisation theorem would then imply that it is also uncorrected by the higher-derivative terms. 

In this paper we aim to address both these issues. The key ingredient is that space of generic supersymmetric backgrounds can be viewed as depending on a superpotential and D-terms on a infinite-dimensional space of a particular class of geometrical structures~\cite{Ashmore:2019qii}. The moduli space then depends only on the form of the superpotential, modded out by the action of a complexified group. Using this structure we   
\begin{enumerate}
    \item calculate the exact bulk moduli of generic finite-flux, $\SU3$-structure, type II $d=4$ $\mathcal{N}=1$ Minkowski backgrounds in the large-radius limit;
    \item argue for a general non-renormalisation theorem implying that the result is perturbatively exact.
\end{enumerate}
There are a number of different classes of such large-radius backgrounds and in each case there is a corresponding superpotential that can be written in terms of the $\SU3$ structure and the fluxes, as summarised in~\cite{Grana:2004bg}. Calculating the bulk moduli we find 
\begin{itemize}
    \item \emph{the dimension of the finite-flux moduli space can be calculated using conventional de Rham or Dolbeault cohomology classes;}
    \item \emph{the exact finite-flux calculation matches the naïve small-flux expectation that the superpotential lifts some of the fluxless Calabi--Yau moduli; in particular the presence of flux always reduces the number of moduli.}
\end{itemize}
It is worth emphasising that we only consider bulk moduli, ignoring deformations of orientifold and D-brane sources. Generically this means we are considering the de Rham or Dolbeault cohomologies on spaces where the sources are excised, and strictly it is the moduli of these excised spaces that is always reduced by the presence of flux. That said, the extra ``relative cohomology'' classes that are present on the excised space should be associated to deformations of the sources. Thus restricting to the unexcised space should still capture the dynamics of the bulk moduli.\footnote{The caveat here is that the sourced flux will nonetheless live in the relative cohomology and can still in principle obstruct bulk moduli, a point we discuss briefly in the conclusions.}  

The formalism we will use to calculate the moduli is exceptional generalised geometry~\cite{Coimbra:2011ky,Coimbra:2012af}. This is a reformation of the supergravity on the compact space that unifies the metric and flux degrees of freedom and their symmetries, so that the whole theory becomes geometrical -- specifically a generalised version of pure Einstein gravity. It is well-known that conventional fluxless backgrounds correspond to special holonomy manifolds, or equivalently spaces admitting a torsion-free $G$-structure (for Calabi--Yau threefold compactifications the special holonomy group $G$ is $\SU3$). In exceptional generalised geometry the internal supersymmetric flux background is analogously described by a torsion-free generalised structure~\cite{Coimbra:2014uxa,Ashmore:2015joa,Ashmore:2019qii}, and it is the moduli space of these torsion-free structures which counts the number of massless fields. 

The advantage of using generalised geometry is that it is naturally adapted to the structure of the preserved supersymmetry. Starting from~\cite{Gauntlett:2002sc,Gauntlett:2002fz,Gauntlett:2003wb}, conventional $G$-structure techniques have long been extremely useful for analysing supersymmetric backgrounds, but the presence of flux means the structures have non-vanishing torsion and also are generically not globally defined, making the analysis of moduli difficult. Using generalised geometrical structures, the supersymmetry conditions naturally align with those familiar from supersymmetric theories on the non-compact space. In particular, for $\mathcal{N}=1$ backgrounds, the space of $\SU7$ structures admits a pseudo-Kähler metric and can be viewed as an infinite-dimensional space of scalar fields in $\mathcal{N}=1$, $d=4$ chiral multiplets. The supersymmetry conditions can then be viewed as F- and D-terms~\cite{Ashmore:2019qii}. Remarkably the gauge group for the D-terms is none other than the ``Generalised diffeomorphism group'' $\GDiff$ formed by the combination of diffeomorphisms and flux potential gauge transformations. 

Analysing type II flux backgrounds via generalised geometry goes back to the seminal work of Graña et al.~(GMPT)~\cite{Grana:2004bg,Grana:2005sn}, who used the $\Orth{d,d}$ theory to reformulate the supersymmetry conditions for the generic class of $\mathcal{N}=1$ backgrounds in terms of pure spinors. The interpretation of the GMPT conditions in terms of superpotential and D-terms was stressed in~\cite{Martucci:2006ij,Koerber:2007xk}, and the formalism has been used by several authors to analyse moduli~\cite{Koerber:2006hh,Tomasiello:2007zq,Martucci:2009sf}, leading to the appearance of new cohomology theories~\cite{Tseng:2009gr,Tseng:2010kt,Tseng:2011gv}. The current paper extends this work, with the key point being that by moving to exceptional generalised geometry, all the fluxes become geometrised such that structures become integrable. Thus the RR moduli are incorporated in the formalism from the beginning rather than being an obstruction to an integrable $\SU3\times\SU3$ structure. It also gives a single formalism for describing all $\mathcal{N}=1$ backgrounds including the subclass, where one type II spinor vanishes, not covered by the GMPT analysis.

In exceptional generalised geometry, the supersymmetry conditions can be equated to the existence of an integrable generalised $\SU{7}$-structure, and the physical moduli fields are given by deformations of this structure preserving its integrability.  This description of the moduli space assumes nothing of the internal space beyond it preserving $\mathcal{N}=1$ supersymmetry, and so is capable of describing the moduli space of all supersymmetric type II backgrounds (as well as M-theory backgrounds~\cite{Ashmore:2019qii,Smith:2022baw}).  We find that for $\SU3$-structure type II backgrounds the moduli are counted by cohomologies of a flux twisted differential operator
\begin{equation}\label{intro:simple_moduli_diagram}
\begin{tikzcd}
\text{Gauge} \ \ \arrow[r,"\dd_{\Delta}+\CF"] &  \ \ \text{Deformations} \ \ \arrow[r,"\dd_{\Delta}+\CF"]& \ \ \text{Torsion}.
\end{tikzcd}
\end{equation}
Here $\Delta\subseteq TM$ is either the tangent bundle $\Delta=TM$ with $\dd_\Delta = \dd$, or the anti-holomorphic vector bundle $\Delta = T^{0,1}$ with $\dd_\Delta = \delb$ and the complex flux $\CF$ is a particular combination of complex differential forms. The corresponding cohomology groups are then counted using spectral sequences from a double complex built from the action of $\dd_\Delta$ and $\CF$. Under the assumption that $M$ is topologically a Calabi--Yau manifold (that is that the manifold also admits a fluxless solution), we find that only the first and second pages contribute, and this is what leads to the full finite-flux calculation agreeing with the naïve small-flux expectation. Moreover, one can then see that the presence of $\CF$ in the first and second maps in the exact sequence,  leads to moduli lifting via D-terms and F-terms respectively. Finally, we also prove that the infinitesimal moduli are isomorphic anywhere along a $\GDiff_\bbC$ gauge orbit. This allows us to compute the moduli of fully supersymmetric backgrounds using any background on their orbit, which gives an elegant tool for calculating the moduli in, for example, the Graña--Polchinski backgrounds. 

We should note that this kind of structure~\eqref{intro:simple_moduli_diagram}, with a flux twisted differential operator is familiar from both the deformation of generalised complex structures in $\Orth{d,d}$ generalised geometry~\cite{Gualtieri:2003dx,Cavalcanti:2005hq} and from the heterotic string compactifications~\cite{Rohm:1985jv,Anderson:2010mh,Anderson:2011ty}. The latter case was in fact extended by analogy to $\SU3$-structure type IIB backgrounds by Gray and Parsian~\cite{Gray:2018kss}, focusing on the complex structure moduli. In this particular case, our analysis gives an extension of their description to the full set of moduli, derived directly from the exceptional generalised geometry. 

The paper is arranged as follows. In section~\ref{sec:SU3_structure_review} we review $\SU{3}$ structures as backgrounds preserving $\mathcal{N}=1$ supersymmetry, and how fluxes encode their intrinsic torsion, summarising the different possibilities in IIA and IIB. In section~\ref{sec:EGG_SUSY_Backgrounds} we rephrase the supersymmetry conditions for each case in terms of $\ExR{7(7)}$ generalised geometry and $\SU{7}$ structures. First reviewing the integrability conditions of $\SU{7}$ structures presented in~\cite{Ashmore:2019qii} we show how they encode the different $\SU3$-structure IIA and IIB backgrounds in section~\ref{subsec:ECS_SU(3)}. Section~\ref{sec:Moduli_Stabilisation} contains a comparison between the conventional and $\ExR{7(7)}$ moduli space descriptions, and derives how the dimensions of the infinitesimal moduli are counted by certain cohomology groups. In section~\ref{sec:SU3_ECS_moduli} the infinitesimal moduli of flux backgrounds with $\SU{3}$ structures are computed explicitly for both IIA and IIB. We then, in section~\ref{sec:corrections_and_obstructions} address the question of higher-derivative perturbative corrections and obstructions. We argue that, under assumptions of locality and gauge and diffeomorphism invariance, the superpotential for the $\SU7$ structures is uncorrected by higher-derivative terms and so, perturbatively, the moduli calculation is unchanged from the leading order supergravity problem considered in the previous section. We then show in several cases there are also no obstructions to the finite moduli problem. We end with a short discussion of our conclusions and further possible directions. There are also three appendices of formulae and details supplementing the main text.

\section{Review of $\protect\SU 3$ structure flux backgrounds}\label{sec:SU3_structure_review}

In this paper we are interested in type II supergravity compactifications of the form $M_{10}=\bbR^{3,1}\times M$ to four-dimensional Minkowski space, preserving $\mathcal{N}=1$ supersymmetry. We will further assume that the internal space $M$ has an $\SU{3}$ structure, or equivalently admits a globally non-vanishing spinor. These $\SU{3}$ structures are not arbitrary, but are constrained by supersymmetry. In this section we briefly review these structures on the internal space at the level of the leading supergravity theory, that is for finite flux but without higher-derivative corrections, closely following the review of~\cite{Grana:2005jc}.

\subsection{Torsion classes and types of Minkowski backgrounds}

The most general background metric for compactification to Minkowski space takes the form
\begin{align*}
\dd s^{2}\left(M_{10}\right) & =\ee^{2\warp}\dd s^{2}\left(\bbR^{3,1}\right)
   +\dd s^{2}\left(M_{6}\right)
\end{align*}
where $\dd s^{2}(\bbR^{3,1})$ is the metric for Minkowski spacetime and the warp factor $\warp$ is a function on $M$, while to preserve the Poincaré symmetry, all the type II fluxes must be purely internal or pure external. The supersymmetry variations are parameterised by a pair of Majorana--Weyl spinors $\varepsilon^\alpha=(\varepsilon^+,\varepsilon^-)$, which have opposite chirality for type IIA and the same for type IIB. Using this notation the fermionic variations can be written in the succinct form (see for example~\cite{Coimbra:2011nw})
\begin{equation}
\label{eq:susy-var}
\begin{aligned}
    \delta \psi^\pm_M 
       &= \left( \nabla_M \mp \tfrac18 H_{MNP}\gamma^{NP}\right) \varepsilon^\pm
            + \tfrac1{16}\ee^\varphi \sum_n (\pm)^{[(n+1)/2]} \slashed{F}_n \gamma_M \varepsilon^\mp , \\
    \delta\left( \lambda^\pm - \gamma^M \psi_M^\pm \right)
      &= - \gamma^M \left( \nabla_M \mp \tfrac{1}{24} H_{MNP}\gamma^{NP}
            - \del_M \varphi \right) , 
\end{aligned}
\end{equation}
where $\psi^\alpha_M$ and $\lambda^\alpha$ are the pairs of gravitini and dilatini respectively and the sum is over even $n$ for type IIA and odd $n$ for type IIB. The fluxes satisfy the Bianchi identities 
\begin{equation}
    \dd F_{n} = H \wedge F_{n-2} , 
\end{equation}
and we are using the democratic formalism of~\cite{Bergshoeff:2001pv}.

$\mathcal{N}=1$ supersymmetry in four dimensions implies that there is a Killing spinor, for which the variations of the fermions vanish, of the form 
\begin{equation}
\label{def:IIA_10D_Killing_spinor}
\begin{aligned}
    \varepsilon^+ &= \theta \otimes \epsilon^+ + \theta^c \otimes \epsilon^{c+} , \\
    \varepsilon^- &= \theta \otimes \epsilon^{c-} + \theta^* \otimes \epsilon^- , 
\end{aligned}
\end{equation}
for type IIA and 
\begin{equation}
\label{def:IIB_10D_Killing_spinor}
     \varepsilon^\alpha = \theta \otimes \epsilon^\alpha + \theta^c \otimes \epsilon^{c\,\alpha}
\end{equation}
for type IIB, where $\theta$ and $\epsilon^\alpha$ are positive chirality four-dimensional and six-dimensional spinors respectively, and $\theta^c$ and $\epsilon^{c\,\alpha}$ denote the complex conjugate spinors. Generically the two internal spinors $\epsilon^\alpha$ define a local $\SU2$ structure. The $\SU3$ structure backgrounds correspond to the special case where $\epsilon^\alpha$ are proportional to the same six-dimensional spinor $\eta$, namely
\begin{equation}
\label{def:functions_a_and_b}
    \epsilon^+ = a \eta , \qquad 
    \epsilon^- = b \eta ,
\end{equation}
where we normalise such that $\bar{\eta}\eta=1$. for some complex functions $a$ and $b$. The Killing spinor equations imply 
\begin{equation}
\label{eq:spinor-norm}
    |a|^2 + |b|^2 = \ee^\Delta
\end{equation}
and relate the fluxes to the intrinsic torsion of the $\SU3$ structure. The latter splits into torsion classes transforming in $\SU{3}$ representations
\begin{equation}
    \begin{aligned}
        W_1\sim \rep{1}_{\bbC}, \quad W_2\sim \rep{8}_{\bbC}, \quad W_3\sim \re (\rep{6}\oplus \bar{\rep{6}}), \quad W_4 ,W_5 \sim \re(\rep{3}\oplus\bar{\rep{3}})
    \end{aligned}
\end{equation}
which are encoded by the exterior derivatives of the 2- and 3-form spinor bilinears $\omega$ and $\Omega$ defined by the $\SU 3$-structure~\cite{Chiossi:2002tw} as
\begin{equation}
\begin{aligned}\dd\omega & =\tfrac{3}{2}\im\bar{W}_{1}\Omega+W_{4}\wedge\omega+W_{3}\\
\dd\Omega & =W_{1}\wedge\omega^{2}+W_{2}\wedge\omega+\bar{W}_{5}\wedge\Omega.
\end{aligned}
\end{equation}
From~\eqref{def:functions_a_and_b} one notes one can absorb an arbitrary phase in $\eta_+$ in the $a$ and $b$ coefficients, such that there is a a $\Uni 1$ gauge freedom $(a,b)\to\ee^{\ii\gamma}(a,b)$ corresponding to $\eta_{+}\to\ee^{-\ii\gamma}\eta_{+}$ and $\Omega\to\ee^{-2\ii\gamma}\Omega$. This has consequences on the intrinsic torsion components $W_{i}$, which transform as
\begin{equation}
\begin{aligned}\left(W_{1},W_{2}\right) & \to\ee^{2\ii\gamma}\left(W_{1},W_{2}\right)\\
W_{5} & \to W_{5}+2\ii\dd\gamma.
\end{aligned}
\end{equation}
We can fix the overall phase of $ab$ by rotating $W_{2}$ so that only the combination $W_{2}^{+}=\tfrac{1}{2}(W_{2} + \bar{W}_{2})$ component enters into the torsion-flux conditions.

The explicit relations between the flux and torsion classes implied by the Killing spinor equations are given in table~\ref{table:IIA_backgrounds} for IIA and table~\ref{table:IIB_backgrounds} for IIB, following~\cite{Grana:2004bg,Grana:2005jc} and arranged by $\SU3$ representations for both the torsion classes and fluxes.\footnote{The IIA conditions summarised in~\cite{Grana:2005jc} have a typo carried forward from the original paper~\cite{Grana:2004bg} that is corrected in table~\ref{table:IIA_backgrounds}. An extra $F_{4}^{(8)}$ component in $W_{2}^{+}$ for special values of $\beta$ in the (BC) backgrounds was included. In fact this term vanishes in all cases.} The (ABC) interpolating backgrounds for type IIB have relationships between flux and torsion with more complicated functions of $a$ and $b$ between them, given in \cite[section 3.3]{Grana:2005jc}.
\begin{table}[h]
\begin{center}%
\begingroup
\setlength{\tabcolsep}{12pt} 
\renewcommand{\arraystretch}{1.4} 
\begin{tabular}{@{}lccc@{}}
\toprule[0.1em]
\textbf{IIA} & $a=0$ or $b=0$ (A) & $a=b\ee^{\ii\beta}$ (BC) \tabularnewline
\midrule[0.1em]
\multirow{2}{*}{$\rep 1$} & \multicolumn{2}{c}{$W_{1}=H_{3}^{(1)}=0$} \tabularnewline
\cmidrule[0.005em](l{8em}r{8em}){2-3}
 & $F_{0}=\mp F_{2}^{(1)}=F_{4}^{(1)}=\mp F^{(1)}_{6}$ & $F_{2n}^{(1)}=0$\tabularnewline
\midrule[0.02em]
$\rep 8$ & $W_{2}=F_{2}^{(8)}=F_{4}^{(8)}=0$ & $W_{2}^{+}=\ee^{\dil}F_{2}^{(8)}$, $W_{2}^{-}=0$
\tabularnewline
\midrule[0.02em]
$\rep 6$ & $W_{3}=\mp*H_{3}^{(6)}$ & $W_{3}=H_{3}^{(6)}=0$\tabularnewline
\midrule[0.02em]
$\rep 3$ 
& $\begin{gathered} 
    \bar{W}_{5} = 2W_{4}=\mp2\ii H_{3}^{(3)}=\delb\dil \\
    \delb\warp =\delb a=0
\end{gathered}$
& $\begin{gathered}
    F_{2}^{(\bar{3})} =2\ii\bar{W}_{5}=-2\ii\delb\warp=\tfrac{2}{3}\ii\delb\dil\\
    W_{4} =0
\end{gathered}$
\tabularnewline
\bottomrule[0.1em]
\end{tabular}\endgroup\end{center}
\caption{Possible $\mathcal{N}=1$ vacua in IIA, taken from \cite{Grana:2005jc,Grana:2004bg}.}
\label{table:IIA_backgrounds}
\end{table}
The A-type solutions are common to (non-massive) IIA, IIB and heterotic supergravity and only have non-trivial purely NSNS fields. For massive IIA one can also include singlet RR flux. For the BC solutions of IIA, the only non-trivial flux is the RR two-form $F_2$ and they can be viewed as the dimensional reduction of eleven-dimensional supergravity on a $G_2$ manifold. For type IIB, the constant dilaton B-type background is the well-known conformal Calabi--Yau solution of~\cite{Grana:2000jj,Grana:2001xn}. Note that there is also an F-theory-like B-type solution which is not Calabi--Yau (for more discussion on the F-theory geometry see for example~\cite{Collinucci:2008pf,Maharana:2012tu}). Finally we recall that provided neither $a$ or $b$ vanishes the $\SU3$ backgrounds are special cases of a general class of geometries described by a pair of pure spinors in $\Orth{6,6}$ generalised geometry~\cite{Grana:2004bg,Grana:2005sn}. 

In each case, to have a compact solutions, one needs orientifold planes in order to provide negative tension sources and so avoid the standard no-go theorems. Supersymmetry of the orientifolds restricts the class of solutions. In particular one has 
\begin{equation}
\begin{aligned}
    \text{IIA:} & \qquad & \text{O6-planes} &\sim \text{BC-type} \\*[3pt]
    \multirow{2}{*}{\text{IIB:}} & \qquad & \text{O3-, O7-planes} &\sim \text{B-type} \\ && 
        \text{O5-, O9-planes} &\sim \text{C-type} 
\end{aligned}
\end{equation}
As we see, there are no suitable orientifold sources for the pure NSNS A-type solutions and so in type II these backgrounds will be non-compact. However, when viewed as heterotic string solutions solving the Hull--Strominger system~\cite{Hull:1986kz,Strominger:1986uh}, one has additional negative tension sources from curvature-squared terms and they can also describe compact geometries.

\begin{table}[h]
\begin{center}%
\begingroup
\setlength{\tabcolsep}{9pt} 
\renewcommand{\arraystretch}{1.4} 
\begin{tabular}{@{}lcccc@{}}
\toprule[0.1em]
\textbf{IIB} & $a=0$ or $b=0$ (A) & $a=\pm\ii b$ (B) & $a=\pm b$ (C) & (ABC)\tabularnewline
\midrule[0.1em]
$\rep 1$ & \multicolumn{4}{c}{$W_{1}=F_{3}^{(1)}=H_{3}^{(1)}=0$}\tabularnewline
\midrule[0.02em] 
$\rep 8$ & \multicolumn{4}{c}{$W_{2}=0$}\tabularnewline
\midrule[0.02em]
$\rep 6$ & 
$\begin{gathered}
    W_{3} =\mp*H_{3}^{(6)}\\
    F_{3}^{(6)} =0
\end{gathered}$ & 
$\begin{gathered}
    \ee^{\dil}F_{3}^{(6)} =\mp*H_{3}^{(6)}\\
    W_{3} = 0
\end{gathered}$ & 
$\begin{gathered}
    W_{3} = \ee^{\dil}*F_{3}^{(6)}\\
    H_{3}^{(6)} =0
\end{gathered} $ & $\cdots$ \tabularnewline
\midrule[0.02em]
$\rep 3$ & 
$\begin{gathered}
\begin{aligned}
    & \bar{W}_{5} = 2W_{4} = \delb\varphi \\ & \quad
    = \mp2\ii H_{3}^{(3)}
\end{aligned} \\*[5pt]
    \delb\warp =\delb a=0
\end{gathered}$ & 
$\begin{aligned}
    & \text{constant $\dil$:} \\ & \quad 
    \ee^{\dil}F_{5}^{(\bar{3})} =\tfrac{2}{3}\ii\bar{W}_{5} =\ii W_{4} \\ & \qquad 
    =-2\ii\delb \warp =-4\ii\delb\log a \\*[5pt]   
    & \text{F-theory-like:} \\ & \quad 
    \ee^{\dil}F_{1}^{(\bar{3})} =2\ee^{\varphi}F_{5}^{(\bar{3})} \\ & \qquad
    = \ii\bar{W}_{5} =\ii W_{4}=\ii\delb\dil
\end{aligned}$ & 
$\begin{aligned}
    & \pm\ee^{\dil}F_{3}^{(\bar{3})} =2\ii W_{5} \\ & \quad
    =-2\ii\delb\warp =-\ii\delb\dil \\ & \quad
    =-4\ii\delb\log a 
\end{aligned}$ & $\cdots$
\tabularnewline
\bottomrule[0.1em]
\end{tabular}
\endgroup
\end{center}\caption{Possible $\mathcal{N}=1$ vacua in IIB, taken from \cite{Grana:2005jc,Grana:2004bg}.}
\label{table:IIB_backgrounds}
\end{table}

\subsection{Moduli and the superpotential}
\label{subsec:superpotential_stab}

We will primarily be interested in this paper in the moduli of these backgrounds, that is how they can be deformed while preserving the conditions of supersymmetry. In the case where the fluxes vanish and the internal geometry is Calabi--Yau, the answer is well known: there are $h^{2,1}$ complex structure moduli and $h^{1,1}$ Kähler moduli, together with moduli from the form-field potentials. The type II background actually preserves $\mathcal{N}=2$ supersymmetry and the moduli fall into either vector multiplets or hypermultiplets with, for type IIA, the Kähler moduli in vector multiplets and the complex structure moduli in hypermultiplets and vice versa for type IIB. There is also always a single additional ``universal'' hypermultiplet that includes the dilaton. Under the orientifold action the $\mathcal{N}=2$ supersymmetry is broken to $\mathcal{N}=1$, and some of the moduli are projected out, leaving $h^{1,1}_-+h^{2,1}+1$ chiral multiplets in the case of IIA O6-planes and $h^{1,1}+h^{2,1}_-+1$ and $h^{1,1}+h^{2,1}_++1$ chiral multiplets in the case of O3/O7- and O5/O9-planes in type IIB respectively. (For details see for example~\cite{Grana:2005jc}.) Here the $\pm$ subscripts refer to the harmonic forms that are even or odd respectively under the orientifold action.

From tables~\ref{table:IIA_backgrounds} and~\ref{table:IIB_backgrounds}, we see that flux backgrounds break the integrability of the Calabi--Yau so that the geometry may not be complex ($W_{2}\neq 0$ or $W_1\neq 0$) and/or not Kähler ($W_{3}\neq 0$, $W_4\neq0$ or $W_1\neq0$). As such, a priori, there is no reason to associate the moduli of the flux background with those of the underlying Calabi--Yau space and no obvious way to calculate them. 

That said, in the standard small-flux scenario, the idea is that geometry must remain close to Calabi--Yau so that the perturbation is too small to make a massive Kaluza--Klein mode become massless. The presence of flux can potentially then lift or ``stabilize'' some of the massless Calabi--Yau moduli~\cite{Giddings:2001yu,Giryavets:2003vd,Giryavets:2004zr,DeWolfe:2005gy,Conlon:2004ds,Balasubramanian:2005zx,Conlon:2005ki}. The standard way to argue for this effect is via the effective superpotential. Compactifying on the Calabi--Yau space in the small-flux limit, leads to a $\mathcal{N}=1$ supersymmetric effective four-dimensional theory with a potential for the Calabi--Yau moduli that comes from F- and D-terms induced by the fluxes. In particular, one finds the familiar Gukov--Vafa--Witten type superpotentials~\cite{Taylor:1999ii,Gukov:1999ya}
\begin{align}
W_{\text{IIA, O6}} & \propto \int_{\text{CY}_{3}}H_{3}\wedge\Omega +\tilde{F}_{\text{IIA}}\wedge\ee^{B+\ii\omega} \label{eq:W_IIA_O6}\\
W_{\text{IIB, O3/O7}} & \propto \int_{\text{CY}_{3}}G_{3}\wedge\Omega \label{eq:W_IIB_O3/O7}\\
W_{\text{IIB, O5/O9}} & \propto \int_{\text{CY}_{3}}\tilde{F}_{3}\wedge\Omega, \label{eq:W_IIB_O5/O9}
\end{align}
where in the first line we define the polyform $\tilde{F}_{\text{IIA}}=\tilde{F}_0+\tilde{F}_2+\tilde{F}_4+\tilde{F}_6$, where $\tilde{F}$ is the non-gauge-invariant, closed, RR flux related to our conventions by $F=\dd C = \tilde{F}+ H\wedge C$, and it is implied that we take the top form in the expression when integrating. In the second line we introduce the standard complex flux
\begin{equation*}
G_{3} = \tilde{F}_3 + \tau H_3 = F_3 + \ii \ee^{-\dil} H_3 , 
\end{equation*}
with $\tau=C_0+\ii\ee^{-\varphi}$. The linearised flux equations, relevant in the small-flux limit, imply that the flux is harmonic and so can be expanded in the same basis of harmonic forms on the undeformed Calabi--Yau space as the K\"ahler and complex structure moduli. For example, in the classic case of type IIB O3/O7-planes, we can write (before the orbifold projection)
\begin{equation}
\begin{aligned}
    \Omega &= Z^K \alpha_K - \mathcal{F}_K \beta^K , \\
    G_{(3)} &= \left(m^K_{\text{RR}}+\tau m^K\right) \alpha_K
        - \left(e_{\text{RR}\,K}+\tau e_K\right) \beta^K ,
\end{aligned}
\end{equation}
where $(\alpha_K,\beta^K)$ are a symplectic basis for the $2(h^{2,1}+1)$ harmonic three-forms. The constant flux coefficients $(m^K_{\text{RR}},m^K,e_{\text{RR}\,K},e_K)$ generate a linear superpotential for the complex structure moduli $Z^K$, that, generically, stabilises them (subject to bounds on the coefficients from the tadpole constraint). By contrast, the Kähler moduli do not enter the superpotential $W$ and correspond to flat directions in the potential, remaining massless in the presence of flux. One can apply the analogous argument in each of the other cases, expanding the flux in a suitable basis and hence inducing a linear superpotential that lifts some of the moduli once the superpotential is extremised.

It is important to note that typically the NSNS two-form in the effective four-dimensional supergravity also becomes massive, due to the gauging of the theory, and in addition there can be D-terms that contribute to the effective potential. These lead to further stabilisation of moduli, beyond those coming the analysis of the superpotential. However, as we discuss in more generality in section~\ref{sec:Moduli_Stabilisation}, they arise from moment maps and, in the standard way, their effect can be viewed as quotienting the moduli space by the complexification of the gauge group of the gauged supergravity. 

As we have stressed, this method is limited to the small-flux limit, $N_{\text{flux}}\ll(R/\ell_s)^{p-1}$. Away from this limit the deformation from Calabi--Yau may be large and focusing only the Calabi--Yau moduli is not justified. There may also be finite-flux supersymmetric non-Kähler and/or non-complex backgrounds that are not even perturbations of a Calabi--Yau geometry. Even in the small flux limit, there may be regions of the compactification where the local flux and deformation are large, even if averaged over the manifold the effect is small, and so the argument about restricting to Calabi--Yau moduli may again break down. Thus it would be very helpful to have a reliable calculation of the moduli that goes beyond the small-flux approximation, and fortunately we will see that this is precisely what generalised geometry allows us to do.  

\subsection{Branes, Chains and Orientifold Planes}\label{sec:sources_note}

It is well known that to have a supergravity background with non-trivial flux on a compact internal space we must include sources, with both positive and negative tension sources, that is both branes and orientifold planes~\cite{Candelas:1984yd,Candelas:1985ux,deWit:1986mwo,Maldacena:2000mw, Giddings:2001yu,Gauntlett:2003cy}.  For a compactification to flat space, the brane and orientifold planes world volumes must fill the external space, and so we only include sources with worldvolume dimension greater than four. A source of more than four dimensions will then wrap cycles $\Sigma$ on the internal space which must be of specific type in order to preserve supersymmetry. In this paper, we will assume that we do not have sources with dimension greater than seven for which a proper treatment would require F-theory.

The wrapped cycles $\Sigma$ represent homology classes $[\Sigma]\in H_*(M)$. These cycles have dual cochains $\rho(\Sigma)$ which have support localised to $\Sigma$ called currents. If a brane is magnetically charged under one of the RR fields of supergravity these currents appear in the Bianchi identity for $F_n$ as a source distribution term
\begin{equation}\label{eq:bianchi}
    \dd F_{n} - H\wedge F_{n-3} = j_{n+1} 
    \qquad 
\end{equation}
where $j_{n+1} = (2\pi \sqrt{\alpha'})^{n-1} \rho(\Sigma_{5-n})$, so that the flux $F_{\text{even/odd}}$ is no longer globally $\dd_H$ closed. (Since we are using the democratic formalism~\cite{Bergshoeff:2001pv}, all sources can be thought as magnetic charges.) There is a similar term in the stress-energy tensor. In general, the sources mean that the fully backreacted supergravity geometry is singular as one approaches the cycle, or opens out an infinite throat. One might also consider smoothing the sources so the $j_{n+1}$ is no longer a distribution. Although the backreacted geometry may then be no longer singular there is still a region where the Bianchi identity fails to hold and naive source-free supergravity calculations are not valid. 

The straight-forward way to deal with this issue is excision.  Suppose $\Sigma\subset M$ represents the union of all cycles on which we place sources, and $N$ a neighbourhood of $\Sigma$ in $M$ which deformation-retracts onto $\Sigma$ (or in the case of smoothed sources $N$ contains the support of $j_{n+1}$). We can then define the excised manifold 
\begin{equation}
    M' := M-N\subset M
\end{equation}
such that the backreacted geometry on $M'$ will be non-singular and the flux Bianchi identity and stress energy will have no source terms.  The price for such a simplification is that in general the cohomology ring of $M'$ is in general different to that of $M$ making the identification of the moduli more challenging~\cite{Figueroa-OFarrill:2000lcd,Alvarez:2003fw,Moore-minicourse}.

The difference between $H^{*}(M)$ and $H^{*}(M')$ is measured by the relative cohomology $H^*(M,M')$ which may be understood by a long exact sequence, given the embedding $i:M'\to M$,
\begin{equation}
\label{eq:long-exact}
    \begin{tikzcd}
        \dots\arrow[r,"\zeta^{*}"]& H^{n}(M)\arrow[r, "i^*"]&H^{n}(M')\arrow[r,"\eta^{*}"]& H^{n+1}(M,M')\arrow[r,"\zeta^{*}"]&H^{n+1}(M)\arrow[r, "i^*"]&\dots
    \end{tikzcd}
\end{equation}
Details of relative cohomology and the long exact sequence above are given in appendix \ref{app:Cohomologies_for_sources}, where for simplicity\footnote{The analysis is easily extended to the full RR Bianchi identity by considering $\dd_H$- rather than $\dd$-cohomology, or alternatively including an $H^{n=1}(M)$ term in the $j_{n+1}$ that cancels the contribution from the Chern--Simons term $H\wedge F_{n-3}$.} we focus a Bianchi identity of the form $\dd F_n=j_{n+1}$. The key point is that the difference between the cohomology of $M$ and $M'$ is given by
\begin{equation}
    \frac{H^{n}(M')}{i^{*}H^{n}(M)} = \im \eta^{*} \sim \text{deformations of sources } \{\delta j_{n+1} = \dd\gamma\}
\end{equation}
The first equality follows from exactness of the sequence \eqref{eq:long-exact}, while the second relation is shown in appendix \ref{app:Cohomologies_for_sources}. In this paper, we will be interested in the bulk moduli only, i.e. moduli which leave the sources unchanged, and hence it seems reasonable that we can safely work with the cohomology of the original, unexcised manifold $H^{*}(M)$. While our results hold in generality, we will mostly assume that we are working with an $M$ which admits a Calabi--Yau metric so we understand its cohomology. We should bear in mind, however, that the additional terms in $H^{*}(M')$ may have an impact on the moduli by changing the higher pages of the spectral sequences we analyse in sections \ref{sec:IIB_moduli} and \ref{sec:IIA_moduli}. A related point is that, if a flux $F_{n}$ is sourced, then $i^{*}F_{n}\in H^{n}(M')$ has components not in $i^{*}H^{n}(M)$. These additional components may stabilise more moduli than we might naively expect from performing the calculation with respect to $H^{n}(M)$.\footnote{We make more comments about this, along with a toy example for how extra moduli can be stabilised through components in $H^{n}(M')$ in the conclusions.} Indeed, this precise mechanism plays a role in moduli stabilisation in heterotic M-theory \cite{Smith:2022baw}. We will not explore those consequences here and leave it for future work, noting only that for a more accurate count, one should work with the cohomology of $M'$.

We will briefly note that, when the sources contain an orientifold plane, the massless moduli, and hence the associated cohomology, should be restricted to those with definite parity under the involution $\sigma$. Whether we restrict to the moduli which are even or odd depends on the particular bosonic field and the orientifold plane we introduce (details can be found in \cite{Grana:2005jc}). Throughout this paper, whenever we deal with backgrounds with orientifold planes, we will leave this reduction implicit and the cohomology groups we write down should be understood to be those with the correct transformation under $\sigma$.

Finally, we remark that integrating \eqref{eq:bianchi} over various cycles in the manifold gives a bound on the number of orientifold planes in terms of the number of branes and the number of units of flux passing through that cycle. Careful consideration of these conditions in the IIB case led to the tadpole conjecture \cite{Bena:2020xrh}, which states that when a Calabi--Yau has a large number of complex structure moduli, one cannot introduce flux to stabilise all of the moduli without violating the tadpole bound. We will mostly ignore these conditions and simply acknowledge that one should impose the tadpole cancellation conditions on top of the our results to obtain a physical background, although we will make some comments on this in section \ref{sec:corrections_and_obstructions} when we discuss higher order deformations and obstructions.

\section{Supersymmetric backgrounds in generalised geometry}\label{sec:EGG_SUSY_Backgrounds}

Generalised geometry is a powerful tool for studying the properties of generic flux backgrounds of string theory and M-theory. The geometry of generic flux compactifications to four-dimensional Minkowski preserving $\mathcal{N}=1$ supersymmetry were studied in the context of $\ExR{7(7)}$ generalised geometry in \cite{PiresPacheco:2008qik,Coimbra:2014uxa,Coimbra:2016ydd,Ashmore:2019qii}. These compactifications are described by global $\SU{7}\subset \ExR{7(7)}$ structures, the Killing spinor equations, namely the vanishing of the fermionic variations~\eqref{eq:susy-var}, are solved if and only if this structure is integrable.  Here integrability is the vanishing of the (generalised) intrinsic torsion of the $\SU{7}$ structure. Following \cite{Ashmore:2019qii}, we will first review how the global $\SU{7}$ structure can be described in terms of global generalised tensor fields. We will then specialise this to the case where the background admits a conventional $\SU{3}$ structure. Lifting the $\SU{3}$ structure to a generalised $\SU{7}$ structure in this way will then allow us to give an accurate description of the moduli space, and make arguments for the form of perturbative corrections coming from string theory. For a review of $\ExR{7(7)}$ generalised geometry see e.g. \cite{Coimbra:2011ky}. Important formulae in generalised geometry are given in appendix \ref{app:GG}.

\subsection{$\mathcal{N}=1$ backgrounds and exceptional complex structures}\label{subsec:EGG-intro}

It is an old idea that, if we consider a project space $M_{10}=\bbR^{3,1}\times M$ (or more generally if the background admits a local product structure), then, by dropping the manifest local Lorentz symmetry, the ten-dimensional supergravity can be viewed as a four-dimensional supergravity theory with an infinite number of supermultiplets~\cite{deWit:1986mz}. This is made explicit in the reformulation of supergravity using exceptional generalised geometry and exceptional field theory~\cite{PiresPacheco:2008qik,Coimbra:2011nw,Hohm:2013pua,Hohm:2013vpa,Hohm:2013uia}. The local product structure means one can decompose the supergravity fields into $\Spin{3,1}\times\Spin{6}$ representations. Remarkably the $\Spin6$ group then be enhance to $\SU8$, and the degrees of freedom can be arranged into a spin-two $\mathcal{N}=8$ supermultiplet, where the fields depend on all ten coordinates, or equivalently, if one expands in Kaluza--Klein modes, an infinite number of four-dimensional $\mathcal{N}=8$ multiplets. If $\mu$ and $m$ denote $\SO{3,1}$ and $\SO6$ vector indices respectively, the components of the supermultiplet are explicitly
\begin{equation}
\label{eq:N=8multiplet}
\begin{aligned}
    \text{spin-2:} & && g_{\mu\nu} && \rep{1} \\
    \text{spin-$\tfrac32$:} & && \psi^\pm_{\mu} && \rep{8} \\
    \text{spin-1:} & && g_\mu{}^m, B_{\mu m}, \tilde{B}_{\mu m_1\dots m_5}, \tilde{g}_{\mu m,m_1\dots m_6}, C_{\mu m_1\dots m_{p-1}} && \rep{28} \oplus \rep{\Bar{28}} \\
    \text{spin-$\tfrac12$:} & && \psi^\pm_m && \rep{56} \\
    \text{spin-0:} & && g_{mn},\varphi,\tilde{B}_{m_1\dots m_6},C_{m_1\dots m_p} && \rep{35}\oplus\rep{\Bar{35}} 
\end{aligned}
\end{equation}
where the spin refers to the $\Spin{3,1}$ representation, the last line is the $\SU8$ representation,  $\tilde{B}$ is the six-form potential dual to $B$, and $p$ is odd for type IIA and even for type IIB. (The formalism naturally gives the spin-1 fields together with their magnetic duals and so also includes $\tilde{g}$, a vector potential defined to be dual to $g_\mu{}^m$ and transforming in $T^*M\otimes\det T^*M$.) The bosonic fields further admit a natural action of an $\Ex{7(7)}\times\bbR^+$ structure group, with the spin-1 vectors transforming in the $\rep{56}$ representation and the spin-0 scalars defining a generalised metric that parameterises an $\Ex{7(7)}/(\SU8\times\bbZ_2)$ coset space. 
 
In this paper, we want to focus on backgrounds admitting four-dimensional $\mathcal{N}=1$ supersymmetry. Hence we consider a product space $M_{10}=\bbR^{3,1}\times M$  and then pick out preferred spinors on $M$. Selecting four of the 32 supercharges with respect we take a decomposition of the form~\eqref{def:IIA_10D_Killing_spinor} and~\eqref{def:IIB_10D_Killing_spinor} which determines a pair of chiral $\Spin6$ spinors $\epsilon^\pm$. In the case that the internal manifold has an $\SU{3}$ structure~\eqref{def:functions_a_and_b}, $\epsilon^+$ and $\epsilon^-$ are proportional to the same spinor $\eta$ selected by the $\SU{3}$. More generally the two spinors may be distinct and one or other may even vanish on the manifold, in which case the background may not have a description in terms of conventional global $G$-structures. While the individual spinors $\epsilon^\pm$ may not define a conventional $G$-structure, one can construct a combined eight-component object
\begin{equation}
    \zeta = \begin{pmatrix}
        \epsilon^+ \\ \epsilon^{c\,-}
    \end{pmatrix} \quad \text{(IIA)} \qquad \text{and} \qquad
    \zeta = \begin{pmatrix}
        \epsilon^+ \\ \epsilon^{-}
    \end{pmatrix} \quad \text{(IIB)}
    \end{equation}
Supersymmetry implies that the $\SU8$ norm is given by
\begin{equation}
    \zeta^\dag \zeta = \ee^\Delta ,
\end{equation}
(one can check this explicitly for $\SU3$ structures from~\eqref{eq:spinor-norm}) and so never vanishes (since the warp factor cannot vanish). Thus $\zeta$ defines a global \emph{generalised} $\SU7 \subset \SU8 \subset \ExR{7(7)}$-structure. The representations in~\eqref{eq:N=8multiplet} decompose under $\SU7\subset\SU8$ as  
\begin{equation}
    \rep8 = \rep1 \oplus \rep7 , \qquad
    \rep{28} = \rep7 \oplus \rep{21} , \qquad
    \rep{56} = \rep{21} \oplus \rep{35} , \qquad
    \rep{35} = \rep{35} .
\end{equation}
The $\mathcal{N}=8$ multiplet then decomposes into $\mathcal{N}=1$ multiplets with at gravity multiplet in the $\rep{1}$ representation of $\SU7$, a set of spin-$\frac32$ multiplets in the $\rep{7}$, vectors multiplets in $\rep{21}$ and chiral multiplets in $\rep{35}$. 

The scalar fields in the chiral multiplets parameterise the $\SU7\subset\ExR{7(7)}$ structure as follows. In generalised geometry, conventional tensors are packaged together to form representations of a larger generalised structure group, in this case $\ExR{7(7)}$. Examples of such generalised tensor bundles are given in Appendix~\ref{app:GG}. Given a global $G$-structure, with $G\subset\ExR{7(7)}$ one can decompose any generalised tensor bundle into irreducible representations of $G$. Bundles transforming in the singlet representation admit globally non-vanishing sections which can be used to define the $G$-structure, provided they are not also a singlet of any $G'\supset G$. For $G=\SU{7}$, the structure is equivalent to the existence of a globally non-vanishing complex tensor
\begin{equation}
    \psi \in \Gamma(\tilde{K}_{\bbC}) \qquad \ext^{3}E_{\bbC} \supset\tilde{K}_{\bbC} \sim \rep{912}_{\rep{3}} ,
\end{equation}
where (see appendix~\ref{app:GG}) 
\begin{equation}
    \tilde{K} = \begin{cases}
        \bbR\oplus \ext^{2}T^{*} \oplus  \ext^3 T^*\oplus\dots & \text{type IIA} \\
        T^{*} \oplus (S\otimes \ext^{3}T^{*}) \oplus \dots & \text{type IIB}
        \end{cases}
\end{equation}
and $\rep{912}$ denotes the $\ExR{7(7)}$ representation the fibres of $\tilde{K}_{\bbC}$ transform in, with the subscript denoting the weight under $\bbR^{+}$. Similar to analysis of supersymmetric backgrounds in conventional geometry, this generalised tensor can be expressed as a bilinear in the internal spinors $\zeta$. Any section of $\tilde{K}_{\bbC}$ can be written in $\SU{8}$ indices as
\begin{equation}
    \kappa = (\kappa^{\alpha\beta},\kappa^{\alpha\beta\gamma}{}_{\delta},\bar{\kappa}_{\alpha\beta},\bar{\kappa}_{\alpha\beta\gamma}{}^{\delta})
\end{equation}
The complex field $\psi$ defining the $\SU{7}$ structure then takes the form 
\begin{equation}\label{eq:psi_spinor_bilinear}
    \psi^{\alpha\beta} = \zeta^{\alpha}\zeta^{\beta} \qquad \psi^{\alpha\beta\gamma}{}_{\delta} = \bar{\psi}_{\alpha\beta} = \bar{\psi}_{\alpha\beta\gamma}{}^{\delta} = 0
\end{equation}
where $\vol_{G}$ is the $\ExR{7(7)}$ invariant volume form and $\lambda$ is some non-zero complex number. Note that if we decompose $\tilde{K}_{\bbC}$ into natural bundles we find for, e.g. IIA 
\begin{align}\label{eq:K_decomp}
\begin{array}{rcccccccc}
    \tilde{K}_{\bbC} & \simeq& \bbC& \oplus& \ext^{2}T^{*}_\bbC& \oplus& \ext^{3}T^{*}_\bbC &\oplus& ... \\
    \psi & \sim& \bar\epsilon^-\epsilon^+ &+& \bar\epsilon^- \gamma_{(2)} \epsilon^+ &+& 
         (\epsilon^{+T}\gamma_{(3)}\epsilon^+ + \epsilon^{-T}\gamma_{(3)}\epsilon^-) &+& ...
\end{array}
\end{align}
Hence, we see that $\psi$ contains all of the information about the conventional spinor bilinears. Similar results hold for M-theory and IIB backgrounds as well, with a different decomposition in \eqref{eq:K_decomp}. The $\SU8$ subgroup is picked out by a choice of generalised metric, which encodes $g$, $B$, $\tilde{B}$, $C_p$ and dilaton $\varphi$ and ``dresses'' the spinor bilinear, so that $\psi$ determines all the supergravity fields on $M$ as well as all the bilinears. The key observation here is that, while individually the spinor bilinears constructed from $\epsilon^\pm$ need not give a well-defined $G$-structure, when collected together in this form they always give a well-defined \emph{generalised} $G$-structure. 

We note that this formulation of supersymmetric backgrounds provides an interesting classification called \emph{type}~\cite{Tennyson:2021qwl,Smith:2022baw}. The type of the background given by $\psi$ is defined to be the form degree of the first non-vanishing component under the decomposition~\eqref{eq:K_decomp}. Thus the IIA structures can be type 0, 2 or 3. Note that for the A solutions in table~\ref{table:IIA_backgrounds}, either $\epsilon^+$ or $\epsilon^-$ vanishes and so they are of type 3. We see that, in contrast to the fact that these solutions do not have a description in terms of pure spinors in $\Orth{6,6}$ generalised geometry, by going to exceptional generalised geometry we get a uniform treatment of all $\mathcal{N}=1$ backgrounds. 

In summary, all the supergravity fields on $M$ and the spinor bilinears are encoded in $\SU7$ structure $\psi$, so that, for example in the special case of an $\SU3$-structure
\begin{equation}
    \psi \qquad \Longleftrightarrow \qquad \{ \omega, \Omega, a, b, \varphi, B, \tilde{B},C_p \} , 
\end{equation}
where $a$ and $b$ are the scalars in~\eqref{def:functions_a_and_b}. Depending on the values of these scalars we get different types of structures, as we will see explicitly in the next section. Furthermore, from the $\mathcal{N}=1$ perspective, $\psi$ parameterises the complex scalar fields in an (infinite) set of chiral multiplets. 

We can then also identify the usual set of objects -- Kähler potential, superpotential and D-terms -- the enter in the $\mathcal{N}=1$ action. The $\SU7$ structures live in a subspace $\psi\in \mathcal{Z} \subset \Gamma(\tilde{K}_{\bbC})$, which can be viewed as the space of sections $\Gamma(\mathcal{Q}_{\SU7})$, where $\mathcal{Q}_{\SU7}$ is the bundle
\begin{equation}\label{eq:Q_bundle}
    \frac{E_{7(7)}\times \bbR^{+}}{\SU{7}} \longrightarrow \mathcal{Q}_{\SU7} \longrightarrow M \, .
\end{equation}
We can then use the following
\begin{equation}\label{eq:sym_space}
    \frac{E_{7(7)}}{\SU{7}} \quad  = \quad  \underbrace{\frac{E_{7(7)}}{\SU{8}}}_{\text{scalars}} \quad \times \underbrace{\frac{\SU{8}}{\SU{7}}}_{\text{supersymmetry}}
\end{equation}
to see that, as discussed above, the space $\mathcal{Z}$ parameterises the scalar fields in the effective theory, along with the space of possible spinors $\zeta$ defining the supersymmetry.\footnote{The additional $\bbR^{+}$ factor in \eqref{eq:Q_bundle} corresponds to conformal rescalings of $\zeta$ which can be absorbed into the warp factor of the four-dimensional metric. Note also that, under $\SU7$, deformations of the scalars transform in the $\rep{35}$ representation and deformations of $\zeta$ in $\rep{7}\oplus\rep{1}$. In the full theory, these extra components are non-physical in that they are be removed by local $\SU8$ transformations. However, both they and the $\bbR^+$ factor are specified once one fixes a particular supersymmetric background. This is exactly analogous to the fact the a Calabi--Yau background specifies more information then simply the physical metric.} The space of structures admits a formal Kähler structure since it is parameterising the space of chiral multiplets. The complex structure is selected uniquely by the off-shell $\mathcal{N}=1$ supersymmetry in four dimensions and simply corresponds to identifying $\psi$ as a complex coordinate on $\mathcal{Z}$. The metric on $\mathcal{Z}$ is inherited from the symmetric space metric on \eqref{eq:sym_space}. In particular, we can write the Kähler potential on $\mathcal{Z}$ as 
\begin{equation}\label{eq:Kahler_potential}
    K = \int_{M} (\ii\, s(\psi,\bar{\psi}))^{1/3}
\end{equation}
where $s$ is the natural extension of the symplectic invariant of $E_{7(7)}$ to $\ext^{3}E_{\bbC}$ (see appendix \ref{app:GG} for more details). In terms of rewriting the two-derivative ten-dimensional theory in a four-dimensional $\mathcal{N}=1$ language, this is the Kähler potential which controls the four-dimensional kinetic terms of chiral multiplet scalars. 

The rewriting means we should also be able to define a superpotential for the chiral fields. It turns out that this is equal to the singlet part of the intrinsic torsion for the $\SU{7}$ structure defined by $\psi$~\cite{Ashmore:2019qii}. More explicitly, up to an overall constant, we can write
\begin{equation}\label{eq:superpotential}
    W \propto \int_{M} \ii \, s(\psi, T) = \int_{M} \frac{ s(\bar{\psi}, (D\times_{\textrm{ad}}\psi)\cdot \psi) }{ s(\bar{\psi},\psi) }
\end{equation}
where $D$ is any generalised connection compatible with the $\SU{8} \supset \SU{7}$ structure. As we will see in the next section, given a conventional $\SU3$ structure there are different ways it can define a generalised $\SU7$ structure, which correspond to the different A/B/C-types in tables~\ref{table:IIA_backgrounds} and~\ref{table:IIB_backgrounds}. Recall that provide neither $a$ or $b$ vanish, the background is a special case of the GMPT pure-spinor geometries~\cite{Grana:2004bg,Grana:2005sn}. For this latter class the superpotential~\eqref{eq:superpotential} takes the form~\cite{Ashmore:2019qii}
\begin{equation}
    W \propto \int_M \left< \Phi_{\pm},F-8\ii\dd (\ee^{-3\Delta}\im \Phi_{\mp})\right>,
\end{equation}
where $\Phi^\pm$ are the pure spinors and $\langle\,,\,\rangle$ is the Mukai pairing, in agreement with~\cite{Grana:2005ny,Koerber:2007xk,Koerber:2008sx}. Specialising further to the different $\SU3$-structure types gives
\begin{align}
    W_{\text{IIA, O6}} & \propto \int_{M}\ee^{3\Delta-\dil} \left[ 
        \ee^{-\dil}\left(H_3 + \ii \dd\omega\right)\wedge\im (\ee^{-\ii\beta}\Omega)
        +\ii\tilde{F}_{\text{IIA}}\wedge\ee^{B+\ii\omega} \right] , \label{eq:W_IIA_O6-full} && \text{(BC-type)}\\
    W_{\text{IIB, O3/O7}} & \propto \int_{M}\ee^{3\Delta-\dil}\, G_{3}\wedge\Omega , && \text{(B-type)} \label{eq:W_IIB_O3/O7-full}\\
    W_{\text{IIB, O5/O9}} & \propto \int_{M}\ee^{3\Delta-\dil}\bigl(\tilde{F}_{3}+\ii\ee^{-\dil}\dd\omega\bigr) \wedge\Omega, && \text{(C-type)} \label{eq:W_IIB_O5/O9-full} 
\end{align}
giving the ``torsionful'' generalisations of the flux superpotentials~\eqref{eq:W_IIA_O6}--\eqref{eq:W_IIB_O5/O9}, where the $\SU3$ structure is no longer assumed to be Calabi--Yau (including the overall warp-factor and dilaton dependence). If $a$ or $b$ vanish then we are outside the GMPT ansatz. However the general expression~\eqref{eq:superpotential} still holds and, on substituting for the $\SU3$ structure one finds the Hull--Strominger-like superpotential~\cite{Becker:2003yv,LopesCardoso:2003dvb,Gurrieri:2004dt,Benmachiche:2006df,delaOssa:2015maa,McOrist:2016cfl} (that also applies to the heterotic string) 
\begin{align}
\label{eq:W_HS}
    W_{\text{HS}} &\propto \int_{M}\ee^{3\Delta}
        \ee^{-2\dil}\left(H_3+\ii\dd\omega\right)\wedge\Omega , && \text{(A-type)}
\end{align}
Thus~\eqref{eq:superpotential} is a universal form that captures each of the familiar Gukov--Vafa--Witten-type superpotentials as special cases of $\SU7$ structures. Furthermore, it is important to note that it is also not restricted to a function of the Calabi--Yau moduli (as was the case in section~\ref{subsec:superpotential_stab}) but rather is a function of completely generic structures and internal fields $\{ \omega, \Omega, a, b, \varphi, B, C_p \}$. 

The final ingredient in the rewriting is to identify the set of gauge fields and $D$-terms. It turns out that the gauge group is precisely the ``generalised diffeomorphism'' group $\GDiff$, which is the group of combining conventional diffeomorphism with form-field gauge transformation. Infinitesimally, $\GDiff$ transformations are parameterised by a generalised vector $V\in\Gamma(E)$ which is a combination of vector and tensor fields transforming in the $\rep{56}$ representation of $\ExR{7(7)}$, as given in~\eqref{A:IIA-vector-bundle} and~\eqref{A:IIB-vector-bundle}, and are generated by the corresponding generalised Lie derivative (or Dorfman derivative) $L_V$. Thus set get an infinite set of vector multiplet fields $A_\mu$ which are sections of $E$. Furthermore, $\GDiff$ preserves the Kähler structure on $\mathcal{Z}$, and for each gauge transformation $V\in\Gamma(E)$ we get a corresponding D-term
\begin{equation}
    D \sim \mu(V):= \tfrac{1}{3}\int_{M} \ii\, s(L_{V}\psi,\bar{\psi}) 
        (\ii s(\psi,\bar{\psi}))^{-2/3} .
\end{equation}
This somewhat messy expression is, as always for D-terms, none other than the moment map for the action of $\GDiff$ on the space of structures $\mathcal{Z}$ \cite{Ashmore:2019qii}. As for the superpotential~\eqref{eq:superpotential}, on assuming we have a conventional $\SU3$-structure background, defining a particular class of $\SU7$ structure, the moment map reduces to the standard D-term expressions~\cite{Grimm:2004uq,Koerber:2007xk}. 

The structure of the $\mathcal{N}=1$ reformulation is extremely useful for understanding the conditions for supersymmetry. From the four-dimensional perspective the superpotential and Kähler potential, together with any $D$-terms, determine the potential for the scalar fields $\psi$ as in \eqref{eq:4d-EFT-potential}. This potential is the ``internal'' part of the theory, that is the part of the ten-dimensional Lagrangian that contains no four-dimensional derivative, integrated over $M$. The conditions for supersymmetry should then correspond to 
\begin{equation}
\begin{aligned}
    \text{F-terms:} \qquad && \delta W/\delta \psi &= 0, \\
    \text{D-terms:} \qquad && \mu &= 0 ,
\end{aligned}
\end{equation}
(since $W$ is homogeneous, $W=0$ follows from $\delta W/\delta\psi=0$). These conditions give differential constraints on $\psi$ that are equivalent to the Killing spinor equations, and to integrability of the $\SU{7}$ structure defined by $\psi$, i.e. the existence of a torsion-free compatible generalised connection~\cite{Ashmore:2019qii}.

For the F-terms, instead of writing the differential constraints directly in terms of $\psi$, as was noted in \cite{Ashmore:2019qii} it is more convenient to first consider a slightly weaker structure. Instead of the $\SU{7}$ structure, one considers a $\Uni{7}\times \bbR^{+}$ structure which can be defined by generalised tensor $J \in \ad \tilde{F}$ transforming in the adjoint representation of $\Ex{7(7)}$. This defines a decomposition of the (complexified) generalised tangent bundle $E_{\bbC}$ as
\begin{equation}
\begin{aligned}
    E_\bbC &= L_{+3}\oplus L_{+1} \oplus L_{-1} \oplus L_{-3} . \\
    \rep{56} &= \repsub{7}{3} \oplus \repsub{\Bar{21}}{1} \oplus \repsub{\Bar{7}}{-1} \oplus \repsub{21}{-3} ,
\end{aligned}
\end{equation}
where $L_{n}$ is the $n\ii$-eigenbundle  under the action of $J$, $L_{-n} = \overline{L_{n}}$, and in the second line we give the decomposition under $\SU7$. The important bundle is $L_{3}$ since a choice of $L_{3}$ is equivalent to a choice of $J$, provided $L_{3}$ satisfies certain algebraic constraints. The weaker $\Uni{7}\times \bbR^{+}$ structure is related to the full supersymmetric $\SU{7}$ structure by
\begin{equation}\label{eq:SU(7)_vs_U(7)}
    V\bullet \psi = 0 \quad \Leftrightarrow \quad V \in \Gamma(L_{3})
\end{equation}
where $\bullet: \tilde{K}\otimes E\to C$ is a projection map to the generalised bundle $C$ transforming in the $\rep{8645}_{\rep{4}}$ representation of $\ExR{7(7)}$.  One can use this equation to find $L_{3}$ uniquely from $\psi$, or $\psi$ up to a scale from $L_{3}$. This also provides an alternative definition of the type of a background as the codimension of $a(L_{3})$, where $a:E\rightarrow TM$ is the anchor map.

Since the structure defined by $L_{3}$, or equivalently $J$, is a strictly weaker structure than the $\SU{7}$ structure, the integrability of the latter should be expressible in terms of integrability of the $\Uni{7}\times \bbR^{+}$ structure along with some additional constraints. One finds 
\begin{equation}\label{eq:integrable SU(7)}
    \Uni{7}\times \bbR^{+} \text{ structure integrable} \qquad \Leftrightarrow \qquad L_{V}W \in \Gamma(L_{3}) \quad \forall \, V,W \in \Gamma(L_{3}) ,
\end{equation}
that is, the $\Uni{7}\times \bbR^{+}$ structure is integrable if and only if the vector bundle $L_{3}$ is involutive with respect to Dorfman derivative. This structure is analogous to complex structures in conventional geometry,\footnote{Or indeed generalised complex structures in $\Orth{d,d}$ generalised geometry.} with $L_{3}$ playing the role of $T^{1,0}$ and $\psi$ playing the role of $\Omega$, and hence the structure was called an \emph{exceptional complex structure}. Furthermore, integrability of $J$ is implied by the F-term conditions $\delta W/\delta\psi=0$, although the latter is a slightly stronger condition. Integrability of the full $\SU{7}$ structure, which implies that the supergravity background preserves at least $\mathcal{N}=1$ supersymmetry, is equivalent to
\begin{equation}\label{eq:SU(7)_int}
    \Uni{7}\times \bbR^{+} \text{ integrable} \qquad \text{and} \qquad \mu = 0 , 
\end{equation}
that is, imposing, in addition to involutivity, the D-term conditions.

It is instructive to understand how the different F- and D-term conditions for integrability are related to the intrinsic torsion of the $\SU7$ structure. The latter decomposes into $\SU{7}\times\Uni1$ representations as
\begin{align}
    W^{\text{int}}_{\SU{7}} \sim \repsub{1}{-7}\oplus \repsub{\Bar{7}}{-3}\oplus \repsub{21}{-1}\oplus \repsub{35}{-5}\oplus c.c. 
\end{align}
Involutivity of the $L_3$ bundle is necessary and sufficient to set the $\repsub{1}{-7}\oplus \repsub{35}{-5}$ components of $W^{\text{int}}_{\SU{7}}$ to zero. The full F-terms $\delta W/\delta \psi=0$ are slightly stronger setting $\repsub{1}{-7}\oplus \repsub{\bar{7}}{-3} \oplus \repsub{35}{-5}$ components to zero. Finally the vanishing of a moment map sets the $\repsub{\bar{7}}{-3}\oplus \repsub{21}{-1}$ components to zero.\footnote{The sharing of the $\repsub{\bar{7}}{-3}$ component between the F- and D-term conditions, is a consequence of the spin-$frac32$ multiplets in the $\mathcal{N}=1$ formulation.}. 

In summary, we have reviewed how one can reformulate the conditions for a four-dimensional supersymmetric background in terms of F- and D-terms for a infinite-dimensional space $\mathcal{Z}$ of chiral fields. A point in $\mathcal{Z}$ correspond to a generalised $\SU7$ structure $\psi$, which completely determines the supergravity fields on the compact space $M$. Supersymmetry imposes involutivity and the moment map constraints on $\psi$, corresponding to 
\begin{equation}
    \begin{array}{ccc}
        \text{F-terms} & \quad \sim \quad & L_{3} \text{ Involutivity} \\
         \text{D-terms} & \sim & \mu = 0
    \end{array}
\end{equation}
We will now turn to how these conditions appear in each of the different classes $\SU3$ structure backgrounds.

\subsection{$\SU{3}$ structure backgrounds as exceptional complex structures}\label{subsec:ECS_SU(3)}

As we saw in section \ref{sec:SU3_structure_review}, there are different  families supersymmetric flux backgrounds with conventional $\SU3$ structure, depending on whether we are in type IIB or IIA, what fluxes are turned on, and two complex functions $a,b$ which define how the internal spinor $\epsilon$ defining the $\SU{3}$ structure uplifts to a ten-dimensional spinor. From our previous discussion, they can all however be described by an exceptional complex structure, with the complicated differential constraints translating into involutivity and vanishing moment map. 

In this section, we will show how to write the generalised tensor $\psi$ and the associated vector bundle $L_3$, which define the exceptional complex structure, in terms of the $\SU{3}$ tensors $\omega, \Omega$, along with the complex functions $a,b$. We will also see explicitly how involutivity recovers a subset of the supersymmetry conditions given in tables \ref{table:IIA_backgrounds} and \ref{table:IIB_backgrounds}.

\subsubsection{$\SU{3}$ structure backgrounds in IIB}

We will start by constructing generic $\SU{3}$ structure backgrounds in type IIB in the language of exceptional complex structures. That is, we will construct $\psi \in \Gamma(\tilde{K}_{\bbC})$, and the associated $L_{3}$, such that the algebraic constraints set out in \cite{Ashmore:2019qii} are satisfied. Decomposing the bundle $\tilde{K}_{\bbC}$ under the IIB structure group $\GL{6,\bbR}\times \SL{2,\bbR}$, we find
\begin{equation}
    \tilde{K}_{\bbC} = T^{*} \oplus S\otimes \ext^{3}T^{*} \oplus \dots
\end{equation}
where $S$ is the $\SL{2,\bbR}$ doublet. This implies that the \emph{type} of an exceptional complex structure in IIB, as defined above, can only be 1 or 3. A type 1 structure implies the existence of a 1-form on the manifold and hence a further reduced structure group. A genuine $\SU{3}$ structure background can therefore only be type 3.

The most general type 3 structure built from $\omega$, $\Omega$, $a$ and $b$ has the form 
\begin{equation}\label{eq:IIB_ECS}
    \psi = \ee^{2\Delta-\dil}\ee^{\Sigma}\cdot s^{i}\tfrac{1}{8}\Omega \qquad \Rightarrow \qquad L_{3} = \ee^{\Sigma}\cdot [T^{0,1}\oplus s^{i}T^{*1,0}\oplus \ext^{3,0}T^{*}]
\end{equation}
where
\begin{equation}\label{eq:IIB_twist}
    \Sigma = r^{i}\omega + \tfrac{1}{2}\alpha \omega^{2} + \tfrac{1}{6}c^{i}\omega^{3}, 
\end{equation}
and $\ee^\Sigma$ is the adjoint action of $\Sigma$ viewed as an element of the Lie algebra $\ex{7,\bbC}$. The $r^{i},s^{i},c^{i}$ are complex vectors in the $\SL{2,\bbR}$ doublet, and $\alpha$ is some complex function which all depend on $a$, $b$ and the dilaton $\dil$. In general, $\psi$ will also have dependence on the form-field potentials $B$, $\tilde{B}$ and $C_p$. However, we will use a formulation where we always ``untwist'' by the potentials (via an $\Ex{7(7)}$ action) and instead modify the Dorfman derivative $L_V$ to one with explicit flux terms denoted $L^F_V$. (For details, see appendix~\ref{app:GG}.)

The complex vectors and functions can be found by expanding the spinor bilinear \eqref{eq:psi_spinor_bilinear} and using the redundancy in the definition of $\Sigma$ to put them into a convenient form.\footnote{Note that in defining $L = \ee^{\Sigma}\cdot \tilde{L}$, the twist $\Sigma$ is only defined up to $\mathrm{Stab}(\tilde{L})$. } We find that
\begin{equation}
    s^{i} = \left( \begin{array}{c}
        ab \\
        \tfrac{1}{2}\ee^{-\dil}(b^{2}-a^{2})
    \end{array} \right)
\end{equation}
while the other coefficients depend on whether $ab=0$ or not. In the case that $ab \neq 0$, so we are in the case studied in \cite{Grana:2005sn,Tomasiello:2007zq}, we find that
\begin{equation}\label{eq:twist_ab=/=0}
    r^{i} = \left( \begin{array}{c}
         0 \\
         \ii\ee^{-\dil}\frac{a^{2}+b^{2}}{2ab}
    \end{array} \right) \qquad \alpha = \ee^{-\dil} \frac{a^{2} - b^{2}}{2ab} \qquad c^{i} = \ii \ee^{-\dil} \frac{a^{2}+b^{2}}{2ab} \left( \begin{array}{c}
        1 \\
        \ee^{-\dil}\frac{b^{2}-a^{2}}{4ab}  
    \end{array} \right).
\end{equation}
In the case where $a=0$ or $b=0$, the expressions for $r^{i}$, $\alpha$ and $c^{i}$ are clearly singular. However, both $\psi$ and the bundle $L_{3}$ are regular in this limit. To see this, we note that
\begin{equation}
    \ee^{\Sigma}\cdot [T^{0,1}\oplus s^{i}T^{*1,0} \oplus \ext^{3,0}T^{*}] = \ee^{\Sigma}\ee^{k\,s^{i}\omega}\cdot [T^{0,1}\oplus s^{i}T^{*1,0} \oplus \ext^{3,0}T^{*}]
\end{equation}
for any function $k$. We can therefore equivalently define our structure using the twist
\begin{equation}\label{eq:change_of_twist}
\begin{split}
    \tilde{\Sigma}(k) &= (r^{i}+k\, s^{i})\omega + \tfrac{1}{2}(\alpha - k\,\epsilon_{ij}r^{i}s^{j}) \omega^{2} + \tfrac{1}{6}\tilde{c}^{i}\omega^{3} \\
    &= \tilde{r}^{i} \omega + \tfrac{1}{2}\tilde{\alpha}\omega^{2} + \tfrac{1}{6}\tilde{c}^{i} \omega^{3}
\end{split}
\end{equation}
Setting $k=\ii/ab$ we find
\begin{equation}
    \tilde{r}^{i} = \left( \begin{array}{c}
        i \\
        \ii\ee^{-\dil}\tfrac{b}{a}
    \end{array} \right) \qquad \tilde{\alpha} = -\ee^{-\dil}\tfrac{b}{a}  \qquad \tilde{c}^{i} = \ii\ee^{-\dil}\left( \begin{array}{c}
        \frac{3b}{2a} \\
        \ee^{-\dil}(\frac{b^{2}}{4a^{2}} - 1)
    \end{array} \right)
\end{equation}
which has a well-defined limit as $b\rightarrow 0$. In this limit we can take the exceptional complex structure as in \eqref{eq:IIB_ECS}, with now\footnote{We have dropped the tildes in this equation to allow for more uniform notation.}
\begin{equation}\label{eq:twist_ab=0}
    r^{i} = \left( \begin{array}{c}
        i \\
        0 
    \end{array} \right) \qquad \alpha = 0 \qquad c^{i} = \left( \begin{array}{c}
        0 \\
        -\ii\ee^{-2\dil}
    \end{array} \right).
\end{equation}
To obtain an expression which has a manifestly well-defined limit under $a\rightarrow 0$, one can take $k = -\ii/ab$.

Now that we know how $\SU{3}$ structure backgrounds in IIB can be expressed in this language, we would like to find the conditions that arise from involutivity of the structure \eqref{eq:IIB_ECS}, or equivalently the F-terms for the associated background. We will not assume any properties of the intrinsic torsion of the $\SU{3}$ structure or the flux a priori, but will instead see how involutivity constrains the background geometry. Taking generic sections $V,V'\in \Gamma(L_{3})$, we find that $L_{V}^{F}V' \in \Gamma(L_{3})$ if and only if
\begin{equation}\label{eq:IIB_involutivity}
    [T^{0,1},T^{0,1}] \subseteq T^{0,1}\, , \quad \CF_{(0,1)} = 0\, , \quad \CF^{i}_{(0,3)} = 0 \, , \quad \epsilon_{ij}s^{i}\CF^{j}_{(1,2)} = 0 \, , \quad \CF_{(2,3)} = 0
\end{equation}
where we define the complex differential polyform $\CF_{\text{IIB}}$, that arises from the (complex) twisting $\Sigma$, as combinations of real flux and geometry of the form\footnote{The non-standard normalisation of $\CF_1$ comes from shifting $\Sigma$ by a term in $\mathrm{Stab}(\tilde{L})$ is chosen to ensure that we have a well-defined $b\to0$ limit.}
\begin{align}
    \CF_{1} &= -F_{1}(\epsilon_{ij}s^{i} p^{j})^{2} + \epsilon_{ij}s^{i}\dd s^{j} \label{eq:IIB_CF_1}\\
    \CF^{i}_{3} &= F^{i} + \dd(r^{i} \omega) + p^{i}(\epsilon_{jk}r^{j}p^{k}) F_{1}\wedge \omega \label{eq:IIB_CF_3}\\
    \CF_{5} &= F_{5} + \tfrac{1}{2}\dd(\alpha\omega^{2}) \label{eq:IIB_CF_5} + \epsilon_{ij}r^{i}\omega\wedge F^{j} + \tfrac{1}{2}\epsilon_{ij}r^{i}\omega \wedge \dd(r^{j}\omega) + \tfrac{1}{2}(\epsilon_{ij}r^{i}p^{j})^{2}F_{1}\wedge \omega^{2}
\end{align}
We have used the notation $F^{i} = (H_{3}, F_{3})$ is the $\SL{2,\bbR}$ doublet of 3-form fluxes, $p^{i} = (0,-1)$ is a constant vector and the subscripts denote the real degree (or in~\eqref{eq:IIB_involutivity} the complex $(p,q)$-degree) of the differential form. The RR fluxes are the gauge-invariant RR fluxes, i.e. those satisfying $\dd F_{n} = H\wedge F_{n-3}$. In the definition of the type 3 structure~\eqref{eq:IIB_ECS}, there is a kernel for the action of $\Sigma$, meaning that only certain parts of the complex flux $\CF_{\text{IIB}}$ actually appear, and these are the ones that are constrained by~\eqref{eq:IIB_involutivity}. 

Note that the equations in \eqref{eq:IIB_involutivity} imply the background is always complex, consistent with $W_1=W_2=0$ in table~\ref{table:IIB_backgrounds}. Given that the involutive bundle is invariant under change of twist \eqref{eq:change_of_twist}, the involutivity conditions should be invariant under $r^{i}\rightarrow \tilde{r}^{i}$, $\alpha \rightarrow \tilde{\alpha}$, for arbitrary $k$. It is a quick calculation to verify that this is indeed the case. Using this fact, we can analyse how the involutivity conditions constrain the flux and intrinsic torsion of the background in the two cases $ab \neq 0$ and $ab=0$.

\paragraph{Case 1 -- $ab \neq 0$:} In this case, we can use the tensors $r^{i}, \alpha$ defined in \eqref{eq:twist_ab=/=0}. The involutivity conditions \eqref{eq:IIB_involutivity} then reduce to
\begin{align}
    [T^{0,1},T^{0,1}] & \subset T^{0,1} \label{eq:IIB_GMPT_invol_1} \\
    F_{(0,1)} - \delb \left( \tfrac{\ee^{-\dil}(a^{2}-b^{2})}{2ab} \right) &= 0 \\
    F_{(0,3)} = H_{(0,3)} &= 0 \\
    F_{(1,2)} - \ee^{-\dil}\left(\tfrac{b^{2}-a^{2}}{2ab} \right) H_{(1,2)} + \ii \delb \left(  \ee^{-\dil} \tfrac{a^{2}+b^{2}}{2ab} \omega \right)&= 0 \\
    F_{(2,3)} + \delb \left( \ee^{-\dil}\tfrac{a^{2}-b^{2}}{4ab}\omega^{2} \right) - \ii \ee^{-\dil} \tfrac{a^{2}+b^{2}}{2ab} \omega \wedge H_{(1,2)} &= 0 \label{eq:IIB_GMPT_invol_5}
\end{align}
Encouragingly, setting $a=\pm b$, we recover some of the equations that appear in column (C) of table \ref{table:IIB_backgrounds}, while setting $a=\pm \ii b$, we recover some of the equations in column (B). More generally, we recover the conditions for backgrounds in the column (ABC), reviewed in \cite{Grana:2005jc}.

\paragraph{Case 2 -- $ab=0$:} We will work with the case $b=0$ and use the twist parameters in \eqref{eq:twist_ab=0}.\footnote{The case $a=0$ works similarly, with the twist parameters given by \eqref{eq:change_of_twist} with $k = -\ii/ab$, and then taking the limit $a\rightarrow 0$.} The involutivity conditions \eqref{eq:IIB_involutivity} translate to the following.
\begin{align}
    [T^{0,1},T^{0,1}] &\subset T^{0,1} \\
    F_{(0,3)} = H_{(0,3)} &= 0 \\
    H_{(1,2)} + \ii \delb \omega &= 0 \\
    F_{(2,3)} + \ii F_{(1,2)} \wedge \omega - \tfrac{1}{2}F_{(0,1)} \wedge \omega^{2} &= 0
\end{align}
We see that these are consistent with the equations in column (A) of table \ref{table:IIB_backgrounds}.

\subsubsection{$\SU{3}$ structure backgrounds in IIA}

To find the possible exceptional complex structures which can be constructed from (conventional) $\SU{3}$ structures in IIA, we first decompose the bundle $\tilde{K}_{\bbC}$ under the IIA structure group $\GL{6,\bbR}\subset \GL{7,\bbR}$, we find
\begin{equation}
    \tilde{K}_{\bbC} = \bbC \oplus \ext^{2}T^{*} \oplus \ext^{3}T^{*} \oplus ...
\end{equation}
We see that, in this case, the structure can be type 0, 2 or 3. For type 2 structures, we require a 2-form on the manifold which again implies a further reduction of the structure group from $\SU{3}$. Genuine $\SU{3}$ structure backgrounds can therefore only be type 0 or type 3. We will start with the type 0 case.

The most general type 0 background we can construct using only the $\SU{3}$ tensors $\omega, \Omega$, two complex functions $a,b$, the dilaton and the warp factor is
\begin{equation}
\label{eq:IIA_ECS}
    \psi = \ee^{2\warp - \dil} \ee^{\Sigma}\cdot f \qquad \Rightarrow \qquad L_{3} = \ee^{\Sigma} \cdot (T\oplus \bbC)
\end{equation}
where
\begin{equation}\label{eq:IIA_twist}
    \Sigma = -\ii\omega + \alpha \Omega + \beta \bar{\Omega}
\end{equation}
The complex parameters $f,\alpha,\beta$ are functions that can be expressed in terms of $a,b,\dil$ as the following.
\begin{equation}
\label{eq:IIA-type0-para}
    f = \tfrac{1}{8}a\bar{b} \qquad \alpha = \ee^{-\dil}\tfrac{\bar{b}}{2a} \qquad \beta = - \ee^{-\dil}\tfrac{ a }{2\bar{b}}
\end{equation}

Checking involutivity of this bundle is involutive under the flux-deformed (massive) IIA Dorfman derivative, as given in \cite{Cassani:2016ncu} and reproduced in \ref{app:GG}, if and only if
\begin{equation}\label{eq:IIA_involutivity}
    \CH = 0, \quad \CF_4 = 0, \quad \CF_6 = 0
\end{equation}
where we define the complex differential polyform $\CF_{\text{IIA}}$, that arises from the (complex) twisting by $\Sigma$, as combinations of real flux and geometry of the form
\begin{equation}\label{eq:IIA_complex_fluxes_type-0}
    \begin{aligned}
        \CF_0 &= F_0 \\
        \CF_2 &= F_2 + \ii\omega F_0 \\
        \CH &= H + \ii\dd\omega \\
        \CF_4 &= F_{4} + \ii  \omega \wedge F_{2} - \tfrac{1}{2}\omega^{2}F_{0} +  \dd(\alpha\Omega +\beta \bar{\Omega})\\
        \CF_6 &= F_6 + \ii\omega\wedge F_4 - \tfrac{1}{2}\omega^2\wedge F_2  + \tfrac{1}{6}\ii\omega^3 F_0 - \CH \wedge (\alpha\Omega +\beta \bar{\Omega})
    \end{aligned}
\end{equation}
There is a kernel for the action of $\Sigma$ in~\eqref{eq:IIA_ECS}, so that only  $\CH$, $\CF_4$ and $\CF_6$ are constrained. , giving
\begin{align}
    H &= 0 \\
    \dd\omega &= 0 \\
    F_{4}
    -  \dd \left( \re\left(\ee^{-\dil} \tfrac{|a|^{2}-|b|^{2}}{ 2ab} \Omega \right) \right) &= 0 \\
    \omega \wedge F_{2} + \dd\left( \im\left( \ee^{-\dil}\tfrac{|a|^{2}+|b|^{2}}{ 2ab}\Omega \right) \right) &= 0 
\end{align}
and 
\begin{equation}
    F_0 = F_6 = \omega \wedge F_4 = \tfrac12 \omega^2\wedge F_2 = 0 ,
\end{equation}
where we have used the vanishing of $W_1$ that follows from $\dd\omega=0$ to simplify the expressions. For these type-0 backgrounds we see the involutivity condition reproduces a subset of the equations which appear in the (BC) column of table \ref{table:IIA_backgrounds}. Note in particular that they set $\CF_0=F_0=0$ and we are necessarily in the non-massive version of type IIA. 

We can ask what happens in the limit $a\rightarrow 0$ (the limit $b\rightarrow 0$ works similarly). In this limit, the coefficients in \eqref{eq:IIA_twist} go like $\alpha \rightarrow \infty$, $\beta \rightarrow 0$. This divergence of the twist factor indicates a change in type, and indeed one can show that in this limit the exceptional complex structure becomes
\begin{equation}\label{eq:IIA_ECS_type3}
    \psi = \ee^{2\warp-\dil}\ee^{\Sigma} \bar{b}^{2} \Omega \qquad \Rightarrow \qquad L_{3} = \ee^{\Sigma}\cdot\left( T^{0,1}\oplus \bbC \oplus \ext^{2,0}T^{*} \right)
\end{equation}
where
\begin{equation}
\label{eq:IIA_twist_3}
    \Sigma = -\ii\omega +\tfrac{1}{8}\ee^{-2\dil}\Omega\wedge \bar{\Omega}
\end{equation}
That is, the case $a=0$ (or $b=0$) is represented by a type 3 structure in IIA. The conditions for involutivity of the $L_{3}$ bundle are then
\begin{equation}
    \begin{aligned}
        [T^{0,1},T^{0,1}] \subset T^{0,1}, \quad \CH_{(1,2)}= \CH_{(0,3)} = 0 , \quad \CF_{(1,3)} = 0,\quad \CF_6 = 0
    \end{aligned}
\end{equation}
where the complex fluxes take the same form as in \eqref{eq:IIA_complex_fluxes_type-0}, except without the $\Omega, \bar{\Omega}$ terms. From these we get the constraints on the geometry:
\begin{align}
    [T^{0,1},T^{0,1}] & \subset T^{0,1} \label{eq:IIA_invol_condition_1}\\
    H_{(1,2)} + \ii \delb\omega &= 0 \label{eq:IIA_invol_condition_2}\\
    F_{(1,3)} + \ii \omega \wedge F_{(0,2)} &= 0 \label{eq:IIA_invol_condition_3}\\
    F_{6} - \tfrac{1}{2}\omega^{2}\wedge F_{2}  &= 0 \label{eq:IIA_invol_condition_4}\\
    \omega \wedge F_{4} - \tfrac{1}{6}\omega^{3}F_{0} &= 0.\label{eq:IIA_invol_condition_5}
\end{align}
These equations reproduce some of the conditions in the A~column of table \ref{table:IIA_backgrounds}, as expected. It is worth noting that while the IIA type 0 (BC~column) solutions and IIB type 3 backgrounds in the B, C and ABC columns, can all famously be described using pure spinors in the $\Orth{6,6}$ generalised geometry~\cite{Grana:2004bg,Grana:2005sn}, the IIA and IIB type 3 backgrounds in the A columns are outside this class. However, by going to $\ExR{7(7)}$ generalised geometry we get a uniform description of all $\mathcal{N}=1$ geometries.

\section{Exact moduli from generalised geometry}\label{sec:Moduli_Stabilisation}

We now turn to our central question of calculating the moduli of $\SU3$-structure flux backgrounds using the generalised structures we have just described. 

In the absence of flux the internal space metric has $\SU{3}$-holonomy, making it a Calabi--Yau manifold and the Kähler and complex moduli are counted by Dolbeault cohomology groups. As we saw summarised in tables~\ref{table:IIA_backgrounds} and~\ref{table:IIB_backgrounds}, the $\SU{3}$ structure metric solving the equations of motion may be profoundly changed by the inclusion of fluxes. This is particularly the case for IIA backgrounds, whose $\SU{3}$ structures may be non-Kähler ($W_3\neq 0$) or even non-complex ($W_2\neq 0$). A priori much less is known about such backgrounds and their moduli, when compared to the Calabi--Yau geometry. 

As reviewed in section \ref{subsec:superpotential_stab}, using the small-flux limit, one can get an approximate understanding of how the Calabi--Yau moduli are lifted by presence of flux by analysing the effective superpotential (and in some cases D-terms). However, this approximation breaks down if the flux is of order the Calabi--Yau curvature at any point on the manifold, so we cannot be sure that previously heavier moduli do not become massless in the presence of flux. In recent years there has been some work on understanding non-complex IIA backgrounds using new cohomology groups~\cite{Tseng:2011gv,Tseng:2009gr,Tseng:2010kt,Tsai:2014ela,Tanaka_2018}.However, despite this progress a full and direct analysis of the moduli of these non-Kähler and non-complex backgrounds is still to be completed.

As we will now discuss, we can derive the exact moduli from the exceptional complex structure. In particular, we will demonstrate that the result is (i) purely holomorphic, independent of the K\"ahler potential on the space of structures (section \ref{subsec:ECS-moduli}) and (ii) invariant under certain deformations of the background $\psi$ called $\GDiff_{\bbC}$ flows (section \ref{sec:GDiff_C_invariance}). The latter point is important because it means that we can calculate the exact moduli either on the fully supersymmetric geometry with all the fluxes back-reacted, or equivalently on a non-supersymmetric background without the full back-reaction but related to the actual solution by a $\GDiff_{\bbC}$ flows. In each case we will see that the moduli are calculated by a cohomology naturally defined by the exceptional complex structure that extends the de Rham or Dolbeault cohomology of the Calabi--Yau space. Under the extension, deformations that were previously closed can become non-closed, and ones that were previously non-exact can become exact. Physically, as we will see, the former correspond to moduli lifted by F-terms and the latter mooduli lifted by D-terms. This structure proves that the full calculations with finite flux in fact agree with the intuition of conventional moduli stabilisation that comes from the small-flux limit. In this section we discuss the general formalism and then in section \ref{sec:SU3_ECS_moduli} we calculate the exact moduli of all $\SU{3}$ structure backgrounds.

\subsection{Moduli space of exceptional complex structures}\label{subsec:ECS-moduli}

The description of moduli of generic $\mathcal{N}=1$ supersymmetric backgounds using exceptional complex structures was first given in~\cite{Ashmore:2019qii} are is equally applicable to M-theory or type II. Since any such background has a unique integrable $\SU{7}$ structure on the internal space, and any two backgrounds are equivalent if they are related by a combination of diffeomorphisms and guage transformations, we can identify the supersymmetric moduli space with
\begin{equation}
    \mathcal{M}_{\psi}:= \{\psi \in \mathcal{Z} \,|\, \psi \text{ an integrable } \SU{7} \text{ structure} \}\quotient \GDiff
\end{equation}
In identifying the physical moduli space, one needs to note that backgrounds defined by $\psi$ and $\psi'=\lambda\psi$ for some non-zero \textit{constant} $\lambda \in \bbC^{*}$ are physically equivalent, since constant rescaling of $\psi$ just corresponds to a constant rescaling of warp factor. Hence, the physical moduli space is given by\footnote{From \eqref{eq:Q_bundle} and \eqref{eq:sym_space}, the space $\mathcal{M}_{\psi}$ could also contain deformations of the Killing spinor which do not change the generalised metric, or equivalently the physical fields. Such deformations can only occur if the background preserves more than minimal supersymmetry and correspond to changing the precise $\mathcal{N}=1$ algebra expressed within an $\mathcal{N}=2$ theory. If we exclude this possibility then $\mathcal{M}_{\psi}/\bbC^{*}$ corresponds to the massless chiral scalars. }
\begin{equation}
    \mathcal{M}_{\text{phys}} = \mathcal{M}_{\psi}/\bbC^{*}
\end{equation}
Noting the discussion around \eqref{eq:SU(7)_vs_U(7)}, a choice of $L_{3}$ defines $\psi$ up to a \textit{local} scale, and it was shown in \cite{Ashmore:2019qii} that we may describe the physical moduli space as
\begin{equation}
\label{eq:moduli_space_kahler_quotient}
\begin{aligned}
    \mathcal{M}_{\text{phys}} &= \{L_{3}\,|\, \psi \text{ an integrable $\SU{7}$ structure}\}/\GDiff \\
    &= \{ L_{3} \,|\, L_{3} \text{ involutive, } \mu= 0\}/\GDiff  \\
    &= \ECSinv\qquotient \GDiff 
\end{aligned}
\end{equation}
In the final line we have defined $\ECSinv=\{ L_{3} \,|\, L_{3} \text{ involutive} \}$, the space of integrable $\Uni{7}\times \bbR^{+}$ structures.\footnote{Note that throughout we are only considering ECSs $L_3$ that admit corresponding $\SU7$ structures $\psi$. That is to say, for which the line bundle $\mathcal{U}_J\subset \tilde{K}_\bbC$ defined by $J$ is trivial.} Since $\mu$ is a moment map for the action of $\GDiff$, we see that the second line is precisely the definition of a K\"ahler quotient of $\ECSinv$ which we denote using the double-slash in the final line. This picture of F-terms imposing a holomorphic condition (here the involutivity) and D-terms giving a K\"ahler quotient is very familiar from the analysis of conventional $\mathcal{N}=1$ backgrounds, the only difference being that here everything is infinite-dimensional. 

The final observation made in \cite{Ashmore:2019qii} was that, by Geometrical Invariant Theory (GIT), K\"ahler quotients by a real group are equivalent to a normal quotient by the complexified group (for reviews see for example~\cite{Woodward,Thomas}), that is 
\begin{equation}\label{eq:M_phys_complex_quotient}
    \mathcal{M}_{\text{phys}} \simeq \ECSinv/\GDiff_{\bbC}
\end{equation}
There is really an important subtlety here in that one should really restrict to a special set of ``polystable points'' $\ECSinv^{\text{ps}}\subset \ECSinv$ in the complex quotient. For the moduli space calculation, assuming the quotients are all well-behaved (which may be a strong assumption), we can ignore this subtlety since the stable points form an open subset of $\ECSinv$ and so assuming we are starting with a background that is supersymmetric (and hence stable), every point in its infinitesimal neighbourhood will also be stable. We should also explain a bit more about what is meant by the complexified group. At a point $p\in M$ the space of $\Uni{7}\times \bbR^{+}$ structures is parameterised by the coset 
\begin{equation}
    J|_p \in \Ex{7(7)}/ \Uni7 \simeq \Ex{7,\bbC}/ P_{\Uni{7}} ,
\end{equation}
where $P_{\Uni{7}}\subset\Ex{7,\bbC}$ is the parabolic subgroup that stabilises $L_3|_p$. As in~\eqref{eq:Q_bundle} the space of $\UniR7$ structures $J$ is given by $\mathcal{X}=\Gamma(\mathcal{Q}_{\UniR{7}})$ where $\mathcal{Q}_{\UniR{7}}$ is that bundle
\begin{equation}\label{eq:Q_bundle_J}
    \frac{E_{7,\bbC}}{P_{\Uni7}} \longrightarrow \mathcal{Q}_{\UniR{7}} \longrightarrow M \, .
\end{equation}
Given an $\UniR7$ structure $J\in\mathcal{X}$, the infinitesimal $\gdiff$ action defines a vector $\rho_V\in T_{J}\mathcal{X}$ by $\imath_{\rho_V}\delta J=L^F_VJ$. Then by definition the action of an element the complexified Lie algebra $\gdiff_{\bbC}\simeq\Gamma(E_\bbC)\ni U=V+\ii W$ is by a vector field of the form 
\begin{equation}
     \rho_{V+\ii W} := \rho_{V}+\mathcal{I}\rho_{W} \in T\mathcal{X}  \qquad \text{where $V,W\in \Gamma(E)$} ,
\end{equation}
where $\mathcal{I}$ is the complex structure on $\mathcal{X}$. The corresponding action on the subbundle $L_3$ is $L^F_{V+\ii W}L_3$. This forms a closed Lie algebra because the complex structure $\mathcal{I}$ is invariant under $\GDiff$. Formally one cannot exponentiate this algebra into a group $\GDiff_{\bbC}$ but one can still define on object as the flows generated by such (real) vector fields on $\mathcal{X}$ and this is what we mean by $\GDiff_{\bbC}$ throughout.\footnote{This is exactly analogous to the definition of the complexified ``group'' that appears in describing Kähler--Einstein geometries~\cite{Thomas}}. From the Leibniz property of $L^F_V$ it is easy to see that $\GDiff_\bbC$ maps involutive structures into involutive structures and hence $\GDiff_\bbC$ acts on the space of involutive structures $\ECSinv$.

\subsection{Counting moduli by cohomologies}\label{sec:GDiff_C_invariance}

One of the benefits of the formulation of the moduli space given in \eqref{eq:M_phys_complex_quotient} is that it naturally gives the infinitesimal moduli of a given background in terms of a cohomology, as we now discuss. Infinitesimally, the physical moduli are counted by the deformations of the integrable $\Uni{7}\times \bbR^{+}$ up to infinitesimal $\GDiff_{\bbC}$ transformations. Thus given an involutive $L_{3}$, under a deformation we get the  
\begin{equation}
    L'_{3} = (1+A)\cdot L_{3} ,
\end{equation}
where $A$ is a section $A\in\Gamma(\ad\mathcal{Q}_{\UniR{7}})$ where $\ad\mathcal{Q}_{\UniR{7}}$ is the (complex) Lie-algebra-valued vector bundle associated to $\mathcal{Q}_{\Uni{7}}$ with fibres $\ex{7,\bbC}/\mathfrak{p}_{\Uni7}$ where $\mathfrak{p}_{\Uni7}$ is the Lie algebra associated to $P_{\Uni7}$. We then require that the deformed bundle $L'_{3}$ is involutive to linear order in $A$. This defines a linear differential operator $\dd_{1}$ under which $A$ must be closed, i.e. $\dd_{1}A = 0$. Since we want to mod out by $\GDiff_\bbC$ We consider a deformation to be exact, or trivial, if it can be written as
\begin{equation}
\label{eq:L_3Gdiff}
    L'_{3} = (1+L_{U}^{F})L_{3} \quad \text{some} \quad U \in \Gamma(E_{\bbC})
\end{equation}
This defines a second linear differential $\dd_{2}$, and we say that $A$ defines a trivial deformation if $A=\dd_{2}V$. We therefore see that the moduli of the involutive background are counted by the cohomology of the following complex.
\begin{equation}\label{eq:moduli_complex}
    \Gamma(E_{\bbC})\xrightarrow{\quad \dd_{2} \quad} \Gamma(\ad \mathcal{Q}_{\UniR7}) \xrightarrow{\quad \dd_{1} \quad} \Gamma(W^{\text{int}}_{\UniR7}) ,
\end{equation}
where $W^{\text{int}}_{\UniR7}$ is the space of the intrinsic torsion of the $\UniR7$ structure. 

An important aspect of this formulation of the moduli space is that the moment map, and hence the Kähler metric on $\mathcal{Z}$, entirely drops out from the right hand side of \eqref{eq:M_phys_complex_quotient}. This means that the moduli depend only on the complex structure of $\mathcal{Z}$, which is determined by the off-shell supersymmetry in four-dimensions. 

In general, as discussed in~\cite{Ashmore:2019qii}, this complex can be related to the Chevalley--Eilenberg complex $(\ext^*L^*_3,\dd_{L_3})$ that arises from the fact that $L_3$ is a Lie algebroid. However, in practice, as we will see in section \ref{sec:SU3_ECS_moduli}, one can use the structure of the particular supersymmetric background to write this cohomology in terms of conventional cohomology groups. The operators $\dd_1$ and $\dd_2$  take the form 
\begin{equation}
    \dd_i \sim \dd_\Delta + \CF
\end{equation}
where $\dd_\Delta$ is the exterior derivative $\dd$ for type 0 backgrounds and the Dolbeault operator $\delb$ for type 3 backgrounds and $\CF$ is the general complex flux that appeared in section~\ref{subsec:ECS_SU(3)}. Thus turning on finite flux deforms the operators that appear in calculating the Calabi--Yau moduli by $\CF$. 

We then find a nice geometric interpretation for F-term and D-term lifting. Suppose we have some naive modulus $R \in \Gamma(\ad \mathcal{Q}_{\UniR7})$ of a Calabi--Yau background. After introducing flux, the differentials $\dd_{1}$ and $\dd_{2}$ can a $\CF$ piece and the naïve modulus may now be obstructed in two possible ways: either it is no longer closed ($\dd_{1} R \neq 0$) or it is now exact ($R=\dd_{2} V$ for some $V \in \Gamma(E_{\bbC})$). In the first case, this would correspond to performing a deformation which breaks involutivity. In the latter case, the deformed background is related to the undeformed via some complex $\GDiff_{\bbC}$. Generically, these will take us away from the $\mu=0$ surface and hence will generate some non-zero D-terms. We therefore see that we have the following interpretation.
\begin{equation}
\begin{aligned}
        \text{F-term lifting} && & \qquad \leftrightarrow \qquad && R \text{ not closed} \\
        \text{D-term lifting} && & \qquad \leftrightarrow  && R \text{ exact}      
\end{aligned}
\end{equation}
Crucially, unlike the usual small-flux argument, our analysis is valid for finite flux, the only assumption being that we have a supersymmetric background. In particular, as we will see in the next section, this method clarifies the question as to how the moduli of the non-Kähler backgrounds are counted. A priori, still requires the validity of the two-derivative supergravity approximation and hence the same restrictions outlined in section \ref{subsec:superpotential_stab}. However, in section \ref{sec:corrections_and_obstructions}, we will argue that it actually captures the moduli even when all the perturbative corrections to the supergravity limit are included.

\begin{figure}
\begin{center}
\begin{tikzpicture}[y={(0.5cm,0.5cm)},x={(1cm,0cm)}, z={(0cm,1cm)}]
\begin{scope}[canvas is xy plane at z=1.25]
\draw[very thick] (0.0,0.0) -- (6.0, 0.0) --(6.0, 6.0) -- (0.0,6.0) -- (0.0,0.0) node[anchor = east] at (0,0) {$\ECSinv$};
\node (A) at (0,1) {};
\node (C) at (0,5) {};
\node (D) at (6,0.3) {};
\node (E) at (-5, 10)[anchor = west] {Calabi--Yau with flux};
\node (F) at (1, 9) [anchor = west] {Graña--Polchinski Background};
\draw[black, thick] (0,1) ..controls (3,1) and (3,5).. (6,5) node[pos = 0.51, shape= circle, scale = 0.5, fill = black](GP){} node[anchor = west] at (6.0, 5.0) {$\mu = 0$} ; 
\draw[blue] (4,6) .. controls (4,3) and (5,3)..(6,3);
\draw[blue] (0,5) .. controls (3,5) and (3,1.5).. (6,0.3) node[pos = 0.1, shape= circle, scale = 0.5, fill = blue](CYF) {};
\draw[blue] (5,6) .. controls (5,5) .. (6,4);
\draw[blue] (0,3) .. controls (3,4) and (3,1)..(5,0);
\draw[blue] (0,1.5) .. controls (1,2)..(3,0);
\draw[black, thick, ->] (E) .. controls (0,7) .. (CYF);
\draw[black, thick, ->] (F) .. controls (1,8).. (GP);
\draw[blue, very thick, ->] (CYF) ..controls (2,4.4).. (GP) node[pos = 0.67, anchor = east]{$\ee^{t L_U}$};
\end{scope}
\end{tikzpicture}
\end{center}
\caption{An illustration of the $\SU{7}$ moduli space around the GP background. The space $\ECSinv$ is the space of all involutive exceptional complex structures. The black line denotes the space of supersymmetric backgrounds, where $\mu=0$. The blue lines denote the orbits under the $\GDiff_{\bbC}$ group. The GP background is connected to a `Calabi--Yau with flux' geometry in which the back-reaction of the flux is not taken into account. Such a geometry is not a solution to the equations of motion (it does not sit on the $\mu=0$ surface). The invariance of the moduli along $\GDiff_{\bbC}$ orbits means that one can equally calculate the moduli at the GP point or the `Calabi--Yau with flux' point.}
\label{fig:GP_GDiff_C_orbits}
\end{figure}
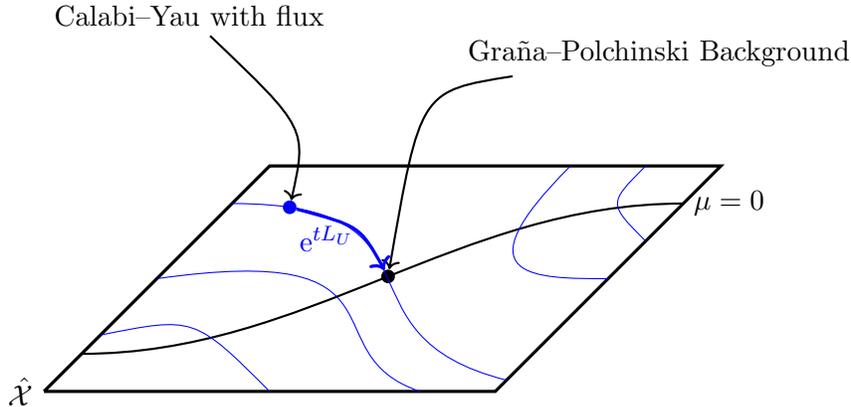

Finally we note that the fact that the moduli space appears as a complex quotient~\eqref{eq:M_phys_complex_quotient} also implies that we do not have to be at the supersymmetric point, where the moment map $\mu$ vanishes, to calculate the moduli but can be anywhere along a $\GDiff_\bbC$ orbit. In other words, the procedure of computing the geometric moduli~\eqref{eq:moduli_complex} would work equally well when deforming around some involutive $L_{3}$ with $\mu \neq 0$ as it would for an involutive $L_{3}$ with $\mu = 0$ provided they are in the same $\GDiff_\bbC$ orbit. Backgrounds which have an involutive bundle $L_{3}$ but which do not satisfy the moment map constraint are ones with vanishing F-terms but not necessarily D-terms. This situation is familiar from some of the cases in section~\ref{subsec:superpotential_stab}, where one calculates the moduli of some Calabi--Yau geometry with flux, rather than the fully back-reacted supersymmetric geometry. In particular, the type IIB B-type Graña--Polchinski supersymmetric backgrounds are conformally Calabi--Yau, and via a $\GDiff_\bbC$ transformation one can set the conformal factor to a constant, so that the non-supersymmetric Calabi--Yau geometry with non-trivial flux lies in the same orbit. Thus in this case we immediately see that the naïve moduli space calculation where one turns on flux on the Calabi--Yau and ignores the backreaction actually gives the correct moduli for the fully back-reacted solution. See figure \ref{fig:GP_GDiff_C_orbits} for an illustration of this point. We will use this equivalence to our advantage in the following calculations.

\section{Calculating the moduli of $\SU{3}$-structure flux backgrounds}\label{sec:SU3_ECS_moduli}

We now turn to the detailed calculation of the number of moduli for $\SU3$-structure string backgrounds with finite flux in the supergravity limit. We will assume that the underlying complex manifold has a Calabi--Yau topology so that the calculation corresponds to finding which moduli are lifted by the presence of the finite flux. Our tools are in principle equally applicable to supersymmetric flux backgrounds which have no fluxless Calabi--Yau counterpart, but we are not then guaranteed to be able to use properties like the $\del\delb$-lemma (which plays a key role for type-3 backgrounds) or the vanishing of certain cohomology classes. 

\subsection{Flux-deformed differentials and spectral sequences}\label{sec:spec_seq}

The discussion of the moduli has so far been somewhat abstract. However, we can be very concrete about the form of the differentials $\dd_{1}$, $\dd_{2}$ in \eqref{eq:moduli_complex} simply by looking at the form of $L_{3}$. Every exceptional complex structure takes the following form
\begin{equation}
    L_{3} = \ee^{\Sigma} \cdot \left( \Delta \oplus \dots \right) 
\end{equation}
for some distribution $\Delta \subset T_{\bbC}$ and some twist $\Sigma$ which is a sum of differential forms. The involutivity constraints tell us that $\Delta$ is an integrable distribution under the Lie bracket, $[\Delta,\Delta]\subset \Delta$, and hence there exists an associated differential $\dd_{\Delta}$. For backgrounds with an $\SU{3}$ structure, we have\footnote{In fact, this discussion holds for generic exceptional complex structures with arbitrary type. In those cases, the distribution $\Delta$, and the associated differential $\dd_{\Delta}$ are more general.}
\begin{equation}
    \Delta = \begin{cases}
        T & \text{type 0} \\
        T^{0,1} & \text{type 3}
    \end{cases} \qquad \Rightarrow \qquad \dd_{\Delta} = \begin{cases}
        \dd & \text{type 0} \\
        \delb & \text{type 3}
    \end{cases}
\end{equation}
The remaining involutivity conditions, given in section \ref{subsec:ECS_SU(3)}, can then be schematically be written as
\begin{equation}
    \pi(\dd_{\Delta} \Sigma + F) = \pi \CF = 0
\end{equation}
where $\CF = \dd_{\Delta}\Sigma + F$ and $\pi$ is a projection onto certain $\GL{6,\bbR}/\GL{3,\bbC}$ representations, given in section \ref{subsec:ECS_SU(3)}, in the type 0/type 3 cases respectively. We can further use the $\G$-structure defined by $\Delta$ to decompose each of the bundles in the deformation complex \eqref{eq:moduli_complex}. In doing so, we will see that the differentials are given by a combination of $\dd_{\Delta}$ and $\CF$.
\begin{equation}\label{fig:simple_moduli_diagram}
\begin{tikzcd}
\Gamma(E_{\bbC})\arrow[r,"\dd_{\Delta}+\CF"] & \Gamma(\ad \mathcal{Q}_{\Uni{7}\times \bbR^{+}})\arrow[r,"\dd_{\Delta}+\CF"]& \Gamma(W^{\text{int}}_{\UniR7}).
\end{tikzcd}
\end{equation}
In each case the complex flux $\CF$ is closed under the action of $\dd_\Delta$. Furthermore, as we are about to see, the calculation of the moduli only depends on the class $[\CF]$ in $\dd_\Delta$ cohomology\footnote{This is consistent with the independence of the moduli under the form-field transformations in $\GDiff_\bbC$. See also~\cite{Smith:2022baw}.}. One can view the $\dd_\Delta+\CF$ operator as a generalisation of the $\dd+H$ operator that appears in $\Orth{d,d}$ generalised geometry to one that includes all the fluxes, although it is also more special in the sense that it is only defined when one has an integrable $L_3$ bundle. 

We would like to relate the cohomology of such complexes to the cohomology of $\dd_{\Delta}$, that is de Rham cohomology for type 0 and Dolbeault cohomology for type 3.\footnote{Recall that, if orientifold planes are present, we need to further reduce these cohomology groups to those that have definite parity under the involution. We will always assume that this is the case in this paper.} In general, this can be done with spectral sequences, the details of which can be determined by looking at the differential equations arising from~\eqref{fig:simple_moduli_diagram}.  Suppose our moduli fields are written as the set of sections $\left\{a_{1},a_{2},\dots,a_{n}\right\} $ that comes from a section of $\Gamma(\ad \mathcal{Q}_{\Uni{7}\times \bbR^{+}})$ decomposed into $\GL{3,\bbC}$ (type 3) or $\GL{6,\bbR}$ (type 0) representations. Similarly we label the gauge transformations $\left\{ b_{1},b_{2},\dots,b_{n}\right\}$, a section of $\Gamma(E_{\bbC})$ again decomposed into irreducible representations. The conditions for the $\{a_{i}\}$ to lie in the cohomology of \eqref{fig:simple_moduli_diagram} are\footnote{Note that in the type 3 case the complex flux $\CF$ may also decompose under $\GL{3,\bbC}$ and so one gets a linear combination  of $\CF\cdot a_i$ and $\CF\cdot b_i$ terms in each expression.}
\begin{equation}
\label{eq:ab-complex}
\begin{aligned}
\dd_\Delta a_{1} & =0, & \qquad \delta a_{1} & = \dd_\Delta b_{1},\\
\dd_\Delta a_{2}+\CF \cdot a_{1} & =0, & \delta a_{2} & = \dd_\Delta b_2 + \CF\cdot b_1,\\
\dd_\Delta a_{3}+\CF \cdot a_{2} & =0, & \delta a_{3} & = \dd_\Delta b_3 + \CF \cdot b_2, \\
\vdots & & \vdots \\
\dd_\Delta a_{n}+\CF \cdot a_{n-1}, & =0 & \delta a_{n} & = \dd_\Delta + \CF \cdot b_3,
\end{aligned}
\end{equation}
where $\CF \cdot a_i$ and $\CF \cdot b_i$ denote a linear action of the complex flux on the deformation and gauge parameters, given by contractions and wedge products of vectors and differential forms. Note that the $\GL{6,\bbR}$ or $\GL{3,\bbC}$ grading is such that there always exists some parameter whose closure and exactness condition is purely differential, $a_{1}$ in this case. Formally one can view~\eqref{eq:ab-complex} as defining a double complex on the spaces of objects $a_i$ and $b_i$ where $\dd_\Delta$ acts vertically and $\CF$ acts horizontally and $\dd_\Delta+\CF$ acts on the total complex formed by combining terms of the diagonals. The spectral sequence is then a standard way to compute the cohomology of the total complex from the double complex. 

To see how this works in practice, let us focus on the type 0 case for concreteness (the type 3 cases are completely analogous). The $\{a_i\}$ and $\{b_i\}$ are then differential forms of certain degrees $r_i$ and $r_i-1$ respectively, $\dd_\Delta$ is the de Rham differential, and the $\CF\cdot a_i$ etc operations are wedge products. Consider first the case where $n=2$ so that
\begin{equation}
\begin{aligned}
\dd a_{1} & =0, & \qquad \delta a_{1} & = \dd b_{1},\\
\dd a_{2}+\CF \wedge a_{1} & =0, & \delta a_{2} & =\dd b_{2} + \CF\wedge b_{1}.
\end{aligned}
\end{equation}
If we set the flux $\CF$ to zero we see that the moduli are counted by the de Rham cohomology groups $H^{r_1}_\dd$ for $a_1$ and $H^{r_2}_\dd$ for $a_2$. Including the flux, the $a_1$ equations imply again that the $a_1$ moduli are counted by de Rham cohomology $[a_1]\in H^{r_1}_\dd$. However, since $[\CF]$ is a non-trivial element in de Rham cohomology, acting on $a_1$ by wedge project it defines a map between cohomologies $[\CF]\wedge: H^{r_1}_\dd\to H^{r_2}_\dd$. For the $a_2$ equation to have a solution we see that $[a_1]$ is actually restricted to live in the kernel of this map $[a_{1}]\in\ker({\left[\CF \right]}\wedge)\subset H_{\dd}^{r_{1}}$. Using this fact to write $\CF\wedge a_1=\dd a^1_2$ for some fixed $a^1_2$, the solution for $a_{2}$ is then $a_{2}=a_{2}^{1}+\Delta_{2}$ where $\Delta$ is $\dd$-closed. From the $b_2$ gauge transformations, we then have that $[\Delta]\in H^{r_2}_\dd$. However, again we need to account for the flux and the $\CF\wedge b_1$ term. Non-closed $b_1$ contribute to $\delta a_1$ and in $\delta a_2$ generate the corresponding variation of $a^1_2$. Closed $b_1$ on the other hand, do not contribute to $\delta a_1$ (and hence $\delta a^1_2$) but can contribute to $\delta\Delta_2$. Thus we actually have to mod out by elements in the image of $[\CF]\wedge:H^{r_1-1}_\dd\to H^{r_2}_\dd$ so that $[\Delta]\in H^{r_2}_\dd/\ker ([\CF]\wedge)$. In summary we find that the moduli are counted by 
\begin{equation}
    \text{moduli} \sim H^{r_1}_{\dd,\CF} \oplus H^{r_2}_{\dd,\CF}
\end{equation} 
where, in this case, $H^{r_1}_{\dd,\CF}$ and $H^{r_2}_{\dd,\CF}$ are cohomology groups for the complex
\begin{equation}
\begin{tikzcd}
    0 \arrow[r] & H^p_\dd \arrow[r,"\lbrack\CF\rbrack\wedge"] & H^{q}_\dd \arrow[r] & 0
\end{tikzcd}  
\end{equation}
with either $p=r_1$ and $q=r_2+1$ or $p=r_1-1$ and $q=r_2$ respectively. In terms of the spectral sequence this is the cohomology that appears on the second page. We see in this simple case that flux obstructs some of the naive moduli, as one would expect, and the true massless modes are counted by some subset/quotient of the de Rham cohomology groups given by some map defined by $\CF$.

Consider now $n=3$ and assume $\CF$ is an odd-rank form
\begin{equation}
\begin{aligned}
    \dd a_{1} & =0, & \qquad \delta a_{1} & =\dd b_{1} ,\\
    \dd a_{2}+\CF \wedge a_{1} & =0, & \delta a_{2} & =\dd b_{2}+\CF\wedge b_{1},\\
    \dd a_{3}+\CF \wedge a_{2} & =0, & \delta a_{3} & =\dd b_{3}+\CF \wedge b_{2}.
\end{aligned}
\end{equation}
(Note that this is a complex since $\CF\wedge\CF\wedge b_1=0$ as well as $\dd\CF=0$.) For the first two lines the calculation follows as above. However, now, from the third line, two new things must happen. First the the $\dd$-closed part $\Delta$ of $a_{2}$ must lie in $\ker(\left[\CF\right]\wedge)\subset H_{\dd}^{r_2}$. Second the component $a^1_2$ of $a_{2}$ which was determined by $a_{1}$ via $\CF \wedge a_{1}$ must also satisfy a condition in order for there to exist an $a_{3}$ solving the third equation. This condition can be written as $[a_{1}]\in\ker D$ for a map $D:H^{r_1}_{\dd,\CF}\to H^{r_3}_{\dd,\CF}$, and in the case that this $D$ map is non trivial its effect is accounted for on the third page of a spectral sequence, and will give a further refinement of the cohomology of \eqref{fig:simple_moduli_diagram}. Whether $D$ is trivial or not depends on both the topology of the base manifold, through the de Rham cohomology and the action of $\CF$ on the sections $\{a_i\},\{b_i\}$ we are considering.

We see that each page of spectral sequences accounts for these connections between equations as they become further separated in the chain. In the calculations below we will often find that, provided one assumes a Calabi--Yau topology, instead of one long chain of equations we get disconnected single, pairs ($n=2$) and triples ($n=3)$ like those discussed above. We call the length of each chain its nilpotency order, and the largest nilpotency order in a set of equations determines the maximum number of pages we would need to compute in the corresponding spectral sequence to recover the moduli counting groups.

We again briefly comment on our treatment of sources. Whether we excise the sources or ignore them, the calculation can be done in the same way. The main distinction is whether we consider cohomology groups on $M$, with Calabi--Yau topology, or on $M'$, the manifold with sources excised. The cohomology of $M'$ is different from that of $M$, the two being related by relative homology. In particular, while some cohomology groups are zero on a Calabi--Yau manifold, they may be non-vanishing on $M'$. This means that, when calculated on a Calabi--Yau background, we may get some terms $\CF\cdot a_{i}$ trivially vanishing in cohomology, but when calculated on $M'$ they may not. Hence, in general, the nilpotency order of the moduli equations will be greater on $M'$ than on $M$, requiring more pages of the spectral sequence to find an exact answer. However, as we have already argued, if we ignore deformations of the sources but focus on the bulk moduli, we can work simply with the cohomology of $M$ and this is what we will do in the following. The extra conditions from cohomology groups on $M'$ reflect the fact that moduli may be further obstructed when coupling to deformations of the sources. (The sources in the heterotic M-theory compactification of Ho\v{r}ava--Witten are a prime example of this~\cite{Smith:2022baw}.) We should emphasise though that our method works equally well in either case, so that if one wants the full picture of the moduli, one should perform the calculation \emph{with} excision and work with the $M'$ cohomologies. 

\subsection{$\SU{3}$ structure backgrounds in IIB}\label{sec:IIB_moduli}
To find the moduli of $\SU{3}$ structures in IIB we find deformations of the eigenbundle 
\begin{equation}
\begin{aligned}
L_{3}&=\ee^{\Sigma}\cdot\left(T^{0,1}\oplus s^{i}T^{*1,0}\oplus\ext^{3,0}T^{*}\right), & && 
\Sigma &= r^{i}\omega + \tfrac{1}{2}\alpha \omega^{2} + \tfrac{1}{6}c^{i}\omega^{3}.
\end{aligned}
\end{equation}
We require that the deformations preserve involutivity and we find the conditions for a deformation to be trivial, in the sense of section \ref{subsec:ECS-moduli}. As discussed infinitesimal deformations of $L_3$ are parameterised by global sections of the quotient bundle $\ad \mathcal{Q}_{\UniR 7}$. For a IIB background this bundle may be decomposed under $\GL{3,\bbC}\subset \GL{6, \bbR}$ as
\begin{equation}
\label{eq:IIB_SU3_adPperp}
\begin{aligned}
   \ad & \mathcal{Q}_{\Uni{7}\times\bbR^{+}} = \ee^{\Sigma}\Big(
      q^{i}\ext^{2,0}T\oplus(T^{1,0}\otimes T^{*0,1})\oplus s^{i}\ext^{0,2}T^{*} \\
      & \quad \oplus q^{i}\epsilon_{jk}q^{k}\ext^{0,0}T^{*}\oplus q^{i}\ext^{1,1}T^{*}\oplus\ext^{2,2}T^{*}\oplus s^{i}\ext^{3,3}T^{*}\oplus q^{i}\ext^{0,2}T^{*}\oplus\ext^{1,3}T^{*} \\
      & \quad \oplus q^{i}\ext^{3,3}T^{*}\Big)
 \end{aligned}
\end{equation}
Here, we have introduced the $\SL{2,\bbR}$ vector $q^{i}$ which satisfies\footnote{It is a quick check to see that it is always possible to make such a choice of $q^{i}$.} $\epsilon_{ij}q^{i}s^{j} \neq 0$ and, for simplicity, $\dd q^{i}=0$. We write a generic section of this bundle $R=\ee^{\Sigma}(\epsilon+\chi+\Theta)$ where each term is a global section of a new line in the above decomposition. (This parameterisation is the one inherited from the underlying generalised complex structure.) To find the moduli, we find the conditions on $R$ such that it preserves involutivity, meaning $L'_{3} = (1+R)\cdot L_{3}$ is involutive. We then quotient by the transformations generated by $L_{V}^{F}$ for some 
\begin{equation}
    V = \ee^{\Sigma}(v + s^{i}\lambda + (q^{i}\rho_{1} + \rho_{3} + s^{i}\rho_{5}) + q^{i}\sigma + \tau)\in \Gamma(E_{\bbC}).
\end{equation}
These equations are given in full in appendix \ref{App:Gen_moduli_equations}. If one assumes that the complex functions $a,b$ defining the background are non-vanishing, and one imposes the $\del\delb$-lemma (as is the case if the underlying complex structure defines a Calabi--Yau), then these equations greatly simplify. We give a detailed analysis of this case in appendix \ref{App:Gen_moduli_equations}. 

Here, we will present the answer in two special cases which correspond to backgrounds with D5/O5 sources and backgrounds with D3/O3 sources. This latter B-type case includes the Giddings--Kachru--Polchinski backgrounds~\cite{Giddings:2001yu} and the non-compact Graña--Polchinski~\cite{Grana:2000jj,Grana:2001xn} and Klebanov--Strassler solution~\cite{Klebanov:2000hb} in the non-compact case (since there are no corresponding local sources). The A-type case, in which $a=0$ or $b=0$, corresponds to solutions to the Hull--Strominger systems. They are S-dual to the D5/O5 C-type backgrounds and we will briefly comment on this case. Finally, in each case, we project to the relevant cohomology groups that have the correct parity under the involution whenever orientifold planes are present.

\subsubsection{D5/O5 source backgrounds}

D5 branes and O5 planes preserve spinors with $a=\pm b$, and hence we are working with backgrounds described by column C of table \ref{table:IIB_backgrounds}. Here we see that, on-shell, the flux is highly constrained, with only a primitive $F_{3}$ being non-vanishing. We can put these conditions into the moduli equations \eqref{eq:IIB_generic_moduli_closure}, \eqref{eq:IIB_generic_moduli_exactness}, use the $\SL{2,\bbR}$ vector $q^{i} = (0,-1)$, and impose a $\del\delb$-lemma to obtain the following moduli equations.
\begin{equation}\label{eq:D5/O5_moduli_eq}
    \begin{aligned}
        \delb \varepsilon^{2,0} &= 0, \\
        \delb \varepsilon^{1,0}_{0,1} &=0, \quad & \delta \varepsilon^{1,0}_{0,1} &= \delb v^{1,0} \\
        \delb\varepsilon_{0,2} &= 0, \quad & \delta \varepsilon_{0,2} &= \delb \lambda_{0,1} \\
        \delb \chi_{0,0} + ab\, \epsilon^{2,0}\cdot \CF_{2,1} &=0, \\
        \delb \chi_{1,1} - \epsilon^{1,0}_{0,1}\cdot \CF_{2,1}&=0, & \delta \chi_{1,1} &= \delb \rho_{1,0} + \imath_{v} \CF_{2,1} \\
        \delb \chi_{0,2} &=0 , & \delta \chi_{0,2} &= \delb \rho_{0,1} \\
        \delb \chi_{2,2} + \varepsilon_{0,2}\wedge \CF_{2,1} &= 0, & \delta\chi_{2,2} &= \delb\rho_{2,1} - \lambda_{0,1}\wedge \CF_{2,1} \\
        && \delta \chi_{1,3} &= \delb \rho_{1,2} \\
        & & \delta \chi_{3,3} &= \delb \rho_{3,2} \\
        && \delta \Theta &= \delb \sigma_{3,2} + \rho_{1,2}\wedge \CF_{2,1}
    \end{aligned}
\end{equation}
where $\CF_{2,1}$ can be derived from \eqref{eq:IIB_generic_moduli_closure}, \eqref{eq:IIB_generic_moduli_exactness}, and \eqref{eq:IIB_CF_3}. We find that
\begin{equation}
    \CF_{2,1} = (F_{3} + \ii\dd(\ee^{-\dil}\omega))_{2,1}
\end{equation}
where, in this case, this flux is a primitive $(2,1)$-form.

Analysing \eqref{eq:D5/O5_moduli_eq}, we can see that the maximum nilpotency order is two, and hence we can immediately read off the moduli. First, note that the fourth line implies that $\epsilon^{2,0}$ is obstructed by $\CF_{2,1}$. Similarly, that the complex structure moduli $\varepsilon^{1,0}_{0,1}$ are obstructed, and hence must lie in
\begin{equation}
    \varepsilon^{1,0}_{0,1}\in H^{0,1}_{\delb,\CF}(T^{1,0}):=\ker [\CF_{2,1}:H^{0,1}(T^{1,0}) \to H^{1,2}]
\end{equation}
which may in general be non-trivial. Finally, we see that the 6-form deformation $\Theta$ can always be made trivial through some appropriate choice of $\rho_{1,2}$ provided the complex flux $\CF_{2,1}$ is non-trivial. Recall that this six-form degree of freedom is dual to the $B$-field on the external space. The triviality of the six-form is a reflection of the well-known fact that the $B$-field becomes massive in flux compactifications, as mentioned previously.

Working through these equations in the way set out in section \ref{sec:spec_seq}, and imposing the vanishing of certain cohomology groups on topological Calabi--Yau manifolds, we find that the bulk moduli are given by 
\begin{equation}\label{eq:D5/O5_moduli}
    \text{moduli} \sim H^{0,1}_{\delb,\CF}(T^{1,0}) \oplus H^{0,0}\oplus H^{1,1} \oplus H^{2,2} \oplus H^{3,3}
\end{equation}
Finally, to obtain the true physical moduli we must restrict the above cohomology groups to the ones which transform appropriately under the involution. For O5-planes, this gives
\begin{equation}
    \text{moduli} \cong H^{0,1}_{\delb,\CF}(T^{1,0})_{+}\oplus H^{0,0}\oplus H^{1,1}_{+} \oplus H^{2,2}_{-}
\end{equation}
where the $\pm$ denotes the parity under the involution defined by the orientifold.

\subsubsection{Hull--Strominger-like backgrounds}

We briefly comment on the S-dual backgrounds to the case above. They have $a=0$ or $b=0$ and are characterised by column A of table \ref{table:IIB_backgrounds}. These have vanishing RR flux and can only have non-vanishing primitive $H_{3}$. We can calculate the moduli similarly, using \eqref{eq:IIB_generic_moduli_closure} and \eqref{eq:IIB_generic_moduli_exactness}, this time with $q^{i} = (1,0)$. Since there are no negative tension sources, or orientifold planes in this case, we will find that the moduli are given by precisely the same equations as \eqref{eq:D5/O5_moduli}, except with $\CF_{2,1}\to \CH_{2,1}$ where
\begin{equation}
    \CH_{2,1} = (H_{3} \pm \ii \dd\omega)_{2,1}
\end{equation}
(the $\pm$ sign depends on whether $a$ or $b$ vanishes). This complex flux is again $(2,1)$ primitive and hence we will find the exact same moduli as the previous case~\eqref{eq:D5/O5_moduli} \emph{before} projection onto the even/odd parts under the orientifold involution. This is unsurprising given, without orientifolds, the type A and type C backgrounds of IIB are S-dual.

\subsubsection{D3/O3 Sources and Graña--Polchinski}

D3 branes and O3 planes preserve supersymmetries with $a=\pm \ii b$, and hence we are in the case described by column B in table \ref{table:IIB_backgrounds}. As we can see, these backgrounds can have non-vanishing $F_{1}$, $F_{3}$, $H_{3}$ and $F_{5}$, with the 3-form fluxes combining into a complex $G = F_{3} + \ii\ee^{-\dil}H_{3}$. The complex $G$ must be a primitive, self-dual (2,1)-form.

A notable example of such background is the Gra\~na--Polchinski solution \cite{Giddings:2001yu,Grana:2000jj}. Here, the warp factor acts as a potential for the location of the D3/O3 planes (localised to points $x_{0}$ on the internal space). The Bianchi identity for $F_{5}$ and the self-duality for the total 5-form flux in 10 dimensions then imposes that 
\begin{equation}\label{eq:GP_F1_F5}
    F_{5} = \ee^{-2\warp}*\dd(\ee^{2\warp}) \ , \qquad -2\nabla^{2}\warp = *( H_{3}\wedge F_{3} + j_{6})
\end{equation}
where the second expression comes from the Bianchi identity for $F_{5}$, and $j_{6}$ is localised to the D3/O3 sources. When calculating the moduli, we should excise the sources so that the Leibniz property of the generalised tangent bundle is implied by the Bianchi identity. In the moduli calculation, we will also have contributions coming from all fluxes in the equations \eqref{eq:IIB_generic_moduli_closure}, \eqref{eq:IIB_generic_moduli_exactness}, and one can use the techniques of that appendix to find the moduli exactly.

Rather than spelling out this complicated calculation here, we would instead like to use the results of section \ref{sec:GDiff_C_invariance} to simplify the problem. If we consider the conditions from involutivity \eqref{eq:IIB_GMPT_invol_1}--\eqref{eq:IIB_GMPT_invol_5} and the Bianchi identities, impose $a=\pm\ii b$ and demand that the background geometry is Calabi--Yau (i.e. $\dd\omega = \dd\Omega = \dd\dil = 0$), then we find the following conditions.
\begin{equation}\label{eq:GP_non-integrable_flux}
    F_{1} = 0\ , \qquad G\in \Omega^{2,1}\ , \qquad F_{5} = 0\ ,\qquad G\wedge \bar{G} + 2\ii\ee^{-\dil}j_6 = 0 \ .
\end{equation}
Clearly these constraints do not satisfy the equations of motion, or even the full supersymmetry constraints; for example, they are incompatible with \eqref{eq:GP_F1_F5} even away from sources. However, we can still use these values for the flux to calculate the moduli associated to some supersymmetric background in the same $\GDiff_{\bbC}$ orbit. To make sense of our results, however, we need to argue that the supersymmetric point corresponds to the Graña--Polchinski solution with non-trivial flux.

Without constructing the $\GDiff_{\bbC}$ flow explicitly, one can argue that this is the case in many ways. First, the superpotential \eqref{eq:W_IIB_O3/O7} for O3/O7 backgrounds is \emph{exact} on Calabi--Yau backgrounds. This means that one can derive the correct F-term equations for the background assuming a Calabi--Yau background geometry (the D-terms will then dictate how the fluxes back-react on the geometry). This mirrors the involutivity and moment map conditions in our formulation and suggests that the $\GDiff_{\bbC}$ flow, which enforces the D-term conditions, flows the geometry to a Graña--Polchinski solution with non-trivial flux. In contrast, the superpotentials for the O5/O9 case \eqref{eq:W_IIB_O5/O9}, and the O6 case \eqref{eq:W_IIA_O6}, require breaking the Calabi--Yau conditions to write an exact superpotential. If one were to derive the F-term constraints in these cases, assuming some Calabi--Yau background geometry, one would determine that the fluxes must vanish. This suggests that the supersymmetric point along the $\GDiff_{\bbC}$ orbit of the ``Calabi--Yau with flux'' point in these cases has trivial fluxes.

In order for the exceptional complex structure defined by the conditions \eqref{eq:GP_non-integrable_flux} to flow to the Gra\~na--Polchinski solution, the $\GDiff_{\bbC}$ action must change the non-primitive components of $G$. We can see that this is possible infinitesimally by considering the action of the Dorfman derivative along complex vector fields. If we consider the infinitesimal deformation generated by the complex vector field
\begin{equation}
    V = \ee^{\Sigma}\cdot v \ , \qquad v \in  \Gamma(T^{0,1})
\end{equation}
then the flow will not change the underlying $\SU{3}$ structure, and hence the bundle $L_{3}$, but it will change the fluxes via the following.
\begin{equation}
    G \to G'= G + \delb(v\lrcorner G) \ , \qquad F_{5} \to F_{5}'= \frac{1}{2\ii \im \tau}((v-\bar{v})\lrcorner G)\wedge \bar{G}
\end{equation}
One can check satisfies the non-trivial Bianchi identity $\dd F'_{5} = (2\ii\im\tau)^{-1}G'\wedge\bar{G}'$. It is therefore clear that the $\GDiff_{\bbC}$ flows can change the flux by changing the non-primitive parts of $G$, and by generating a non-trivial $F_{5}$. We find this strong evidence that the non-integrable point \eqref{eq:GP_non-integrable_flux} flows to the Gra\~na--Polchinski solution under $\GDiff_{\bbC}$ flows, and hence we can calculate the exact moduli around this point rather than the full solution. If, however, the reader is uncomfortable with these arguments, they can use the techniques in appendix \ref{App:Gen_moduli_equations} to calculate the moduli with all fluxes turned on and they will find the same answer.

Finding the moduli around the point \eqref{eq:GP_non-integrable_flux} is significantly simpler. From the moduli equations \eqref{eq:IIB_generic_moduli_closure}, \eqref{eq:IIB_generic_moduli_exactness}, using $q^{i} = (0,-1)$, we find the following moduli equations
\begin{equation}
\begin{aligned}
    \delb \varepsilon^{2,0} &= 0, \\
    \delb \varepsilon^{1,0}_{0,1} + \varepsilon^{2,0} \cdot H_{1,2} &= 0, \quad  & \delta \varepsilon^{1,0}_{0,1} &= \delb v \\
    \delb \varepsilon_{0,2} + \varepsilon^{1,0}_{0,1} \cdot H_{1,2} &= 0, & \delta\varepsilon_{0,2} &= \delb \lambda_{0,1} + \imath_{v}  H_{1,2} \\
    \delb \chi_{0,0} - \varepsilon^{2,0}\cdot G &= 0, \\
    \delb {\chi}_{1,1} + H_{1,2}\wedge \chi_{0,0} - \varepsilon^{1,0}_{0,1}\cdot G  &= 0, & \delta \chi_{1,1} &= \delb\rho_{1,0} - \imath_{v} G_{2,1} \\
    \delb \chi_{0,2} &= 0, & \delta \chi_{0,2} &= \delb \rho_{0,1} \\
    \delb {\chi}_{2,2} + H_{1,2}\wedge {\chi}_{1,1} + \varepsilon_{0,2} \wedge G &= 0, & \delta \chi_{2,2} &= \delb \rho_{2,1} + H_{1,2}\wedge \rho_{1,0} + \lambda_{0,1}\wedge G_{2,1} \\
    && \delta \chi_{1,3} &= \delb \rho_{1,2} + H_{1,2}\wedge \rho_{0,1} \\
    && \delta \chi_{3,3} &= \delb \rho_{3,2} + H_{1,2}\wedge \rho_{2,1} \\
    && \delta \Theta &= \delb \sigma_{3,2} - \rho_{1,2}\wedge G_{2,1}
\end{aligned}
\end{equation}
where we have used the $\del\delb$-lemma on the Calabi--Yau, and the closure of $H$ to remove terms involving $\del$ and $H_{2,1}$. We have also used the notation $ \varepsilon^{2,0} \cdot H = \varepsilon^{da}H_{dbc}$.

It is possible to collect these equations such that they have nilpotency order two and hence we can read off the moduli. Removing all the moduli which are trivial on a Calabi--Yau, we see that we can naively have complex structure moduli $\varepsilon^{1,0}_{0,1}$, as well as moduli associated to $\chi_{i,i}$ for all $i=0,1,2,3$ and $\Theta$. We can see that both $\chi_{1,1}$ and $\chi_{2,2}$ are unobstructed (the term $H_{1,2}\wedge \chi_{1,1} \in H^{2,3} = 0$ and hence does not give an obstruction). On the other hand, $\chi_{3,3}$ and $\Theta$ are always trivial deformations (assuming $H\neq 0$), once again reflecting the fact that the $B$-field on the external space is massive in flux compactifications. Finally, we see from the equations involving $\delb\varepsilon_{0,2}$ and $\delb\chi_{1,1}$ that $\varepsilon^{1,0}_{0,1}$ can be obstructed by $G_{2,1}$ \emph{and} $H_{1,2}$, while $\chi_{0,0}$ can obstructed by $H_{1,2}$. (We should stress that these latter two equations and the obstructions they give to the complex structure and axion-dilaton deformations were first written down and analysed in~\cite{Gray:2018kss}.) Let us first consider the case where there is no combination $H_{1,2}\chi_{0,0} - \varepsilon^{1,0}_{0,1}\cdot G_{2,1}$ that is trivial in cohomology, meaning $\chi_{0,0}$ is obstructed and so has to vanish. The complex structure moduli are then obstructed by both $G_{2,1}$ and $H_{1,2}$ giving
\begin{equation}\label{eq:O3/O7_C-structure_moduli}
\begin{split}
    \varepsilon^{1,0}_{0,1} \in H^{0,1}_{\delb,G,H}(T^{1,0}) &:= \ker [G_{2,1}:H^{0,1}(T^{1,0}) \rightarrow H^{1,2}(M)] \\
    & \qquad \qquad \qquad  \cap \ker[H_{1,2}:H^{0,1}(T^{1,0}) \to H^{0,3}(M)] . 
\end{split}
\end{equation}
More generally, we expect to be able to solve for  $\varepsilon^{1,0}_{0,1}=\pi^{1,0}_{0,1}+\Delta^{1,0}_{0,1}$ such that $H_{1,2}\wedge \chi_{0,0}=\pi^{1,0}_{0,1}\cdot G$ and $\Delta^{1,0}_{0,1}\in \ker [G_{2,1}:H^{0,1}(T^{1,0})]$, and naïvely the axion-dilaton deformation $\chi_{0,0}$ is now unobstructed. However, the condition that $\varepsilon^{1,0}_{0,1} \cdot H_{1,2}$ is trivial then translates into $\psi(\pi,\pi,G)=0$, independent of $\Delta$, where $\psi$ is the standard symmetric map $S^3H^{2,1}\to \bbC$ induced by the Calabi--Yau geometry.\footnote{In terms of the special Kähler geometry on the moduli space $\psi$ is just the third derivative of the prepotential $\mathcal{F}_{ABC}$.} Generically, the only solution will be $\pi=0$ so that the axion-dilaton deformation is again obstructed, and the complex structure moduli are again counted by~\eqref{eq:O3/O7_C-structure_moduli}. As stressed in~\cite{Gray:2018kss}, counting the dimension of $H^{0,1}_{\delb,G,H}(T^{1,0})$ can be quite complicated because $G_{2,1}$ and $H_{1,2}$ are not independent, but rather are proportional to complex conjugates of each other (because of the condition $G_{1,2}=0$) and so the full analysis requires knowledge of the underlying special Kähler geometry. It is perhaps worth noting that, from our generalised geometry formulation, the deformation theory is still nonetheless  holomorphic. It is only that the fluxes that the define the Lie algebroid bracket that end up being restricted to be related by complex conjugation.\footnote{This is because they come from a real algebroid structure on $E$. Interestingly, if one starts instead with the algebroid defined on $L_3$, one is actually free to choose the $G_{2,1}$ and $H_{1,2}$ fluxes independently.} 

In total, we see that the physical moduli are generically given by
\begin{equation}\label{eq:GP_moduli}
    \text{moduli} \sim H^{0,1}_{\delb,G,H}(T^{1,0})\oplus H^{1,1}\oplus H^{2,2}
\end{equation}
Once again, we must further impose the correct transformation of the moduli under the involution induced by the O3-planes. We find
\begin{equation}
    \text{moduli}\cong H^{0,1}_{\delb,G,H}(T^{1,0})_{-}\oplus H^{1,1}_{-}\oplus H^{2,2}_{+}
\end{equation}
Note that this matches the expectation of the moduli from previous work~\cite{Giddings:2001yu}. Indeed, this shows that even though there are in principal fluxes of all degree, only the complex combination $G=\tilde{F}_{3} - \tau H_{3}$ appears in the moduli equations and it acts to obstruct the complex structure and axion-dilaton moduli.

\subsection{$\SU{3}$ structure backgrounds in IIA}\label{sec:IIA_moduli}
\subsubsection{Type-0}\label{subsec:IIA_type-0_moduli}
To find the moduli of $\SU{3}$ structures in IIA we consider deformations of the eigenbundle 
\begin{equation}
\begin{aligned}
    L_3 = \ee^{\Sigma}\cdot \left(T\oplus \bbC\right) \qquad \Sigma = -\ii\omega + \alpha \Omega + \beta \bar{\Omega},
\end{aligned}
\end{equation}
where $\alpha$ and $\beta$ are given by~\eqref{eq:IIA-type0-para}. The only allowed complex flux is the two-form $\CF=F_2$, which must also be primitive. The deformations are again parameterised by sections of $\ad \mathcal{Q}_{\Uni{7}\times\bbR^{+}}$, which we can decompose under $\GL{6,\bbR}$ as
\begin{align}
    \ad \mathcal{Q}_{\Uni{7}\times\bbR^{+}}^{\perp} = \ee^{\Sigma}\left(\Lambda^2T^*_\bbC\oplus \Lambda^3T^*_\bbC\oplus \Lambda^5T^*_\bbC\oplus \Lambda^6T^*_\bbC\right) .
\end{align}
Letting $R=\ee^{\Sigma}R_0$ be one of these sections we have
\begin{equation}
    R_0 = \varepsilon+\chi+\Theta,\quad \varepsilon\in\Gamma(\Lambda^2T^*_\bbC), \ \ \chi\in\Gamma(\Lambda^3T^*_\bbC\oplus \Lambda^5T^*_\bbC), \ \ \Theta\in\Gamma(\Lambda^6T^*_\bbC)
\end{equation}
Requiring this deformation preserve involutivity, and identifying components of $R_0$ which are in the image of the generalised diffeomorphism action, parameterised by $V= \ee^{\Sigma}( x + \lambda + \sigma + w)$, we get the system of equations
\begin{equation}
    \label{eq:IIA_invol_deformation}
    \begin{aligned}
    \dd\varepsilon &= 0, \qqq\delta\varepsilon  =\dd\lambda ,\\
    \dd \chi_3 - \frac{1}{2}\varepsilon\wedge \CF_2 &= 0,\qqq\delta \chi_3 = \dd w_2-\lambda\wedge\CF_2 ,\\
    \dd\left(\chi_5 +\tfrac{1}{2}\ii\omega\wedge\chi_3\right) - \tfrac{1}{12}\ii\varepsilon\wedge \omega\wedge \CF_2 &= 0,\qqq  \delta \chi_5 = \dd w_4  - \tfrac{1}{2}\ii\lambda\wedge \omega\wedge \CF_2,\\
     &\qquad \ \qqq\delta\Theta  = -\dd\sigma + w_4\wedge \CF_2 .
    \end{aligned}
\end{equation}
In all cases with non-trivial flux the six form $\Theta$ can be totally removed by gauge transformations, and so does not contribute any non-trivial moduli. In the case that the flux vanishes $F_2=0$, the background is genuinely Calabi--Yau and hence we have an $\mathcal{N}=2$ theory. The modulus generated by $\Theta$ in that case corresponds to changing the $\mathcal{N}=1 \subset \mathcal{N}=2$ we use to write the exceptional complex structure, and is therefore not physical.\footnote{The field in the four dimensional effective field theory generated by $\Theta$ will appear with the wrong sign in from of the kinetic term, indicating it is a ghost field generating the additional symmetry.} 

These equations have nilpotency order two, so we do not need a spectral sequence to compute the cohomology groups.  The moduli are given by 
\begin{equation}
    \text{moduli}\sim H^2_{\dd,\CF}(M) \oplus H^3_{\dd,\CF}(M) \oplus H^5_{\dd,\CF}(M).
\end{equation}
where
\begin{equation}
    H^{2}_{\dd,\CF}(M) =
        \ker[\CF_{2}:H^{2}_\dd\to H^{4}] \ , \qquad H^{p}_{\dd,\CF}(M) =  \frac{H^{p}}{\im[\CF_{2}:H^{p-2}\to H^{p}]} \ , \ p=3,5
\end{equation}
It is worth noting that here we can easily find the cohomology groups counting the moduli with almost no assumptions.  We do not need a $dd^c$-lemma or equivalent, and also do not need to assume any cohomology classes vanish in order to simply write down the cohomology groups. If we did once again constrain ourselves to Calabi--Yau topology then $H^1=H^5=0$ and
\begin{equation}
    \text{moduli}\sim H^{2}_{\dd,\CF}(M) \oplus H^3
\end{equation} 
where $H^3_{\dd,\CF_2}=H^3$ as $H^3_{\dd,\CF_2}$ is the quotient by the image of $H^1$ under the $[\CF_2]$ map, but $H^1=0$. If the background has an O6-plane then we must again impose the correct transformation of the moduli under the involution and we find
\begin{equation}
    \text{moduli}\cong H^{2}_{\dd,\CF}(M)_{-} \oplus H^{3}
\end{equation}
We see that a non-trivial 2-form flux $\CF_2$ is capable of lifting some of the Kähler moduli, in agreement with the results summarised in \cite{Grana:2005jc}.

\subsubsection{Type-3}
For type-3 backgrounds the eigenbundle is
\begin{equation}
    L_3 = \ee^{\Sigma}\cdot \left[T^{0,1}\oplus \bbC\oplus \ext^{2,0}T^*\right]
\end{equation}
with infinitesimal deformations sections of
\begin{equation}
\begin{aligned}
    \ad \mathcal{Q}_{\Uni{7}\times\bbR^{+}} &= \ee^{\Sigma}[ \ext^3T^{1,0}\oplus T^{1,0}\oplus \left(T^{0,1}\otimes \ext^{0,1}T^*\right) \\& \quad \oplus \ext^{0,2}T^*\oplus\ext^{1,1}T^*\oplus \ext^{0,3}T^*\oplus\ext^{1,2}T^*\oplus \ext^5T^*\oplus \ext^6T^*].
\end{aligned}
\end{equation}
We have used the $\GL{3,\bbC}$ structure associated to the bundle $T^{0,1}$ to decompose the space of deformations into natural bundles. Letting $R=\ee^{\Sigma}R_0$ be a section of the above bundle, we can write
\begin{equation}
    R_0=\kappa+v+r+\theta+\phi+\mu_5+\mu_6
\end{equation}
where 
\begin{gather}
    \kappa\in\Gamma(\ext^3T^{1,0}), \ \ v\in\Gamma(T^{1,0}), \ \ r\in\Gamma(T^{0,1}\otimes \ext^{0,1}T^*_\bbC), \ \ 
    \theta\in\Gamma(\ext^{0,2}T^*_\bbC\oplus\ext^{1,1}T^*_\bbC), \\ \phi\in\Gamma(\ext^{0,3}T^*_\bbC\oplus\ext^{1,2}T^*_\bbC), \ \ \mu_5\in\Gamma(\ext^5T^*_\bbC), \ \ \mu_6\in\Gamma(\ext^6T^*_\bbC)\nonumber.
\end{gather}
Requiring $R$ to preserve involutivity and removing components which are in the image of the generalised diffeomorphisms action, parameterised by $V = \ee^{\Sigma}(x+\lambda+\rho+\nu)$, we get the system of equations presented in full in appendix~\ref{App:Gen_moduli_equations}.  These equations are in general challenging to solve without requiring that our background satisfies the $\del\delb$-lemma.  Once we do this we are able to reparameterise such that there is only one derivative operator, $\delb$.  If we were to further assume, as we did above, that our background was topologically Calabi--Yau such that we have a global $\Omega$ and $H^{1,0}_{\delb}=H^{2,0}_{\delb}=0$ we then have the remaining conditions
\begin{equation}\label{eq:IIA_type_3_general}
\begin{aligned}
    \delb\kappa &= 0, \\ 
    \delb v + \kappa \cdot \CH &= 0, \\
    \delb r +\kappa\cdot \CF_4 &= 0,\qquad \delta r = \delb x \\
    \delb\phi_{0,3} &= 0, \qquad \delta\phi_{0,3} = \delb\nu_{0,2}\\
    \delb\phi_{1,2} + -r\cdot \CF_4 &= 0,\qquad \delta\phi_{1,2} = \delb \nu_{1,1} - x\cdot \CF_4 \\
    \delb\theta_{1,1} - r\cdot \CH - v\cdot \CF_4 &= 0, \qquad \delta\theta_{1,1} = \delb\lambda_{1,0}  - x\cdot \CH \\
    \delb\mu_5-\theta_{1,1}\wedge\CF_4-\phi_{1,2}\wedge \CH &= 0, \qquad\delta\mu_5 = \delb\nu_4 - \CF_4\wedge\lambda - \CH\wedge\nu,\\
    & \ \qquad \qquad \delta\mu_6 = \delb\rho_5 - \CF_4\wedge\nu.
\end{aligned}
\end{equation}
Due to the requirement that the undeformed background is involutive, the only non-zero components of $\CF$ and $\CH$ are the $(2,2)$ and $(2,1)$ respectively, which can be read off from~\eqref{eq:IIA_complex_fluxes_type-0}.
We restrict to backgrounds which also satisfy the D-terms, which imposes that the only non-vanishing RR fluxes are singlets and are proportional to the Romans mass $F_{0}=m$.  Moreover for all IIA supergravity backgrounds the bianchi identity tells us that when the Romans mass $m\neq 0$
\begin{equation}
    H_3=\frac{1}{m}\dd F_2.
\end{equation}
In particular, this implies that $\CH_{(2,1)}$ is $\delb$-exact.\footnote{It actually tells us that $\CH_{(2,1)}$ is $\del$-exact. However, $\CH_{(2,1)}$ is also $\delb$-closed and hence by the $\del\delb$-lemma, it must be $\delb$-exact.} Being trivial in cohomology, this flux will not act as an obstruction to the above equations and can be absorbed through a redefinition of the deformation parameters. We therefore find two cases depending on whether $m$ vanishes or not.

Firstly for $m=0$, all RR fluxes vanish but $[\CH]$ may be non-trivial. The moduli satisfy 
\begin{equation}\label{eq:IIA_type_3_moduli_massless}
\begin{aligned}
    \delb\kappa &= 0, \\ 
    \delb v + \kappa \cdot \CH &= 0 \\
    \delb r  &= 0,\qquad \delta r = \delb x \\
    \delb\theta_{1,1} - r\cdot \CH  &= 0, \qquad \delta\theta_{1,1} = \delb\lambda_{1,0} - x\cdot \CH \\
    \delb\phi_{0,3} &= 0, \qquad \delta\phi_{0,3} = \delb\nu_{0,2}\\
    \delb\phi_{1,2} &= 0,\qquad \delta\phi_{1,2} = \delta \nu_{1,1} \\
    & \ \qquad \qquad \delta\mu_6 = \delb\rho_5 .
\end{aligned}
\end{equation}
These equations have second-order nilpotency and so do not require a spectral sequence to compute.  In the case that $[\CH]\neq 0$ the second of the above equations imposes $\kappa=0$, and in the case that $[\CH]=0$ the modulus $\kappa$ generates unphysical moduli as $\Theta$ did in the type-0 case~\ref{subsec:IIA_type-0_moduli}. We can see the cohomology groups counting the physical moduli just by inspecting the above equations
\begin{equation}
    \text{moduli}\cong  H^{0,1}_{\delb,\CH}\left(T^{1,0}\right)\oplus H^{1,1}\oplus H^{1,2}\oplus H^{0,3}\oplus H^{3,3}
\end{equation}
We have defined 
\begin{equation}
    H^{0,1}_{\delb,\CH}\left(T^{1,0}\right) = \ker\left(\CH:H^{0,1}\left(T^{1,0}\right)\to H^{1,2}\right)
\end{equation}

For the $m\neq 0$ case, $[\CH]=0$ but $\CF_{4} =\frac{1}{2}m\omega^{2} \neq 0$. As previously, $[\CF_{4}]$ being non-trivial implies $\kappa=0$, leaving
\begin{equation}
\begin{aligned} 
    \delb v &= 0 \\
    \delb\theta_{1,1} +  \tfrac{1}{2}mv\cdot\omega^2&= 0, \qquad \delta\theta_{1,1} = \delb\lambda_{1,0} \\
    \delb r &= 0,\qquad \delta r = \delb x \\
    \delb\phi_{1,2}+ \tfrac{1}{2}mr\cdot \omega^2&= 0,\qquad \delta\phi_{1,2} = \delta \nu_{1,1} +  \tfrac{1}{2}mx\cdot\omega^2\\
    \delb\phi_{0,3} &= 0, \qquad \delta\phi_{0,3} = \delb\nu_{0,2}\\
    & \ \qquad \qquad \delta\mu_6 = \delb\rho_5 + \tfrac{1}{2}m\omega^2\wedge\nu
\end{aligned}
\end{equation}
again with nilpotency order two.  The six-form can be entirely removed by gauge transformations, so that in this case we are left with
\begin{equation}
    \text{moduil}\cong H^{0,1}_{\delb,\omega^2}\left(T^{1,0}\right)\oplus H^{1,1}\oplus H^{1,2}\oplus H^{0,3}.
\end{equation}
Similar the the previous cases we have defined
\begin{equation}
    H^{0,1}_{\delb,\omega^2}\left(T^{1,0}\right) = \ker\left(\omega^2:H^{0,1}\left(T^{1,0}\right)\to H^{1,3}\right)
\end{equation}

\subsection{Summary}
By simply computing the deformations which preserve the involutivity of $L_3$ we have for IIB, on a manifold with the topology of a Calabi--Yau
\begin{equation}
    \text{IIB moduli}\cong H_{\delb,\CF}^{0,1}(T^{1,0})_{\pm}\oplus H^{0,0}\oplus H^{1,1}_{\pm} \oplus H^{2,2}_{\mp} \\
\end{equation}
where the upper sign is for O5-planes and the lower for O3-planes. For IIA, we have
\begin{equation}
    \text{IIA moduli}\cong \begin{cases}H^{2}_{\dd,\CF_2}(M)_{-} \oplus H^3 & \text{type-0}\\
    H^{0,1}_{\delb,\CH}\left(T^{1,0}\right)\oplus H^{1,2}\oplus H^{1,1}\oplus H^{0,3}\oplus H^{3,3} & \text{massless type-3}\\
    H^{0,1}_{\delb,\omega^2}\left(T^{1,0}\right)\oplus H^{1,1}\oplus H^{1,2}\oplus H^{0,3} & \text{massive type-3}.\end{cases}
\end{equation}
It is worth stressing that in each case the equations had nilpotency or order two or less, and so there was no need to go to higher pages in the spectral sequence, which could, in principle, have led to the stabilisation of further moduli. As a consequence, in each case the naive stabilising argument from the superpotential and possible D-terms in the linearised flux approximation actually matches the full calculation.

\section{Perturbative corrections and obstructions}\label{sec:corrections_and_obstructions}

In the previous section we found the exact infinitesimal moduli of supersymmetric $\SU3$-structure backgrounds assuming finite flux but ignoring higher-derivative corrections. More precisely, our calculation is valid in the regime that $N_{\text{flux}} (\ell_{s}/R)^{p-1}$ is not small (finite flux) but $l_{s}/R \ll 1$ (two-derivative action). Clearly, this requires $N_{\text{flux}} \gg 1$. On the other hand, cancellation of tadpoles puts a bound on the possible values of $N_{\text{flux}}$ in terms of the topology of the internal manifold. Thus, one is typically already in the small-flux regime, and our finite-flux calculation is not justified unless we can somehow include higher-derivative corrections. First one should note that this actually too naïve: the small-flux condition only refers to the average flux. One can still locally have regions of large flux -- in fact one might expect this generically -- in which case one really needs the full power of the finite-flux analysis here to justify the moduli calculation. Notwithstanding this, it remains an important question as to whether our calculation might still be valid for finite average flux regime, that is, with the inclusion of higher-derivative corrections. 

The second caveat to our calculations is that we considered only the linearised perturbations when calculating the moduli. In general, even in the two-derivative theory, there can be obstructions to extending to finite perturbations. These are always measured by cohomology groups of the underlying deformation problem. 

These observations naturally lead us to two questions. Will our moduli calculation be affected by higher-derivative corrections to the effective action? Can higher order-terms in the moduli calculation further obstruct the moduli? We will examine both of these questions in this section, arguing that, under suitable assumptions, (i)~the infinitesimal moduli are unaffected by the higher-derivative corrections, making our calculation exact to all orders in perturbation theory, and (ii)~the preliminary indications are that the obstructions to the finite moduli calculation vanish.

\subsection{Perturbative corrections}

Let us address the question of higher-derivative corrections to the moduli of the supersymmetric background. We start with the basic observation that the moduli counting is controlled by an $\mathcal{N}=1$ superpotential and then ask if this is corrected by perturbative higher-derivative terms. More specifically, our analysis relies the following elements
\begin{enumerate}
\item 
string corrections are captured by a local, supersymmetric, gauge- and diffeomorphism-invariant ten-dimensional higher-derivative effective action $S$;
\item 
for geometries of the form $\bbR^{3,1}\times M_6$ where $M_6$ admits an $\SU3$ structure, the full ten-dimensional action $S$ can be rewritten in an off-shell four-dimensional $\mathcal{N}=1$ language, with a K\"ahler potential $K$ and superpotential $W$ for an infinite number of chiral fields, gauged by the group of diffeomorphisms and form-field gauge symmetries, $\GDiff$;
\item
determining the moduli, when viewed as a complexified quotient by $\GDiff_\bbC$, depends only on the superpotential $W$; 
\item 
holomorphy, and gauge and diffeomorphism invariance on $M_6$ imply that the superpotential $W$ is uncorrected by the higher-derivative terms and hence the infinitesimal moduli space is unchanged from the one calculated in the two-derivative theory. 
\end{enumerate}
Note that the first assumption already implies that we are not including string and brane instanton corrections and so at most this result holds for all orders in perturbation theory. We also are also assuming there are no contributions from non-trivial wound string state on $M_6$, as is reasonable for simply-connected spaces such as those with Calabi--Yau topology. As in the analysis of the previous section, we also excise the sources and ignore their corrections. Finally in the last step we need to invoke the standard $\mathcal{N}=1$ non-renormalisation of the superpotential to argue that there are no corrections from loops in Kaluza--Klein fields on $M_6$. 

Implicitly we are also assuming that turning on the corrections does not destroy the solution. Although we will see that the superpotential is not corrected, the Kähler potential is modified and appears through the moment map $\mu$ (D-term) and so the actual supersymmetric background will be deformed. When deforming a moment map the GIT picture implies that the solution to $\mu=0$ survives in at least an open neighbourhood of the undeformed solution, although for large deformations there can be ``walls'' where a previously stable backgrounds become unstable and so the solution no longer exists. Thus we expect as one turns on the corrections there is at least a neighbourhood of the two-derivative solution where the corrected solution exists, and our claim is that in this neighbourhood the infinitesimal moduli remain the same. 

We note that Burgess et al.~\cite{Burgess:2005jx} have previously shown that the Gukov-Vafa-Witten O3/O7-plane superpotential in type IIB is indeed not corrected in either the $\alpha'$ expansion, or in the string-loop expansion. Their argument first uses holomorphy to show that the two-derivative superpotential is not corrected by string loops. Then, one can use the fact that the dimensionless parameter in the $\alpha'$ expansion is $\alpha'/r^{1/2}$, where $r$ is the breathing mode of the reduction. This is an axion with a shift symmetry and hence the superpotential cannot depend explicitly on it. Therefore there can be no higher-derivative corrections. Note that this argument works directly in four-dimensions, focusing on the superpotential as a function of the finite number of massless moduli of the fluxless Calabi--Yau background.  

Here, we use a different approach, working directly with the full ten-dimensional theory. Let $S$ be the ten-dimensional effective action one gets by integrating out higher string modes and loops. By assumption this is a local higher-derivative theory which can be written as a power series in $\alpha'$. (The locality assumption means we are not including string and brane instantons.) Schematically we have 
\begin{equation}\label{eq:higher_deriv_action}
    S = S_{\text{IIA/B}} + \alpha'^{3} S_{(1)} +  \dots ,
\end{equation}
where the leading term is just the type IIA/B supergravity theory, and the $S_{(1)}$ correction includes both $\alpha'$ and string-loop corrections that are quartic in the curvature tensor. While all the higher-derivative corrections are not known, they should be invariant under both form-field gauge transformations and diffeomorphisms. Now assume the spacetime is product $\bbR^{3,1}\times M_6$ and $M_{6}$ admits an $\SU{3}$ structure, a necessary condition for there to be an $\SU3$-structure supersymmetric background. The $\SU{3}$ structure ensures that we can  rewrite the full ten-dimensional effective theory in a four-dimensional $\mathcal{N}=1$ formalism with an infinite number of fields. Crucially, we can do this with \emph{off-shell} multiplets. This means that the chiral fields for the rewriting of $S$ are the same as for the two-derivative theory $S_{\text{IIA/B}}$. But the latter we already analysed in section~\ref{subsec:EGG-intro}: the chiral fields are parameterised by the complex $\SU{7}$ tensor $\psi$. When considering the rewriting it is important to distinguish higher-derivative corrections that come from external derivatives in $\bbR^{3,1}$ and internal derivatives on $M_6$. Supersymmetry implies that the theory involving up to two external derivatives in the chiral fields $\psi$ will still be described by a superpotential $W$, a Kähler potential $K$ and D-terms. However, generically these will all be corrected by internal higher-derivative contributions. (There will also of course be external higher-derivative terms, but these we can ignore since they are irrelevant when looking for the vacuum moduli space.) In particular we will have an expansion of the superpotential 
\begin{equation}
\label{eq:W-corrections}
    W = W_{\text{IIA/B}} + W_{(1)} + \dots ,
\end{equation}
where $W_{\text{IIA/B}}$ is the superpotential of the two-derivative IIA/B theory~\eqref{eq:superpotential} and $W_{(1)}$ is the leading correction from internal higher-derivative terms. There will be a similar expansion of the K\"ahler potential $K$. Assuming the symmetries of the theory remain diffeomorphisms and gauge transformations, the $D$-terms arise from gauging $\GDiff$ on $M_6$, just as in the two-derivative theory. Given the Kähler potential (and hence metric) deform, so will the corresponding $D$-term moment maps $\mu$. However, as we have stressed, viewed as a quotient by $\GDiff_\bbC$, the exact form of the Kähler potential and D-terms are not relevant to the moduli calculation and hence we can restrict our focus to the superpotential. 

The key question is then what form can the higher-order contributions take? They must be diffeomorphism and gauge invariant, as in the full action. But they must also only depend holomorphically on the fields, namely on the $\SU{7}$ tensor $\psi$. This means that the superpotential can only depend\footnote{We could also, in principal, allow tensors which appear in sections of the bundle $\mathcal{U}_{L}^{*}$ which is dual to the complex line bundle $\mathcal{U}_{L} \subset \tilde{K}_{\bbC}$ of which $\psi$ is a section. This would allow for objects like $\bar{\Omega}^{\#}$ to appear in the superpotential for IIB. We will see that such objects do not change the conclusions drawn here.} on the holomorphic combination of forms which appear in $\psi$. These are given by 
\begin{equation}
    \Sigma_C \, , \qquad 
    \ee^{2\Delta-\dil}f \quad \text{(IIA, type 0)} \qquad 
    \ee^{2\Delta-\dil}\bar{b}^2\Omega \quad \text{(IIA, type 3)} \qquad 
    \ee^{2\Delta-\dil}s^i\Omega \quad \text{(IIB)}
\end{equation}
Here, $\Sigma_C$ is the exponential factor, defined in \eqref{eq:IIB_twist} for IIB and \eqref{eq:IIA_twist} and \eqref{eq:IIA_twist_3} for IIA, twisted the gauge potentials $C$ and $B$ such that $\ee^{\Sigma_C} = \ee^{C}\ee^B\ee^{\Sigma}$ (viewing $C$ and $B$ as a local sections of the adjoint bundle). Since $\Sigma_{C}$ depends explicitly on the gauge potentials, gauge invariance demands that they can only appear in gauge-invariant combinations with derivatives, that is as the complex fluxes \eqref{eq:IIB_CF_1}--\eqref{eq:IIB_CF_5} for IIB and \eqref{eq:IIA_complex_fluxes_type-0} for IIA. To be more precise only the holomorphic projections of these fluxes that can appear. Allowing for the kernel for the action $\Sigma_C$ in the definitions of $\psi$, these are precisely the terms appearing in the involutivity conditions \eqref{eq:IIB_involutivity} for IIB and \eqref{eq:IIA_involutivity} for IIA. Hence, the superpotential can only depend on the following:
\begin{equation}\label{eq:superpotential_fields}
    \begin{array}{lcccc}
        \text{IIA, type 0:} & & \CH \, , \, \CF_{4} \, ,\, \CF_{6} & & \ee^{2\Delta-\dil}f \\[3pt]
        \text{IIA, type 3:} && \CH_{(1,2)+(0,3)} \, , \, \CF_{(1,3)} \, ,\, \CF_{6} && \ee^{2\Delta-\dil}\bar{b}^2\Omega \\[3pt]
        \text{IIB:} &\quad & \epsilon_{ij}s^{i} \CF_{(0,1)}^{j}\, , \, \CF^{i}_{(0,3)} \, , \, \epsilon_{ij}s^{i}\CF^{j}_{(1,2)} \, , \, \CF_{(2,3)}\, , &\qquad & \ee^{2\Delta-\dil}s^i\Omega  
    \end{array}
\end{equation}

We must also included any internal derivatives of the fields listed. Diffeomorphism invariance requires that we use only the exterior derivative and the Levi--Civita connection. Recall that the metric can be written in terms of the $\SU3$ structure as $g_{mn}=\omega_{mp}I^p{}_n$, where $I$ is the complex structure $I^m{}_n=\frac12\ii(\bar\Omega^{\sharp mpq}\Omega_{npq}-\Omega^{\sharp mpq}\bar\Omega_{npq})$, which is clearly not holomorphic in either $B-\ii\omega$ or $\Omega$.\footnote{One might wonder if certain projections of the Levi--Civita connection are holomorphic. The only possibility is in IIB, where acting on an arbitrary $T^{1,0}$ vector $v$ the component of $\nabla v$ in $\Lambda_{0,1}\otimes T^{1,0}$ is uniquely determined by 
$\Omega$ alone. However, the only holomorphic field combinations with components solely in $\ext^nT^{1,0}$ or $\Lambda_{n,0}$ are $\bar{\Omega}^\sharp$ and $\Omega$ and the Levi--Civita derivative of either is completely determined by the intrinsic torsion $\dd\Omega$. Thus we do not get any new terms this way.} Thus the only holomorphic differential combinations that can appear are exterior derivatives of the objects in~\eqref{eq:superpotential_fields}. However, the complex fluxes satisfy the Bianchi identity $\dd\CF_{\text{IIA/IIB}}=\CH\wedge\CF_{\text{IIA/IIB}}$ and so only the exterior derivatives of the terms in the final column of~\eqref{eq:superpotential_fields} are relevant. 

We see that to determine the higher-derivative corrections to the superpotential we simply ask how to build a top-form on $M_{6}$ out of the fields in \eqref{eq:superpotential_fields} and the exterior derivative $\dd$. By inspection shows the only possibilities are precisely the superpotentials we get for the two-derivative action.\footnote{Note that even including $\bar{\Omega}^{\#}$ in the IIB case does not save us as the complex fluxes are of the wrong complex type.} These have the form
\begin{equation}
\label{eq:(non)corrected_superpotentials}
\begin{aligned}
    &\text{IIA, type 0:} && \int_{M} \ee^{2\warp-\dil}f\,\CF_{6} , \\
    &\text{IIA, type 3:} && \int_{M} \ee^{2\warp-\dil}\bar{b}^2\,
        \CH\wedge\Omega , \\
    &\text{IIB:} &&  \int_{M} \ee^{2\warp-\dil}\epsilon_{ij}s^i\,\CF^j\wedge \Omega ,
\end{aligned}    
\end{equation}
and agree with \eqref{eq:W_IIA_O6-full}--\eqref{eq:W_IIB_O5/O9-full} when the appropriate values for $\CF$ and $a$ and $b$ are inserted. Thus we conclude that all the higher-order local corrections to the superpotential~\eqref{eq:W-corrections} vanish. The final contribution we need to consider are possible corrections to the superpotential arising from loops in the Kaluza--Klein modes. But these modes are precisely the infinite set of chiral fields encoded by $\psi$ and so we can use standard $\mathcal{N}=1$ non-renormalisation theorems to argue that the do not correct the superpotential. 

Put together, this analysis means that the F-terms, which determine involutivity of the exceptional complex structure $L$, are unaffected by the local perturbative corrections to the action~\eqref{eq:higher_deriv_action}. The Kähler potential will, of course, have all sorts of corrections. From the perspective of calculating moduli, this will change the $\mu=0$ surface within the space of involutive $L$. However, as we saw in section~\ref{subsec:ECS-moduli}, the precise surface is irrelevant for the infinitesimal moduli since rather than solving the moment map we simply quotient by the complexified $\GDiff_\bbC$ group within the space of involutive $L$, irrespective of where the $\mu=0$ point lives on the orbit. Thus the leading-order calculation of the moduli that we have performed will actually be valid to all orders in perturbation theory, at least in a finite neighbourhood of the two-derivative solution. 

Let us end by making two further clarifying comments. First, we note that there is one more holomorphic combination of fields appearing in $\psi$ that we have not so far mentioned. This is the overall scale of $\psi$. On substituting for $a$ and $b$, one finds that it is given by $\ee^{3\warp-\dil}$, which combines with the phase to define a holomorphic field. In IIB, for example, this phase corresponds to the phase of $\Omega$, while in IIA it is related to a phase of $f$ or $\bar{b}^2$. The phase, however, is not a physical field as it corresponds to the phase of the internal spinor $\zeta$. This means that the holomorphic combination of phase and scale cannot correspond to the scalar field in a physical chiral multiplet. One may naively expect that the superpotential therefore cannot depend on these fields, however the GVW superpotential clearly changes if we make a constant shift in the phase of $\Omega$ and hence there is some non-trivial dependence. The point is that the superpotential is not really a scalar but rather a section of a line bundle over the Kähler moduli space. This line bundle arises from to the fact that the moduli space has a Kähler--Hodge structure, meaning the Kähler form on the moduli space is an integral class and hence is induced from a line bundle. In the language of generalised geometry and exceptional complex structures, this line bundle is precisely the cone geometry given by
\begin{equation}
    \bbC^{*} \rightarrow \mathcal{M}_{\psi} \rightarrow \mathcal{M}_{\psi} \quotient  \bbC^{*}
\end{equation}
where the $\bbC^{*}$ action is constant shifts in the scale and phase of $\psi$. From this, it is clear that the superpotential is fixed to be a homogeneous function of $\psi$ of degree one, so must have exactly a single power of the phase, and hence scale, of $\psi$. Therefore, the expressions in \eqref{eq:(non)corrected_superpotentials} are correct and additional powers of overall scale of $\psi$ cannot appear. 

The second comment is on holomorphy. It is well-known that corrections can modify which combinations of the moduli are holomorphic, as originally observed in the heterotic theory~\cite{Derendinger:1991hq} and for bullk moduli in type II theories, as for example in~\cite{Antoniadis:1997eg,Antoniadis:2003sw,Berg:2005ja,Haack:2018ufg}. This might seem to contradict our statement that the holomorphic field combinations are fixed off-shell independent of the corrections. However it is not. The holomorphy we are referring to is for the full set of fields before imposing the moment map conditions, which is an infinite set. The moduli fields themselves appear only after imposing the moment map condition, which is non-holomorphic. Hence, if one corrects the Kähler potential, and hence the moment map, the expression for how the moduli space embeds in the space of involutive $\SU7$ structures can change in a non-holomorphic way, without the holomorphic structure on full set of fields changing. To consider a very simple example, suppose we have a Kähler potential $K=|z_1|^2 + \alpha |z_2|^2$ on $\bbC^2$ and a $\Uni1$ action $z_i\to \ee^{\ii\theta}z_i$ so that the moment map is also given by $\mu=|z_1|^2 + \alpha |z_2|^2$. Independent of $\alpha$, the Kähler quotient for $\mu=c^2$ is projective space $\bbC P^1$ with, on a patch, the complex coordinate $z=z_2/z_1$. However, if one explicitly solves the moment map and uses the $\Uni1$ action to set the phase of $z_1$ to zero, one gets a non-holomorphic relation $z=z_2/\sqrt{c^2-\alpha|z_2|^2}$. Thus the $z$ coordinates for different $\alpha$ are non-holomorphic functions of each other. To translate, for example, to the Kähler moduli of a compactification on a Calabi--Yau space, one should view $z$ as a coordinate on the $H^{1,1}$ cohomology, while $z_1$ and $z_2$ are coordinates on the space of closed forms. The relation $z=z_2/\sqrt{c^2-\alpha|z_2|^2}$ corresponds to fixing a particular representative of the class. To leading order the representative is the harmonic form with respect to the Calabi--Yau metric. Including higher-order corrections, corresponding to changing $\alpha$, the modulus deforms away from the harmonic representative, and does so in a non-holomorphic way.\footnote{In terms of the parameterisation used in~\cite{Haack:2018ufg}, $z_1$ and $z_2$ are the analogue of the ``string theory field variables'' $t_i$ and $z$ is the the analogue of the ``supergravity field variables'' $T_i$.}

\subsection{Higher-order deformations and obstructions}\label{sec:obstructions}

The calculations performed in sections \ref{sec:IIB_moduli} and \ref{sec:IIA_moduli} gave the infinitesimal moduli space of the theory. To find the finite deformations, one needs to find solutions to a non-linear Maurer--Cartan equation. In general, it may not be the case that all infinitesimal deformations can be completed to a finite deformation which solves the full non-linear equation. If this is the case, we say that the deformation is obstructed by higher-order terms.

As an example, consider conventional complex structures on a complex manifold, defined by some choice of (anti)holomorphic tangent bundle $T^{0,1} \subset T_{\bbC}$. A deformation of this structure is given by\footnote{The complex type of $r$ is defined in terms of the original complex structure.}
\begin{equation}\label{eq:CS_deformation}
    T'^{0,1} = (1+r)\cdot T^{0,1} \, , \qquad r \in \Omega^{0,1}(T^{1,0})
\end{equation}
It is an integrable deformation if $T'^{0,1}$ is involutive. We find
\begin{equation}
    [T',T'] \subset T' \quad \Leftrightarrow \quad \delb r + [r,r] = 0 
\end{equation}
On the right hand side, we find the Maurer--Cartan equation for integrable deformations of complex structures. The bracket is given in terms of the Lie bracket on vector components, and exterior product on form components. Working to linear order, we find the familiar result that $\delb r = 0$, or that $r\in H^{0,1}(T^{1,0})$.

Suppose now we have $r = \epsilon r_{0} + \epsilon^{2} r_{1} + ... + \epsilon^{n} r_{n-1}$, where $\epsilon$ is a small but finite parameter. Working order by order in epsilon, we find that $r$ is an integrable deformation if
\begin{equation}\label{eq:finite_deformation}
    \begin{aligned}
        \delb r_{0} &= 0 \\
        \delb r_{1} +[r_{0},r_{0}] &= 0 \\
        & \ \; \vdots\\
        \delb r_{n-1} + \sum_{i=0}^{n-2} [r_{i},r_{n-2-i}] &= 0
    \end{aligned}
\end{equation}
In particular, $r_{0}$ is a solution of the infinitesimal deformation problem. Suppose we start with such an $r_{0}$ and ask whether we can find $r_{1},...,r_{n}$ such that $r$, as above, solves the full Maurer-Cartan equation. Equivalently, we need to find $r_{1},...,r_{n}$ that solve \eqref{eq:finite_deformation}. Starting with $\delb r_{0} = 0$, working iteratively, it is easy to show that all of the non-linear terms above are $\delb$-closed. That is
\begin{equation}
    \sum_{i=0}^{k}[r_{i},r_{n-2-k-i}] \in H^{0,2}(T^{1,0})
\end{equation}
Solving the Maurer-Cartan equation then amounts to showing that the non-linear terms in fact vanish in cohomology. Of course a sufficient, but not necessary, condition for no obstructions is that $H^{0,2}(T^{1,0}) = 0$.

Deformations of an exceptional complex structure $L$ follow largely the same pattern. Infinitesimally, deformations are given by $R \in \Gamma(\ad \mathcal{Q}_{\Uni{7}\times \bbR^{+}})$. A finite deformation is then given by
\begin{equation}
    L' = \ee^{R}\cdot L
\end{equation}
and it is integrable (i.e. the F-terms are satisfied) if the deformed bundle is involutive with respect to the Dorfman derivative. We therefore end up with a Maurer-Cartan equation which will, in general, have quadratic, cubic, and higher order terms. We can write this is as
\begin{equation}
    (d_{\Delta} + \CF) R + l_{2}(R,R) + l_{3}(R,R,R) + ... = 0
\end{equation}
The higher order brackets $l_{n}$ form an $L_{\infty}$ structure, the precise details of which depend on how we parameterise the deformations $R$.\footnote{This also occurs for conventional complex structures. We just happened to choose a nice parameterisation in \eqref{eq:CS_deformation} in which all brackets $l_{n\geq 3} = 0$. No matter how we parameterise the deformations, all of the $L_{\infty}$ algebras will be quasi-isomorphic. } In any case, we can expand $R = \epsilon R_{0} + ... + \epsilon^{n}R_{n-1}$ as before and show that the non-linear terms must be $(\dd_{\Delta}+\CF)$-closed. That is, extending the complex \eqref{fig:simple_moduli_diagram}, the obstructions lie in the following cohomology group.
\begin{equation}
    \Gamma(\ad \mathcal{Q}_{\Uni{7}\times \bbR^{+}}) \xrightarrow{\dd_{\Delta} + \CF} \Gamma(W^{\text{int}}_{\bbC}) \xrightarrow{\dd_{\Delta} + \CF} \Gamma(R_{4})
\end{equation}
where $R_{4}$ is a (particular quotient of a) bundle transforming in the $\rep{8645}$, the next step in the tensor hierarchy, the details of which are not important for the present discussion. Decomposing into natural bundles, in the same way we have throughout the paper, we find that the obstructions lie in Dolbeault or de Rham cohomology groups of the background, and there are no obstructions if these groups vanish in cohomology. Rather than do the full analysis, in the follow we show how this works in two key cases, namely IIA type O and IIB Graña--Polchinski backgrounds.

\subsubsection*{IIA Type 0}

Recall, in this case, the differential $\dd_{\Delta} = \dd$, and hence the obstructions will be measured by de Rham cohomology groups. More specifically, we can see from section \ref{sec:IIA_moduli} that
\begin{equation}\label{eq:IIA_type0_deformation_space}
\begin{aligned}
    \ad \mathcal{Q}_{\Uni{7}\times \bbR^{+}} &= \ext^{2}T^{*} \oplus \ext^{3}T^{*} \oplus \ext^{5}T^{*} \oplus \ext^{6}T^{*} \\
    W^{\text{int}}_{\bbC} &= \ext^{3}T^{*} \oplus \ext^{4}T^{*} \oplus \ext^{6}T^{*}
\end{aligned}
\end{equation}
A simple extension of the spectral sequence used to find the moduli of type 0 IIA backgrounds shows that the obstructions must be in the following cohomology groups.\footnote{In the language of \cite{Ashmore:2019qii,Tennyson:2021qwl}, the deformations lie in $H^{4}_{L} \oplus H^{7}_{L}$, where $H^{*}_{L}$ is the Lie algebroid cohomology associated to the exceptional complex structure $L$.}
\begin{equation}
    \text{Obstructions} \in H^{3}_{\dd,\CF}(M) \oplus H^{4}_{\dd,\CF}(M) \oplus H^{6}_{\dd,\CF}(M)
\end{equation}
where
\begin{equation}
    H^{3}_{\dd,\CF}(M) =
        \ker[\CF_{2}:H^{3}_\dd\to H^{5}_\dd] \ , \qquad H^{p}_{\dd,\CF}(M) =  \frac{H^{p}}{\im[\CF_{2}:H^{p-2}\to H^{p}]} \ , \ p=4,6
\end{equation}
Assuming the cohomology groups are those of a Calabi--Yau geometry, we have $\ker[\CF] = H^{3}$ and hence generically the space of obstructions is non-vanishing.

Thus naïvely, one might then argue that some infinitesimal deformations are obstructed and the \emph{true} moduli space is in fact smaller than expected. This is not the case however, since with the parameterisation of the deformations as in \eqref{eq:IIA_type0_deformation_space}, it is easy to show that the brackets $l_{2},l_{3},...$ all vanish identically, and the finite deformation problem is actually the same as the infinitesimal one. Hence, even though the cohomology groups do not vanish, the obstructions are always trivial. We conclude that the moduli of IIA type 0 backgrounds are always unobstructed.

\subsection*{IIB Graña--Polchinski Backgrounds}

For type 3 structures we will consider only the obstructions arising in the Gra\~na--Polchinski backgrounds of IIB, i.e. those sources by D3/O3 planes. The other cases work in a similar manner and this will present enough of the salient features for us to comment. Further, obstructions in this case are important because they could provide a way to evade the tadpole conjecture of~\cite{Bena:2020xrh}. If some of the non-flux-lifted moduli are obstructed by higher-order deformations, then the true number of moduli would be lower than the linear calculation would suggest.

In IIB, the deformations and obstructions can be decomposed into the following natural bundles, under the $\GL{3,\bbC}$ structure of the background.
\begin{equation}
    \begin{aligned}
        \ad \mathcal{Q}_{\Uni{7}\times\bbR^{+}} & =\ee^{\Sigma}\left[(q^{i}\ext^{2,0}T\oplus(T^{1,0}\otimes T^{*0,1})\oplus s^{i}\ext^{0,2}T^{*})\right. \\
        & \qquad \oplus(q^{i}\epsilon_{jk}q^{k}\ext^{0,0}T^{*}\oplus q^{i}\ext^{1,1}T^{*}\oplus\ext^{2,2}T^{*}\oplus s^{i}\ext^{3,3}T^{*}) \\
        & \qquad \oplus(q^{i}\ext^{0,2}T^{*}\oplus\ext^{1,3}T^{*}) \\
        & \qquad \left. \oplus q^{i}\ext^{3,3}T^{*}\right] \\[3pt] 
        W^{\text{int}}_{\bbC} &= \ee^{\Sigma}\left[(\ext^{3,0}T\oplus q^{i}(\ext^{2,0}T\otimes T^{*0,1})\oplus(T^{1,0}\otimes \ext^{0,2}T^{*})\oplus s^{i}\ext^{0,3}T^{*})\right. \\
        & \qquad \oplus(q^{i}\epsilon_{jk}q^{k}\ext^{0,1}T^{*}\oplus q^{i}\ext^{1,2}T^{*}\oplus\ext^{2,3}T^{*}) \\
        &  \qquad\left. \oplus(q^{i}\ext^{0,3}T^{*}) \right]
    \end{aligned}
\end{equation}
In this case, we have $\dd_{\Delta} = \delb$ and hence the obstructions will be counted by the Dolbeault cohomology groups. As we did in section \ref{sec:IIB_moduli}, we will use the $\GDiff_{\bbC}$ invariance of the moduli and work around the non-integrable point defined by \eqref{eq:GP_non-integrable_flux}. An extension of the spectral sequence analysed there, or of the complex \eqref{eq:GMPT_moduli_complex}, tells us that the obstructions lie in the following cohomology groups.
\begin{equation}\label{eq:GP_obstructions}
    \text{Obstructions} \in  H^{0,2}(T^{1,0}) \oplus H^{0,1}(\ext^{2,0}T) \oplus H^{0,3} \oplus H^{1,2}_{\delb,G}
\end{equation}
where
\begin{equation}
    H^{1,2}_{\delb,G}= \frac{H^{1,2}}{\im[G:H^{0,1}(T^{1,0}) \to H^{1,2}]}
\end{equation}
Note that in arriving at \eqref{eq:GP_obstructions}, we have assumed, as above, that the Dolbeault cohomology groups align with those of a Calabi--Yau manifold, and have made some simple assumptions about the form of $G$ (e.g. it is non-vanishing). Note also that, if $G$ is generic, then $H^{1,2}_{\delb,G}$ vanishes. However, we do not want to assume this a priori here as we want to see if higher-order obstructions can be used to evade the tadpole conjecture. Generic values of $G$ will fix \emph{all} the complex structure moduli but will typically violate the tadpole bound.

As in the IIA case, just because the cohomology groups in \eqref{eq:GP_obstructions} do not vanish, it does not necessarily mean that some of the moduli will be obstructed. For that, we need to understand the precise form of the higher order brackets $l_{2},l_{3},...$. We will leave a full analysis of the obstructions to future work. For now, we simply observe that, with the parameterisation of the deformations that we have chosen, the only non-vanishing components of the brackets contain either a complex structure deformation $\epsilon^{1,0}_{0,1}\in \Omega^{0,1}(T^{1,0})$, or $\epsilon^{2,0} \in \Omega^{0,0}(T^{2,0})$. This suggests that only the complex structure obstructions will be relevant in the non-linear perturbation theory.

To check this, consider a finite deformation by $\epsilon = \epsilon^{1,0}_{0,1}\in \Omega^{0,1}(T^{1,0})$. We will find the usual Maurer-Cartan equation for complex structure deformations with the bracket
\begin{equation}
    [\epsilon,\epsilon]\in H^{0,2}(T^{1,0})
\end{equation}
However, recall that we are working around the non-integrable point \eqref{eq:GP_non-integrable_flux} which corresponds to a `Calabi--Yau-with-flux' geometry. In particular, the Tian-Todorov lemma applies \cite{1987mast.conf..629T,1989CMaPh.126..325T}. This states that the bracket is a derived bracket generated from the holomorphic divergence operator $\delta$.\footnote{To be more precise, under the isomorphism $\Omega^{0,q}(\ext^{0,p}T) \simeq \Omega^{3-p,q}$ formed by contraction with the holomorphic 3-form $\Omega$, the operator $\del$ corresponds to an operator
\begin{equation}
    \delta: \Omega^{0,q}(\ext^{p,0}T) \rightarrow \Omega^{0,q}(\ext^{p-1,0}T)
\end{equation}} We have
\begin{equation}
    [\epsilon_{1},\epsilon_{2}] = (-1)^{p_{1}+q_{1} + 1}(\delta(\epsilon_{1}\wedge \epsilon_{2}) + \delta \epsilon_{1}\wedge \epsilon_{2} + (-1)^{p_{1}+q_{1}}\epsilon_{1}\wedge \delta \epsilon_{2})
\end{equation}
for $\epsilon_{i}\in \Omega^{0,q_{i}}(\ext^{p_{i},0}T)$. It then follows from the fact that this must be $\delb$-closed and the following corollary of the $\del\delb$-lemma for K\"ahler manifolds
\begin{equation}
    \im\delta\cap\ker \delb = \ker \delta \cap \im \delb = \im \delta\delb 
\end{equation}
that the bracket always gives an element trivial in cohomology and hence the deformations are unobstructed.

This would suggest that higher-order deformations of the complex structure are not obstructed for Graña--Polchinski backgrounds, and hence it does not provide a way to evade the tadpole conjecture. It should be noted that there are actually more fields in the deformation space than just complex structure deformations which may lead to other obstructions and so strictly a more detailed analysis is required. This result also uses the Tian-Todorov lemma which relied on the fact that our background was `Calabi--Yau-plus-flux'. For other IIB backgrounds, such as those arising from D5/O5 planes, this is not the case and the Tian-Todorov lemma does not apply. In these cases, obstructions could in principle play a role in fixing the complex structure moduli.

\section{Conclusions}

In this paper, we have found the exact bulk moduli of flux backgrounds of type IIA and IIB with an $\SU{3}$ structure. In particular, our results go beyond the linearised, small-flux approximation typically used previously, and are also valid in cases where the geometry is not conformally Calabi--Yau (such as non-complex and non-K\"ahler backgrounds). By using the formalism of exceptional complex structures \cite{Ashmore:2019qii,Ashmore:2019rkx,Tennyson:2021qwl,Smith:2022baw} we found that the moduli could be expressed in terms of flux-obstructed Dolbeault cohomology groups (for IIB) or de Rham cohomology groups (for the generic IIA case). The final expressions are derived from a spectral sequence associated to a flux-obstructed differential $\dd_{\Delta} + \CF$, where $\dd_{\Delta} = \dd$ or $\delb$, and $\CF$ is a complex combination of flux and torsion. Our results extend thus the previous analysis in three significant ways:
\begin{enumerate}
    \item we allow finite flux, that is, $N_{\text{flux}}$ can be of order $R/\ell_s$;
    \item we consider the effect of higher-derivative perturbative corrections to the supergravity action, to all orders in both $\alpha'$ and string-loop expansions;
    \item and we consider finite deformations of the background, calculating the possible obstruction to the infinitesimal. 
\end{enumerate}
Note that the first two points are not really independent, since, without fine tuning, the regime in which we can ignore higher-derivative corrections is precisely the regime in which the flux is relatively small. 

For point 1, one might have expected that for finite flux the moduli would be obstructed by higher-order terms. This would appear in the spectral sequences as non-trivial terms in the third, or higher page. We found, however, that in every case the spectral sequence terminated at the second page and hence reproduced the result obtained from the linear approximation. For point 2, we used the fact that the moduli are independent of the precise form of the K\"ahler potential on the space of structures and only depend on the vanishing locus of the superpotential (i.e. involutivity of the exceptional complex structure \cite{Ashmore:2019qii}). We then showed in section \ref{sec:corrections_and_obstructions} that, under reasonable assumptions of locality, diffeomorphism and gauge invariance, the superpotential is not corrected by perturbative, higher-derivative terms and hence the moduli are unaffected by higher-derivative corrections. Hence, our results are exact to all orders in perturbation theory. For point 3, we initiated an analysis of the obstructions to infinitesimal moduli arising from higher-order perturbations in section \ref{sec:corrections_and_obstructions}. We found that for type 0 IIA backgrounds, all obstructions vanish because the higher-order brackets in the corresponding Maurer--Cartan equation are all trivial. For IIB, it was possible that there could be obstructions to e.g. the complex structure moduli. Nonetheless, preliminary results suggested that one could apply a Tian--Todorov-type lemma to show that these obstructions also vanish. We verified that this is indeed the case for the Gra\~na--Polchinski solutions.

Throughout this paper we have have focused only on bulk moduli, ignoring moduli of the flux sources and how these might interact with the bulk moduli. If $M'$ is the compactification manifold $M$ with the local sources excised, then bulk moduli live the cohomology $H(M)$ while source moduli are captured by classes in the relative cohomology $H(M,M')$. However, even if one focuses on the bulk moduli, as noted in section~\ref{sec:sources_note}, if a particular flux $F$ is sourced, then it has a component which corresponds to a non-trivial class in 
the relative cohomology $H(M,M')$, that is, there is a non-zero component
\begin{equation}
    [F] \neq 0 \in \frac{H^{*}(M')}{i^{*}H^{*}(M)}
\end{equation}
This non-vanishing component of $[F]$ in $H^{*}(M')$ which does not come from any class in $H^{*}(M)$ can in principle fix additional moduli. This indeed occurs in heterotic M-theory~\cite{Smith:2022baw} where the bulk complex structure moduli couple to bundle moduli on the $S^1/\bbZ_2$ orbifold fixed planes, reproducing the obstructions that appear from the Atiyah algebroid in the heterotic theory~\cite{Anderson:2010mh,Anderson:2011ty}, and similar couplings appear at tree-level between open- and closed-string moduli for branes wrapping holomorphic cycles~\cite{Kachru:2000ih}. As a toy example of how such stabilisation can occur, consider a one-form flux on $M=S^{2}$ sourced at the north and south poles. The excised manifold $M'=S^{1}\times (-1,1)$ has the topology of a cylinder and hence a non-trivial $H^{1}(M')$ which is generated by the class of the one-form flux $F$. Suppose one then has the moduli equations
\begin{equation}
    \dd b = 0 \ , \quad \dd a + F b = 0 \ , \quad a,b\in C^{\infty}(M)\ .
\end{equation}
In particular, we need that $F b$ is exact. If we just take this calculation on $M$, then we would argue that all closed one-forms are exact so there is no obstruction. However, on $M'$, the non-trivial class generated by $F$ says that we need $b=0$. Hence, sourced fluxes may fix more moduli than one would expect from performing the calculation on the unexcised manifold. Related to this is the fact that, throughout our calculation, we used the vanishing of certain cohomology groups on Calabi--Yau manifolds to simplify the spectral sequence analysis. If one performs the calculation with respect to the cohomology $H^{*}(M')$, it is possible that higher pages in the spectral sequence will appear and hence fix more moduli. It would be interesting to explore this in the future.

One of the motivations for this work is that having a complete understanding of the moduli of flux backgrounds is important for model building and the swampland programme. In particular, it is relevant to the tadpole conjecture \cite{Bena:2020xrh} which says that the tadpole should grow approximately linearly with the number of moduli being fixed, with proportionality $\gtrsim \tfrac{1}{3}$. Much of the evidence for this conjecture has been performed in the large complex structure regime using the linear approximation \cite{Plauschinn:2021hkp,Lust:2021xds,Bena:2021qty,Grimm:2021ckh,Grana:2022dfw}.One might have thought that you could circumvent the conjecture by including higher-order effects to fix more moduli than one might naively expect. Our work shows, however, that this is not the case.

In principal, our techniques can be applied to more general backgrounds than just those with an $\SU{3}$ structure. Backgrounds with a local $\SU{2}$ structure have been described in $\Orth{6,6}$ geometry in terms of two pure spinors $\Phi_{\pm}$ \cite{Grana:2005sn}. These can then be lifted to an exceptional complex structure, as in \cite{Ashmore:2019qii}. We can then apply the techniques of this paper to find the moduli in terms of flux obstructed \emph{generalised} Dolbeault cohomology groups, associated to the pure spinors $\Phi_{\pm}$. This could enlarge the search for realistic models of string theory. In practice, this may be more difficult given the relatively few explicit examples of backgrounds admitting such a structure. Even of the examples we know, calculating their generalised Dolbeault cohomology groups is hard.

Another interesting application of our work may be to understand how mirror symmetry might work in the presence of generic NSNS and RR flux. Restricting to three-folds, there has been considerable interest in extending mirror symmetry  to include $\SU{3}$-structure manifolds with torsion \cite{Fidanza:2003zi,Grana:2005ny,Benmachiche:2006df,MGraña_2007}. It has been argued that such a correspondence should exchange $H$-flux and torsion classes \cite{Fidanza:2003zi} and attempts to construct such a duality have appeared in~\cite{BENBASSAT2006533,BENBASSAT20061096,Bouwknegt:2003vb,Bouwknegt:2003zg,Bunke:2005sn,Tomasiello:2005bp,Cavalcanti:2011wu}. Most work, however, has focused on backgrounds with $H$-flux, although work on trying to define the duality for non-K\"ahler backgrounds with specific RR flux appeared in \cite{Lau:2014fia}, so that a full understanding of mirror symmetry with generic fluxes remains unknown. If such a correspondence exists, one should be able to match the moduli on each side. Analysing the moduli we calculated in section \ref{sec:SU3_ECS_moduli}, we note a qualitative matching between the IIA and IIB moduli, at least in the case $ab\neq 0$. Indeed, one might conjecture that for every flux background $(M,H,F)$, there exists a mirror background $(\tilde{M},\tilde{H},\tilde{F})$ whose moduli satisfy the following
\begin{equation}\label{eq:type_0_mirror_corresp}
    \begin{array}{ccc}
    H^{2}_{\dd,\CF_2}(M)&\longleftrightarrow& H^{0,1}_{\delb,\CF_3}\left(\Tilde{M},T^{1,0}\right)\\[6pt]
    H^3_{\dd}(M)& \longleftrightarrow &H^{1,1}_{\delb}(\Tilde{M})\oplus H^{2,2}_{\delb}(\Tilde{M}).
    \end{array}
\end{equation}
We expect that understanding mirror symmetry in the context of exceptional complex structures should help answer this question.

Mirror symmetry is intimately linked with topological theories, in particular the topological A and B-models. The role that generalised complex structures plays in these theories has also been well studied \cite{Kapustin:2004gv,Pestun:2005rp,Pestun:2006rj}. Recently, extensions of these ideas have been used to build the target space theory of topological heterotic strings \cite{Ashmore:2023vji}, and topological strings on $\G_{2}$ and $\Spin{7}$ manifolds \cite{Ashmore:2021pdm}. These ideas were unified so as to understand how one can build topological models solely from information about differential complexes and $G$-structures in Courant algebroids in \cite{Kupka:2024rvl}. We note that the complex we define here is simply the exceptional version of the complexes they study. It would be interesting to see if similar techniques can lead to topological theories for exceptional bundles, possibly describing topological branes in string/M-theory.

Finally, recall that the moduli space had a natural description as a holomorphic constraint (the F-terms) modulo an complex group action $\GDiff_{\bbC}$ (equivalent to imposing the D-terms and modding out by the gauge symmetry). As noted in \cite{Ashmore:2019qii}, when one passes from the Kähler quotient to the complex quotient in \eqref{eq:M_phys_complex_quotient} one should really restrict to the set of ``poly-stable'' points $\hat{\mathcal{X}}^{\text{ps}}\subset \hat{\mathcal{X}}$ before taking the quotient. Polystability is a condition which dictates whether the $\GDiff_{\bbC}$ flow through a given point intersects the $\mu=0$ surface. Understanding the criterion for which a point is polystable is the content of geometric invariant theory (GIT). Similar techniques have been applied to the study of Kähler--Einstein geometries, flat connections on Riemann surfaces and hermitian Yang--Mills equations. In the context of supersymmetric flux backgrounds it is an open and interesting problem. If understood, we would be able to find geometries which admit a flux background of string theory within some given class without having to solve the full set of equations of motion. More specifically, we would only need to solve the involutivity condition, a linear differential constraint, along with a GIT criterion, typically some algebraic conditions, to know that a supersymmetric flux background exists on the manifold. (A caveat here, as throughout, is the r\^ole of the sources, meaning one would need to consider the existence of solutions on spaces with boundary.) Much as for Calabi--Yau manifolds, where one needs only to check Kählerity together with $c_{1}(X)=0$, this would make the search for flux backgrounds of string theory significantly easier. A slightly easier problem may be to try to define the analogue of the Futaki invariant for exceptional complex structures, defining an obstruction to the existence of a supersymmetric geometry.

\acknowledgments

We thank Mariana Graña, Chris Hull and Ruben Minasian for helpful discussions. GRS was supported by an EPSRC DTP studentship, DW is supported in part by the STFC Consolidated Grants ST/T000791/1 and ST/X000575/1 and the EPSRC New Horizons Grant EP/V049089/1, and DT is supported by the NSF grant PHYS-2112859.

\appendix 

\section{$\ExR{7(7)}$-generalised geometry for type II}\label{app:GG}
\subsection{$\ExR{7(7)}$ representation bundles}
In Exceptional Generalised Geometry the spinor, flux and $G$-structure torsion degrees of freedom are repackaged into sections of $\ExR{d(d)}$ representation bundles over $M$, and the killing spinor equations are replaced by the vanishing of a generalised intrinsic torsion for a generalised $G$-structure, with $G\subset \ExR{d(d)}$.  In this work we focus on the case where $d=7$.  For $\ExR{7(7)}$ the ``vector" or defining irreducible representation is the $\repsub{56}{1}$, and the vector bundle $E\to M$ has fibres isomorphic to the real 56 dimensional representation space. This geometry is ``generalised" precisely because the transition functions for a local trivialisation of $E\to M$ are not in $\GL{6,\bbR}$ but instead in $\ExR{7(7)}$ or a subgroup thereof.  Using the subgroup $\GL{6,\bbR}\subset \ExR{7(7)}$ the fibre bundle $E\to M$ can be locally decomposed into natural $\GL{6,\bbR}$ bundles over $M$ (but not globally as the transition functions do not respect this decomposition).  For type IIA and IIB backgrounds, $E$ decomposes under different $\GL{6,\bbR}$ subgroups.  For IIA we decompose under $\GL{6,\bbR}\subset\GL{7,\bbR}\subset \ExR{7(7)}$, while for IIB we decompose under a $\GL{6,\bbR}\times \SL{2,\bbR}\subset\ExR{7(7)}$ subset.  This gives
\begin{align}
    E_{\text{IIA}} &\simeq TM \oplus T^{*}M  \oplus \ext^{5}T^{*}M \oplus (T^{*}M\otimes \ext^{6}T^{*}M)\oplus \ext^{\text{even}}T^{*}M\label{A:IIA-vector-bundle} \\
    E_{\text{IIB}} &\simeq TM \oplus \left(S\otimes T^{*}M\right)\oplus \ext^3 T^*M  \oplus\left(S\otimes \ext^{5}T^{*}M\right)\oplus (T^{*}M\otimes \ext^{6}T^{*}M)\label{A:IIB-vector-bundle}
\end{align}
where $S$ is an $\SL{2,\bbR}$ doublet.  Next we consider the $\ExR{7(7)}$ adjoint bundle $\repsub{133}{0}\oplus\repsub{1}{0}$, which contains the flux gauge field degrees of freedom $C_{\text{odd}}$ and $C_{\text{even}}$ for IIA and IIB respectively along with other components describing the structure forms of the background.  The adjoint bundle similarly decomposes under a  $\GL{6,\bbR}\subset \ExR{7(7)}$ subgroup as \cite{Coimbra:2012af, Coimbra:2011nw,Coimbra:2011ky}:
\begin{align}
\begin{split}
    \ad \tilde{F}_{\text{IIA}} &\simeq \bbR \oplus \bbR\oplus (TM\otimes T^{*}M) \oplus \ext^{2}TM \oplus \ext^{2}T^{*}M \oplus \ext^{\text{odd}}TM \oplus \ext^{\text{odd}}T^{*}M \\
    & \qquad  \oplus \ext^{6}TM \oplus \ext^{6}T^{*}M\label{A:IIA-adj-bundle}
\end{split} \\
\begin{split}
    \ad \tilde{F}_{\text{IIB}} &\simeq \bbR \oplus (TM\otimes T^{*}M) \oplus \ext^{2}TM \oplus \ext^{2}T^{*}M \oplus \ext^{\text{even}}TM \oplus \ext^{\text{even}}T^{*}M \\
    & \qquad \oplus \bbR \oplus \ext^{6}TM \oplus \ext^{6}T^{*}M.\label{A:IIB-adj-bundle}
\end{split}
\end{align}
Finally a bundle $\tilde{K}$ in the $\repsub{912}{3}$ \cite{Ashmore:2019qii,Coimbra:2011ky} which decomposes as
\begin{align}
    \tilde{K}_{\text{IIA}} = (\det T^*)^2\otimes K_{\text{IIA}} &= \bbR\oplus \ext^{2}T^{*}M \oplus  \ext^3 T^*M\oplus\dots \\
    \tilde{K}_{\text{IIB}} = (\det T^*)^2\otimes K_{\text{IIB}} &= T^{*}M \oplus (S\otimes \ext^{3}T^{*}M) \oplus\dots
\end{align}
Where $K$ is the representation bundle containing the generalised intrinsic torsion. We will often express sections of these bundles as formal sums of their $\GL{6,\bbR}$ components.  In general the $\GL{6,\bbR}$ components are mixed by the transition function, not just on double overlaps but on triple and higher overlaps, forming a gerbe structure \cite{Hitchin:1999fh}.  It is possible to absorb this gerbe behaviour into a frame for $E\to M$, called the split frame, such that the frame \emph{components} do have $\GL{6,\bbR}$ tensorial transformations (but still not the whole tensors).

After choosing a local trivialisation, we can use the $\GL{6,\bbR}$ decomposition to describe the fiberwise action of the adjoint bundle on $E$.  Parameterising a section $V$ of $E$ and $R$ of $\ad\tilde{F}$ as
\begin{equation}
    V  = v + \lambda + \sigma + \tau + \omega,\qquad R =  l + \varphi + r + \beta + b + \tilde \beta + \tilde b + \alpha + a
\end{equation}
where the components of these sections are sections of the $\GL{6,\bbR}$ sub bundles in the order they appear on the right hand side of ~\eqref{A:IIA-vector-bundle} and ~\eqref{A:IIA-adj-bundle}.
Denoting the adjoing action of $R$ on $V$ by $V' = R\cdot V$, we have
\begin{align}
\label{A:IIA-adj-vec-action}
\begin{split}
v' &= l v +  r \cdot v - [ \alpha \,\lrcorner\, s(\omega)]_{-1} - \beta \,\lrcorner\, \lambda - \tilde \beta \,\lrcorner\, \sigma \ , \end{split}
\\
\begin{split}
\lambda' \,&=\, l \lambda  +  r \cdot \lambda  - v \,\lrcorner\, b - [\alpha\,\lrcorner\, s(\omega)]_1 - \tilde \beta \,\lrcorner\, \tau \ ,
\end{split}\\
\begin{split}
\sigma' \,&=\,  (l-2\varphi) \sigma  +  r \cdot \sigma +  v \,\lrcorner\, \tilde b - [\omega\wedge s(a)]_5 - \beta \,\lrcorner\, \tau \ ,\end{split}\\
\begin{split}
\tau' \,&=\,  (l-2\varphi) \tau  +  r \cdot \tau + j a \wedge s(\omega) + j \tilde b \wedge \lambda - j b \wedge \sigma \ ,\end{split}\\
\begin{split}
\omega' \,&=\, (l-\varphi) \omega + r \cdot \omega + b\wedge \omega +  v \,\lrcorner\, a + \lambda \wedge a + \beta \,\lrcorner\, \omega + \alpha \,\lrcorner\, \sigma  +\alpha\,\lrcorner \,\tau \ \end{split}
\end{align}
above the function $s$ is the sign operator $s(\omega_n) = (-1)^{[n/2]} \omega_n$ for $\omega_n \in \Lambda^n T^*$, and $[\ldots]_p$
denotes the degree $p$ component of any polyform within $[\dots]$, with $p=-1$ corresponding to the vector component. The $\ExR{7(7)}$ subalgebra is specified by the choice $\frac 12{\rm tr}(r) = l - \varphi$. In particular, the $\Orth{6,6} \subset \ExR{7(7)}$ action is generated by $r, b$ and~$\beta$, while setting $\varphi = -\frac 12{\rm tr}(r)$ and all other generators to zero. Similarly we can compute $R'' = [R,R']$ to be
\begin{align}
\begin{split}
l'' &=   -\tfrac{1}{2} (\alpha_1 \,\lrcorner\, a_1' -  \alpha_1' \,\lrcorner\, a_1)+\tfrac{1}{2} (\alpha_3 \,\lrcorner\, a_3' -  \alpha_3' \,\lrcorner\, a_3)   - \tfrac{1}{2} (\alpha_5 \,\lrcorner\, a_5' -  \alpha_5' \,\lrcorner\, a_5) \\ &\quad + (\tilde \beta' \,\lrcorner\,  \tilde b - \tilde \beta \,\lrcorner\, \tilde b') \end{split} \\ \begin{split}
\phi'' &= \tfrac{3}{2} (\alpha_1' \,\lrcorner\, a_1 -  \alpha_1 \,\lrcorner\, a_1') + \tfrac{1}{2} (\alpha_3 \,\lrcorner\, a_3' -  \alpha_3' \,\lrcorner\, a_3)  - \tfrac{1}{2}
 (\alpha_5' \,\lrcorner\, a_5 -  \alpha_5 \,\lrcorner\, a_5') \\
&\quad    - ( \beta \,\lrcorner\,  b' - \beta' \,\lrcorner\,  b) + (\tilde \beta' \,\lrcorner\,  \tilde b - \tilde \beta \,\lrcorner\, \tilde b') \end{split}  \\ \begin{split}
r'' &= [ r, r'] + j \alpha_1' \,\lrcorner\,  j a_1 -  j  \alpha_1\,\lrcorner\, j a_1'  + j\alpha_3 \,\lrcorner\, j a_3' -  j\alpha_3' \,\lrcorner\, ja_3  - j\alpha_5 \,\lrcorner\, j a_5'   +j \alpha_5' \,\lrcorner\,  j a_5  \\ 
&\quad+ j \beta \,\lrcorner\,  j b'  -  j \beta' \,\lrcorner\, j b - j \tilde \beta \,\lrcorner\, j \tilde b'  +  j  \tilde \beta' \,\lrcorner\, j \tilde b + \tfrac{1}{2}\mathbbm{1} (\alpha_1' \,\lrcorner\, a_1 - \alpha_1 \,\lrcorner\,  a_1') \\
&\quad + \tfrac{1}{2} \mathbbm{1}  (\alpha_3' \,\lrcorner\, a_3 - \alpha_3 \,\lrcorner\,  a_3')   + \tfrac{1}{2} \mathbbm{1}  (\alpha_5' \,\lrcorner\, a_5 - \alpha_5 \,\lrcorner\,  a_5') + \mathbbm{1}(\tilde \beta \,\lrcorner\, \tilde b' - \tilde\beta' \,\lrcorner\, \tilde b) \end{split}\\ \begin{split}
b'' &= r \cdot b'-  r' \cdot b  +\alpha_1 \,\lrcorner\, a_3'  - \alpha_1' \,\lrcorner\, a_3 - \alpha_3 \,\lrcorner\, a_5' + \alpha_3' \,\lrcorner\, a_5  \end{split} \\ \begin{split}
\tilde b'' &= r  \cdot \tilde b' - r' \cdot \tilde b  -2\varphi \tilde b' + 2\varphi'\tilde b  + a_1 \wedge a_5' - a_1' \wedge a_5 - a_3 \wedge a_3' \end{split} \\ \begin{split}
a'' &=  r \cdot a' - r' \cdot a - \varphi a'  + \varphi' a  + b\wedge  a' - b'\wedge a   + \beta \,\lrcorner\, a' - \beta' \lrcorner\, a -\alpha \,\lrcorner\, \tilde b'  + \alpha' \lrcorner\, \tilde b  \end{split} \\ \begin{split}
\beta'' &= r \cdot \beta' - r' \cdot \beta + \alpha_3' \,\lrcorner\, a_1 - \alpha_3  \,\lrcorner\,  a_1'  - \alpha_5' \,\lrcorner\, a_3 + \alpha_5 \,\lrcorner\,  a_3'   \end{split} \\ \begin{split}
\tilde \beta'' &=  r \cdot \tilde \beta' - r' \cdot \tilde \beta + 2\varphi \tilde \beta' - 2\varphi' \tilde \beta   + \alpha_1 \wedge \alpha_5' - \alpha_3 \wedge \alpha_3' + \alpha_5\wedge \alpha_1' \end{split} \\ \begin{split}
\alpha'' &=  r \cdot \alpha'  - r' \cdot \alpha +\varphi \alpha'  - \varphi' \alpha + \beta\wedge\alpha' -\beta'\wedge\alpha  - \alpha \,\lrcorner\, b'  + \alpha' \lrcorner\, b -  \tilde \beta  \,\lrcorner\, a' + \tilde \beta'  \lrcorner\, a.  \end{split}
\end{align}
In the above expressions, we have used the notation
\begin{align}
    (j\alpha\wedge \beta)_{a,a_{1}...a_{p+q-1}} &= \frac{(p+q-1)!}{(p-1)!q!}\alpha_{a[a_{1}...}\beta_{...a_{p+q-1}]}\\
    (ja\lrcorner j\alpha)^{a}{}_{b} &= \frac{1}{(p-1)!}a^{ac_{1}...c_{p-1}}\alpha_{bc_{1}...c_{p-1}}
\end{align}
for $\alpha\in \Omega^{p}(M)$, $\beta\in \Omega^{q}(M)$, $a\in \Omega^{0}(\ext^{p}T)$.

For IIB, taking sections of the \eqref{A:IIB-vector-bundle} and \eqref{A:IIB-adj-bundle} bundles, where each term matches a successive term in the expression above
\begin{equation}
    V = v + \lambda^{i} + \rho + \sigma^{i} + \tau \qquad R = l+r+a+\beta^{i} + B^{i} + \gamma + C + \tilde{\alpha}^{i} + \tilde{a}^{i}.
\end{equation}
The adjoint action $R\cdot V = V'$ is
\begin{align}
    v' &= lv + r\cdot v + \epsilon_{ij}\beta^{i}\lrcorner \lambda^{j} + \gamma\lrcorner \rho + \epsilon_{ij} \tilde{\alpha}^{i}\sigma^{j} \\
    \lambda'^{i} &= l\lambda^{i} + r\cdot \lambda^{i} + a^{i}{}_{j}\lambda^{j} + \imath_{v} B^{i} + \beta^{i}\lrcorner\rho - \gamma\lrcorner \sigma^{i} - \tilde{\alpha}^{i}\lrcorner\tau \\
    \rho' &= l\rho + r\cdot \rho + \imath_{v} C+ \epsilon_{ij}\lambda^{i}\wedge B^{j} + \epsilon_{ij}\beta^{i}\lrcorner\sigma^{j} + \gamma\lrcorner\tau \\
    \sigma'^{i} &= l\sigma^{i} + r\cdot \sigma^{i} +a^{i}{}_{j}\sigma^{j} +  \imath_{v}\tilde{a}^{i} + \rho\wedge B^{i} - C\wedge \lambda^{i} + \beta^{i}\lrcorner \tau \\
    \tau' &= l\tau + r\cdot \tau - \epsilon_{ij}j\sigma^{i}\wedge B^{j} - j\rho\wedge C + \epsilon_{ij}j\lambda^{i}\wedge \tilde{a}^{j}
\end{align}
The Lie algebra bracket, or the adjoint action on the adjoint representation $[R,R'] = R''$, is 
\begin{align}
    l'' &= +\tfrac{1}{4}\epsilon_{ij}( \beta^{i}\lrcorner B'^{j} - \beta'^{i}\lrcorner B^{j}) + \tfrac{1}{2}(\gamma\lrcorner C' - \gamma'\lrcorner C) + \tfrac{3}{4}\epsilon_{ij}(\tilde{\alpha}^{i}\lrcorner \tilde{a}'^{j} - \tilde{\alpha}'^{i}\lrcorner\tilde{a}^{j} ) \\
    \begin{split}
        r'' &= [r,r'] + \epsilon_{ij}(j\beta^{i}\lrcorner j B'^{j} - j\beta'^{i}\lrcorner jB^{j}) - \tfrac{1}{4}\mathbb{I}\epsilon_{ij}(\beta^{i}\lrcorner B'^{j} - \beta'^{i}\lrcorner B^{j}) \\
        & \quad +(j\gamma\lrcorner jC' - j\gamma'\lrcorner jC) - \tfrac{1}{2}\mathbb{I}(\gamma\lrcorner C' - \gamma' \lrcorner C) \\
        & \quad + \epsilon_{ij}(j\tilde{\alpha}^{i} \lrcorner \tilde{a}'^{j} - \tilde{\alpha}'^{i}\lrcorner \tilde{a}^{j}) - \tfrac{3}{4}\mathbb{I}(\tilde{\alpha}^{i}\lrcorner\tilde{a}'^{j} - \tilde{\alpha}'^{i}\lrcorner\tilde{a}^{j})
    \end{split} \\
    \begin{split}
        a''^{i}{}_{j} &= (a\cdot a' - a'\cdot a)^{i}{}_{j} + \epsilon_{jk}(\beta^{i}\lrcorner B'^{k} - \beta'^{i}\lrcorner B^{k}) - \tfrac{1}{2}\delta^{i}{}_{j}\epsilon_{kl}(\beta^{k}\lrcorner B'^{l} - \beta'^{k}\lrcorner B^{l}) \\
        & \quad + \epsilon_{jk}(\tilde{\alpha}^{j}\lrcorner \tilde{a}'^{k} - \tilde{\alpha}'^{j}\lrcorner\tilde{a}^{k}_ - \tfrac{1}{2}\delta^{i}{}_{j} \epsilon_{kl} (\tilde{\alpha}^{k}\lrcorner \tilde{a}'^{l} - \tilde{\alpha}'^{k}\lrcorner \tilde{a}^{l})
    \end{split} \\
    \beta''^{i} &= (r\cdot \beta'^{i} - r'\cdot \beta^{i}) + (a\cdot \beta' - a'\cdot \beta)^{i} - (\gamma\lrcorner B'^{i} - \gamma'\lrcorner B^{i}) - (\tilde{\alpha}^{i}\lrcorner C' - \tilde{\alpha}'^{i}\lrcorner C) \\
    B''^{i} &= (r\cdot B'^{i} - r'\cdot B^{i}) + (a\cdot B' - a'\cdot B)^{i} + (\beta^{i}\lrcorner C' - \beta'^{i}\lrcorner C) - (\gamma\lrcorner \tilde{a}'^{i} - \gamma'\lrcorner \tilde{a}^{i}) \\
    \gamma'' &= (r\cdot \gamma' - r'\cdot \gamma) + \epsilon_{ij}\beta^{i}\wedge \beta'^{j} + \epsilon_{ij}(\tilde{\alpha}^{i}\lrcorner B'^{j} - \tilde{\alpha}'^{i}\lrcorner B^{j}) \\
    C'' &= (r\cdot C' - r'\cdot C) - \epsilon_{ij}B^{i}\wedge B'^{j} + \epsilon_{ij}(\beta^{i}\lrcorner \tilde{a}'^{j} - \beta'^{i}\lrcorner \tilde{a}^{j}) \\
    \tilde{\alpha}''^{i} &= (r\cdot \tilde{\alpha}' - r'\cdot \tilde{\alpha})^{i} + (a\cdot \tilde{\alpha}' - a'\cdot \tilde{\alpha})^{i} - (\beta^{i}\wedge \gamma' - \beta'^{i}\wedge \gamma) \\
    \tilde{a}''^{i} &= (r\cdot \tilde{a}' - r'\cdot \tilde{a})^{i} + (a\cdot\tilde{a}' - a'\cdot \tilde{a})^{i} + (B^{i}\wedge C' - B'^{i}\wedge C)
\end{align}
Here $\epsilon_{ij}$ is the $\SL{2,\bbR}$ invariant antisymmetric tensor with $\epsilon_{12} = -1$.  For both IIA and IIB the $\ExR{7(7)}$ vector bundles $E\to M$ admit a symplectic pairing which is invariant under the action of $\ExR{7(7)}$ (and moreover the complexification $E_{7\bbC}$
\begin{align}
    s(V,V') = 
    \begin{cases}
        -\frac{1}{4}\left(\imath_v\tau' - \imath_{v'}\tau-\lambda\wedge\sigma'+\lambda'\wedge\sigma +[s(\omega)\wedge\omega']_6 \right)&\text{IIA}\\
        -\frac{1}{4}\left(\imath_v\tau' - \imath_{v'}\tau+\epsilon_{ij}(\lambda^i\wedge\sigma'^j - \lambda'^i\wedge\sigma^j)-\rho\wedge\rho' \right) & \text{IIB}
    \end{cases}
\end{align}

\subsection{Dorfman derivatives and the $\gdiff$ algebra}
Sections of $E$ are elements of the generalised diffeomorphism algebra, containing both the diffeomorphism algebra and the algebra of gauge-for-gauge fields.  The generalised diffeomorphism algebra has a bracket 
\begin{equation}
    \llbracket V,V'\rrbracket = \Dorf_V W - \Dorf_W V,
\end{equation}
the antisymmetrisation of the \emph{Dorfman derivative} $\Dorf$.  In IIA the Dorfman derivative of two vectors of the form $V=v+\lambda+\sigma+\tau+\omega$ given by
\begin{equation}
\begin{aligned}
    \Dorf_V V' &=\mathcal{L}_{v}v'+(\mathcal{L}_{v}\lambda'-\imath_{v'}\dd\lambda)+\left(\mathcal{L}_{v}\sigma'-\imath_{v'}(\dd\sigma+mw_{6})+\left[s(w')\wedge(\dd w-m\lambda)\right]_{5}\right)\\
    &\quad +\left(\mathcal{L}_{v}\tau'+j\sigma'\wedge\dd\lambda+\lambda'\otimes\left(\dd\sigma+mw_{6}\right)+js(w')\wedge\left(\dd\omega-m\lambda\right)\right)\\
    &\quad +\left(\mathcal{L}_{v}w'+\dd\lambda\wedge w'-\left(\imath_{v'}+\lambda'\wedge\right)\left(\dd w-m\lambda\right)\right) \\
    &= L_V^{(m=0)}V' + \underline{m}(V)\cdot V'
\end{aligned}
\end{equation}
where we have included the possible Romans mass term for massive IIA
\begin{equation}
\begin{aligned}
\underline{m}(V)&=m\lambda-mw_{6},\\
\underline{m}(V)\cdot V'&=m\left(-\imath_{v'}w_{6}-\lambda\wedge w_{4}'+\lambda'\otimes w_{6}-\lambda\otimes w_{6}'+\imath_{v'}\lambda+\lambda'\wedge\lambda\right).
\end{aligned}
\end{equation}
In IIB the Dorfman derivative for two vectors of the form $V = v + \lambda^{i} + \rho + \sigma^{i} + \tau$ is
\begin{equation}
\begin{aligned}
    L_{V}V' &= \mathcal{L}_{v}v' + (\mathcal{L}_{v}\lambda'^{i} - \imath_{v'} \dd\lambda^{i}) + (\mathcal{L}_{v}\rho' - \imath_{v'}\dd\rho +\epsilon_{ij}\dd\lambda^{i}\wedge \lambda'^{j}) \\
    & \qquad +(\mathcal{L}_{v}\sigma'^{i} - \imath_{v'} \dd\sigma^{i} +\dd\rho\wedge \lambda'^{i} - \dd\lambda^{i} \wedge \rho') \\
    & \qquad + (\mathcal{L}_{v}\tau' - \epsilon_{ij}j\lambda'^{i}\wedge \dd\sigma^{j} + j\rho'\wedge \dd\rho + \epsilon_{ij} j\sigma'^{i}\wedge \dd\lambda^{j})
\end{aligned}
\end{equation}

In these expressions the $\GL{6,\bbR}$ components of $V$ and $V'$ are only locally defined and have non-tensorial patching on overlaps. To define a global expression for the Dorfman derivative, we need to absorb all non-trivial patching into a global twist. We write
\begin{equation}
    V = \ee^{\Sigma}\cdot \hat{V}
\end{equation}
where $\Sigma$ is some section local of the adjoint bundle within the nilpotent subalgebra defined by the differential forms. On overlaps, if we demand that $\Sigma$ transforms in the same way as higher-form gauge potentials in IIA/B, then we find that the $\GL{6,\bbR}$ components of $\hat{V}$ transform as tensors on overlaps and hence can be globally defined. Using this isomorphism in the Dorfman derivative, we find
\begin{equation}
    L_V V' = \ee^{\Sigma}\cdot L_{\hat{V}}^{\CF}\hat{V}'.
\end{equation}
where $L_{V}^{\CF}$ is a `flux-twisted' Dorfman derivative. This shows that the Leibniz algebroid using $L_{V}$ and the twisted generalised tangent bundle is isomorphic to the algebroid defined by $L^{\CF}_{V}$ and the untwisted generalised tangent bundle. Because it allows us to work with globally defined objects, in the paper we are always working with the $L_{V}^{\CF}$ algebroid, and simply write $L_{V}$ for convenience. The fluxes appearing in the twisted Dorfman derivative are globally defined differential forms
\begin{equation}
\begin{aligned}
    \CF_{\text{IIA}} &= \CF_0+\CF_2+\CH_{3}+\CF_4+\CF_6, \qquad \CF_0 = m \\
    \CF_{\text{IIB}} &= \CF_1 + \CF^i_3 + \CF_5
\end{aligned}
\end{equation}
They must satisfy the Bianchi identity for the algebroid to be Leibniz. Often, they are identified with the sum of RR and NSNS fluxes in the theory. However, since we work with complexified tangent bundles, we found it convenient in the paper to allow the $\CF_{\text{IIA/B}}$ to be complex valued.

For massive IIA the flux twisted Dorfman derivative on the tensorial component of the generalised vectors ($V\simeq v+\lambda+\sigma+\tau+\omega$ in the order of~\eqref{A:IIA-vector-bundle}) is \cite{Cassani:2016ncu}
\begin{equation}
\begin{aligned}
    L_{\hat{V}}^{\CF}\hat{V}' & =\mathcal{L}_{v}v'+(\mathcal{L}_{v}\lambda'-\imath_{v'}\dd\lambda+\imath_{v'}\imath_{v}\CH)\\
 & +\mathcal{L}_{v}\sigma'-\imath_{v'}\dd\sigma+\left[\imath_{v'}\left(s(w)\wedge\CF\right)+s(w')\wedge\left(\dd w-\CH\wedge w-\left(\imath_{v}+\lambda\wedge\right)\CF\right)\right]_{5}\\
 & +\mathcal{L}_{v}\tau'+j\sigma'\wedge\left(\dd\lambda-\imath_{v}\CH\right)+\lambda'\otimes\left(\dd\sigma-\left[s(w)\wedge\CF\right]_{6}\right)\\
 & +js(w')\wedge\left(\dd w-\CH\wedge w-\left(\imath_{v}+\lambda\wedge\right)\CF\right)\\
 & +\mathcal{L}_{v}w'+\left(\dd\lambda-\imath_{v}\CH\right)\wedge w'-\left(\imath_{v'}+\lambda'\wedge\right)\left(\dd w-\CH\wedge w-\left(\imath_{v}+\lambda\wedge\right)\CF\right) 
\end{aligned}
\end{equation}
Above, for convenience, we have written $\CF$ for the even parts of $\CF_{\text{IIA}}$. For IIB the flux twisted Dorfman derivative on the tensorial component of the generalised vectors ($V\simeq v+\lambda^i+\rho+\sigma^i+\tau$ in the order of~\eqref{A:IIB-vector-bundle}) is
\begin{equation}
\begin{aligned}
    L^{\CF}_{\hat{V}}\hat{V}' &= \mathcal{L}_{v}v' + \big(\mathcal{L}_{v}\lambda'^{i} - \imath_{v'} \dd\lambda^{i} + \imath_{v'}\imath_v \CF^i\big) \\&\qquad+ \big(\mathcal{L}_{v}\rho' - \imath_{v'}\dd\rho +\imath_{v'} \imath_v\CF +\imath_{v'}( \epsilon_{ij}\lambda^i\wedge\CF^j)\big) \\
    & \qquad +\big(\mathcal{L}_{v}\sigma'^{i} - \imath_{v'} (\dd\sigma^{i} -\imath_{v'}(\lambda^i\wedge\CF)+\imath_{v'}\rho\wedge\CF^i)+\dd\rho\wedge \lambda'^{i} \\&\qquad\quad-\imath_v\CF\wedge \lambda'^{i}-\epsilon_{jk}\lambda^j\wedge\CF^k\wedge \lambda'^{i} - \dd\lambda^{i}\wedge \rho' +\imath_v\CF^i\wedge \rho'\big) \\
    & \qquad + \big(\mathcal{L}_{v}\tau' - \epsilon_{ij}j\lambda'^{i}\wedge \dd\sigma^{j} -\epsilon_{ij}j\lambda'^{i}\wedge\lambda^j\wedge\CF+\epsilon_{ij}j\lambda'^{i}\wedge\rho\wedge\CF^j \\&\qquad\quad + j\rho'\wedge \dd\rho - j\rho'\wedge\imath_v\CF - j\rho'\wedge\epsilon_{ij}\lambda^i\wedge\CF^j + \epsilon_{ij} j\sigma'^{i}\wedge\dd\lambda^{j} -\epsilon_{ij} j\sigma'^{i}\wedge\imath_v\CF^j\big)
\end{aligned}
\end{equation}
These twisted or untwisted Dorfman derivatives furnish the $\gdiff$ algebra with a bracket.  They also allow the $\gdiff$ algebra to act on other $\ExR{7(7)}$ tensors using the form
\begin{align}
    L_V = \mathcal{L}_v - (\del\times_{\ad} V)
\end{align}
where $\del$ is viewed as a differential operator valued section of $E^*$ acting on $V$ to produce adjoint-valued sections.  The adjoint bundle is a subset of $E^*\otimes E$, and as such we have used the projection operator
\begin{equation}
    \times_{\ad}: E^*\otimes E \to \ad\Tilde{F}.
\end{equation}

\section{General moduli equations}\label{App:Gen_moduli_equations}

\subsection{IIB}
In section \ref{sec:IIB_moduli} we gave the equations dictating the moduli of IIB backgrounds in three special cases. In this appendix, we provide the moduli equations in the most general IIB case and show that under certain assumptions they can be solved.

We consider deformations of the bundle
\begin{equation}
\begin{aligned}
L_{3}&=\ee^{\Sigma}\cdot\left[T^{0,1}\oplus s^{i}T^{*1,0}\oplus\ext^{3,0}T^{*}\right] \qquad \Sigma= r^{i}\omega + \tfrac{1}{2}\alpha \omega^{2} + \tfrac{1}{6}c^{i}\omega^{3}
\end{aligned}
\end{equation}
We take the deformation parameter $R = \ee^{\Sigma}(\varepsilon + \chi+ \Theta)$ as in \eqref{eq:IIB_SU3_adPperp}. Checking the closure of $L'_{3} = (1+R)\cdot L_{3}$, we find the following constraints.
\begin{equation}\label{eq:IIB_generic_moduli_closure}
    \begin{aligned}
        0 &= \delb(\epsilon_{ij}q^{i}s^{j} \varepsilon^{2,0}) -  \tilde{\CF}_{0,1}\wedge \varepsilon^{2,0} \\
        0 &= \delb \varepsilon^{1,0}_{0,1} - \epsilon_{ij}q^{i} j \varepsilon^{2,0} \lrcorner j^{2} \CF_{1,2}^{j} \\
        0 &= \delb \varepsilon_{0,2} + \eta^{-1}\epsilon_{ij} q^{i}\varepsilon^{1,0}_{0,1}\cdot \CF_{2,1}^{j} + \eta^{-1}\tilde{\CF}_{0,1}\wedge\varepsilon_{0,2} \\
        0 &= \delb (\eta \chi_{0,0}) - \tilde{\CF}_{0,1}\wedge\chi_{0,0} - \eta^{-1} \varepsilon^{1,0}_{0,1}\cdot \CF_{1,0} + \epsilon_{ij}s^{i} \varepsilon^{2,0}\lrcorner \CF_{2,1}^{j} \\
        0 &= \delb \chi_{1,1}+ \del \chi_{0,2} - \eta^{-1}(\CF_{1,0}\wedge \varepsilon_{0,2} + \epsilon_{ij} s^{i} \varepsilon^{1,0}_{0,1}\cdot \CF_{2,1}^{j}) - \varepsilon^{2,0}\lrcorner \CF_{3,2} + \chi_{0,0}\epsilon_{ij}q^{i}\CF_{1,2}^{j} \\
        0 &= \delb \chi_{0,2} \\
        0 &= \delb \chi_{2,2} + \del \chi_{1,3} + \epsilon_{ij} q^{i}(\chi_{0,2}\wedge \CF_{2,1}^{j} + \chi_{1,1} \wedge \CF_{1,2}^{j}) - \varepsilon^{1,0}_{0,1}\CF_{3,2} + \epsilon_{ij}s^{i} \varepsilon_{0,2}\wedge \CF_{2,1}^{j}
    \end{aligned}
\end{equation}
Here we have used the notation
\begin{equation}
    \eta = \epsilon_{ij}s^{i}q^{j}\ , \qquad j\varepsilon^{2,0}\lrcorner j^{2} \CF^{j} = \varepsilon^{ab}\CF^{j}_{acd}\ , \qquad \varepsilon^{1,0}_{0,1}\cdot \CF = -k \varepsilon^{a}_{[b_{1}|}\CF_{a|b_{2}...b_{k}]}
\end{equation}
and where\footnote{Note that this slightly awkward notation is used to ensure that the equations are valid for all values of $a$ and $b$. As we shall see later, the formulae simplify significantly if we assume $a$ and $b$ are non-vanishing.}
\begin{equation}
    \tilde{\CF}_{1} = \epsilon_{ij}q^{i}p^{j}(\epsilon_{ij}p^{i}s^{j})F_{1} + \epsilon_{ij}q^{i}\dd s^{j}
\end{equation}
The trivial deformations are given by those such that $L'_{3} = (1+L_{V}^{F})\cdot L_{3}$, for some $V \in \Gamma(E_{\bbC})$. Parameterising $V= \ee^{\Sigma}\cdot(v + s^{i} \lambda + (q^{i}\rho_{1} + \rho_{3} + s^{i}\rho_{5})+ q^{i} \sigma + \tau) $, we find that the trivial transformations are given by
\begin{equation}\label{eq:IIB_generic_moduli_exactness}
\begin{aligned}
    \delta \varepsilon^{1,0}_{0,1} &= \delb v^{1,0} \\
    \delta \varepsilon_{0,2} &= \delb \lambda_{0,1} + \eta^{-1}\tilde{\CF_{0,1}}\wedge \lambda_{0,1} -\eta^{-1}\epsilon_{ij}q^{i}v^{1,0}\lrcorner \CF_{1,2}^{j} \\
    \delta \chi_{0,0} &= \eta^{-1}v^{1,0}\lrcorner \CF_{1,0} \\
    \delta \chi_{1,1} &= \delb \rho_{1,0} + \del \rho_{0,1} - \eta^{-1}\CF_{1,0}\wedge \lambda_{0,1} + \eta^{-1}\epsilon_{ij}s^{i} v^{1,0}\lrcorner \CF_{2,1}^{j} \\
    \delta \chi_{0,2} &= \delb \rho_{0,1} \\
    \delb \chi_{2,2} &= \delb \rho_{2,1} + \del\rho_{1,2} - v^{1,0}\CF_{3,2} - \epsilon_{ij}s^{i}\lambda_{0,1}\wedge\CF_{2,1}^{j} - \epsilon_{ij}q^{i}\rho_{1,0}\wedge \CF^{j}_{1,2} - \epsilon_{ij}q^{i}\rho_{0,1}\CF_{2,1}^{j} \\
    \delta\chi_{1,3} &= \delb \rho_{1,2} + \del \rho_{0,3} - \epsilon_{ij}q^{i}\rho_{0,1}\wedge \CF_{1,2}^{j} \\
    \delta \chi_{3,3} &= \delb \rho_{3,2} + \del \rho_{2,3} + \lambda_{0,1}\wedge \CF_{3,2} + \eta^{-1} \tilde{\CF}_{1}\wedge \rho_{5} - \eta^{-1}\epsilon_{ij}q^{i}\rho_{3}\wedge \CF^{j}_{3} \\
    \delta\Theta_{3,3} &= \delb \sigma_{3,2} + \del \sigma_{2,3} - \eta^{-1}\CF_{1,0}\wedge \rho_{2,3} + \eta^{-1}\epsilon_{ij}s^{i}\rho_{1,2}\wedge \CF^{j}_{2,1} + \rho_{0,1}\wedge \CF_{3,2}
\end{aligned}
\end{equation}

The presence of both the $\del$ and $\delb$ operators in these equations makes them difficult to solve in general. However, if we assume that the $\del\delb$ lemma holds then we can make simplifications. We will also assume that the complex functions $a,b$ used in the description of the $\SU{3}$ structure background are both non-vanishing. The case in which one of them vanishes was addressed in the main text. Note that the Killing spinor equations imply that either $a,b$ are globally non-vanishing or one of them is the 0-function and hence it is consistent to consider these cases separately.

Considering these simplifications, we find the following constraints for the closure of the complex (i.e. conditions for involutivity, or $\dd_{1}R=0$).
\begin{equation}\label{eq:ab=/=0_closure}
    \begin{aligned}
        \delb \varepsilon^{2,0} &= 0 \\
        \delb \varepsilon^{1,0}_{0,1} - j\varepsilon^{2,0}\lrcorner j^{2} H_{1,2} &= 0 \\
        \delb\varepsilon_{0,2} - \varepsilon^{1,0}_{0,1}\cdot H_{1,2} &= 0 \\
        \delb \chi_{0,0} - \varepsilon^{1,0}_{0,1}\cdot G_{1,0} - \varepsilon^{2,0}\lrcorner G_{2,1} &= 0 \\
        \delb \chi_{1,1} + H_{1,2}\wedge \chi_{0,0} - G_{1,0}\wedge \varepsilon_{0,2} - \varepsilon^{1,0}_{0,1}\cdot G_{2,1} - \varepsilon^{2,0}\lrcorner G_{3,2} &= 0 \\
        \delb \chi_{0,2} &= 0 \\
        \delb \chi_{2,2} + \chi_{1,1}\wedge H_{1,2} - \varepsilon_{0,2}\wedge G_{2,1} - \varepsilon^{1,0}_{0,1}\cdot G_{3,2}&= 0
    \end{aligned}
\end{equation}
Note that in the above expressions we have redefined some of the parameters to absorb factors of $ab$ and $\del$-exact terms. We have also defined the complex fluxes
\begin{equation}
    G_{1} = (ab)^{-2}\CF_{1}\, , \qquad G_{3} = -(ab)^{-1}\epsilon_{ij}s^{i}\CF^{j}_{3} \, , \qquad G_{5} = \CF_{5}
\end{equation}
The exactness conditions can be expressed as
\begin{equation}\label{eq:ab=/=0_exact}
    \begin{aligned}
        \delta \varepsilon^{1,0}_{0,1} &= \delb v^{1,0} \\
        \delta \varepsilon_{0,2} &= \delb \lambda_{0,1} - v^{1,0}\lrcorner H_{1,2} \\
        \delta \chi_{0,0} &= -v^{1,0}\lrcorner G_{1,0} \\
        \delta \chi_{1,1} &= \delb \rho_{1,0} +\lambda_{0,1}\wedge G_{1,0} - v^{1,0} \lrcorner G_{2,1} \\
        \delta \chi_{0,2} &= \delb \rho_{0,1} \\
        \delta \chi_{2,2} &= \delb \rho_{2,1} + H_{1,2} \wedge \rho_{1,0} + \lambda_{0,1} \wedge G_{2,1} - v^{1,0}\lrcorner G_{3,2} \\
        \delta \chi_{1,3} &= \delb \rho_{1,2} + H_{1,2} \wedge \rho_{0,1} \\
        \delta \chi_{3,3} &= \delb \rho_{3,2} +H_{1,2}\wedge \rho_{2,1} + \lambda_{0,1}\wedge G_{3,2} \\
        \delta \Theta_{3,3} &= \delb \sigma_{3,2} + \rho_{0,1}\wedge G_{3,2} - \rho_{1,2}\wedge G_{2,1} + \rho_{2,3}\wedge G_{1,0} 
    \end{aligned}
\end{equation}

In this case, we can rewrite these expressions in terms of $\Orth{6,6}$ generalised geometry objects (see e.g. \cite{Gualtieri:2003dx} for details). In particular, the complex structure $J$ of IIB backgrounds defines a generalised complex structure $\mathcal{J}_{-} = -J$ which defines some $\Uni{3,3}$ structure. collecting objects into representations of this larger structure group, we find that the deformation parameters can be written
\begin{equation}
    \varepsilon \in \Gamma(\ext^{2}\bar{L})\, , \qquad \chi \in \Gamma(S_{0}\oplus S_{-2})\, , \qquad \Theta \in \Gamma(\ext^{6}T^{*})
\end{equation}
where $L = T^{0,1}\oplus T^{*1,0}$ and $S_{n} = \bigoplus_{p}\ext^{p,p-n}T^{*}$. Similarly, the parameters for trivial deformations can be collected into terms
\begin{equation}
    V = v^{1,0}+\lambda_{0,1} \in \Gamma(\bar{L})\, , \qquad \rho \in \Gamma(S_{1}\oplus S_{-1})\, , \qquad \sigma \in \Gamma(\ext^{5}T^{*})
\end{equation}
we can then write the conditions from \eqref{eq:ab=/=0_exact} and \eqref{eq:ab=/=0_exact} into the following complex.
\begin{equation}\label{eq:GMPT_moduli_complex}
    \begin{tikzcd}
        \bar{L} \arrow[r,"\delb_{H}"] \arrow[dr, "G"] & \ext^{2}\bar{L} \arrow[r,"\delb_{H}"] \arrow[dr, "G"] & \ext^{3}\bar{L} \\
        S_{1}\oplus S_{-1} \arrow[r,"\delb_{H}"] \arrow[dr, "G"] & S_{0}\oplus S_{-2} \arrow[r,"\delb_{H}"] &S_{-1} \oplus S_{-3} \\
        \ext^{3,2}T^{*} \arrow[r,"\delb"] & \ext^{3,3}T^{*} &
    \end{tikzcd}
\end{equation}
The diagonal arrows come from viewing $G \in \Gamma(S_{-}) = \Omega^{\text{odd}}(M)$ as a generalised spinor of negative chirality. The top row of arrows then comes from the Clifford action of $\bar{L}$ on $S_{-}$, while the bottom row of arrows corresponds to the Mukai pairing on $S_{-}$. We can then perform a spectral sequence analysis to obtain the exact moduli from this complex. We will write $H^{p}_{L}$ for the cohomology of $\delb_{H}$ on $\ext^{p}\bar{L}$, and $H^{p}_{S}$ for the cohomology of $S_{p}$. 

In principal, one finds a non-trivial second page of the spectral sequence which relates $\ker(G:H^{1}_{L}\to H^{0}_{S}\oplus H^{-2}_{S})$ and $H^{3,3}_{\delb}(M)/\im(G:H^{1}_{S}\oplus H^{-1}_{S}\to H^{3,3}_{\delb})$. However, provided $G$ is non-trivial, this second term vanishes and hence the spectral sequence terminates. Hence, we find that the physical moduli are given by
\begin{equation}
    \text{moduli} \cong \ker[G:H^{2}_{L}\to H^{-1}_{S}\oplus H^{-3}_{S}] \oplus \frac{H^{0}_{S}\oplus H^{-2}_{S}}{\im[G:H^{1}_{L}\to H^{0}_{S}\oplus H^{-2}_{S}]}
\end{equation}

We would like to further express the $H^{p}_{L}, H^{p}_{S}$ in terms of conventional Dolbeault cohomology groups, for which a further spectral sequence is required. For the $H_{L}$ cohomologies, we decompose the first line of \eqref{eq:GMPT_moduli_complex} into $\GL{3,\bbC}$ bundles and find
\begin{equation}
    \begin{tikzcd}
        \bbC \arrow[r,"\delb"] &T^{*0,1} \arrow[r, "\delb"] & \ext^{0,2}T^{*} \arrow[r, "\delb"] & \ext^{0,3}T^{*}& \\
        & T^{1,0} \arrow[r,"\delb"] \arrow[ur, "H"]  & T^{*0,1}\otimes T^{1,0} \arrow[r,"\delb"] \arrow[ur, "H"] & (\ext^{0,2}T^{*})\otimes T^{1,0} \arrow[r,"\delb"] &... \\
        & & \ext^{2,0}T \arrow[r,"\delb"] \arrow[ur, "H"] & T^{*0,1}\otimes (\ext^{2,0}T) \arrow[r,"\delb"] \arrow[ur, "H"] & ... \\
        &&& \ext^{3,0}T \arrow[r,"\delb"] \arrow[ur,"H"]& ...
    \end{tikzcd}
\end{equation}
To find the moduli, we only need $H^{1}_{L}$ and $H^{2}_{L}$. In principal, one finds a non-tricial second page of the spectral sequence relating $\ker(H_{1,2}: H^{0,0}(\ext^{2,0}T) \to H^{0,2}(T^{1,0}))$ and $H^{0,3}(M)/\im(H_{1,2}: H^{0,1}(T^{1,0})\to H^{0,3})$. However, if we impose the topology of a Calabi--Yau then $H^{0,0}(\ext^{2,0}T)$ vanishes and hence the spectral sequence terminates at the second page. Working through the calculation and imposing the cohomology relations of a Calabi--Yau, we find
\begin{equation}
    \begin{aligned}
        H^{1}_{L} &= 0 \\
        H^{2}_{L} &= \ker[H_{1,2} : H^{0,1}(T^{1,0}) \to H^{0,3}(M)]
    \end{aligned}
\end{equation}

For the $H_{S}$ cohomology groups, we need to analyse the following complex.
\begin{equation}
    \begin{tikzcd}
        ... \arrow[r,"\delb"] & \ext^{3,2}T^{*} \arrow[r,"\delb"] & \ext^{3,3}T^{*} & & & \\
        ... \arrow[r,"\delb"] \arrow[ur, "H"] & \ext^{2,1}T^{*} \arrow[r,"\delb"] \arrow[ur, "H"] & \ext^{2,2}T^{*} \arrow[r,"\delb"]  & \ext^{2,3}T^{*} && \\
         & \ext^{1,0}T^{*} \arrow[r,"\delb"] \arrow[ur, "H"] & \ext^{1,1}T^{*} \arrow[r,"\delb"] \arrow[ur, "H"] & \ext^{1,2}T^{*} \arrow[r,"\delb"] & \ext^{1,3}T^{*} & \\
        && \bbC \arrow[r,"\delb"] \arrow[ur, "H"] & \ext^{0,1}T^{*} \arrow[r,"\delb"] \arrow[ur, "H"] & \ext^{0,2}T^{*} \arrow[r,"\delb"] & \ext^{0,3}T^{*}
    \end{tikzcd}
\end{equation}
Once again, we find that if we impose the identities for Calabi--Yau cohomology groups, the second page of the spectral sequence vanishes and we have
\begin{equation}
    \begin{aligned}
        H^{0}_{S} &= \ker[H_{1,2}:H^{0,0}(M)\to H^{1,2}(M)] \oplus H^{1,1}(M)\oplus H^{2,2}(M) \\
        & \qquad \oplus \frac{ H^{3,3}(M)}{\im[H_{1,2}: H^{2,1}(M)\to H^{3,3}(M)]} \\
        H^{-1}_{S} &= H^{1,2}(M)/\left<H_{1,2}\right> \\
        H^{-2}_{S} &= 0 \\
        H^{-3}_{S} &= H^{0,3}(M)
    \end{aligned}
\end{equation}
If $H\neq 0$ then the $H^{0,0}$ and $H^{3,3}$ terms are removed by the kernel and image of the $H$ map, respectively, and hence the moduli in this case are
\begin{equation}
    \text{moduli} \cong H^{0,1}_{\delb,G,H}(T^{1,0}) \oplus H^{1,1}(M)\oplus H^{2,2}(M)
\end{equation}
where
\begin{equation}
    H^{0,1}_{\delb,G,H}(T^{1,0}) = \ker[G: H^{0,1}(T^{1,0}) \to H^{1,2}(M)/ \left<H_{1,2} \right>] \cap \ker[H_{1,2} : H^{0,1}(T^{1,0}) \to H^{0,3}(M)]
\end{equation}
This is in agreement with the calculation done for the Gra\~na--Polchinski background done around \eqref{eq:GP_moduli}. If $H=0$, which is the case for the D5/O5 source backgrounds, we get an additional factor of $H^{0,0}$ and $H^{3,3}$ in $H^{0}_{S}$ and therefore get
\begin{equation}
    \text{moduli} \cong H^{0,1}_{\delb,G}(T^{1,0}) \oplus H^{0,0}\oplus H^{1,1}\oplus H^{2,2} \oplus H^{3,3}
\end{equation}
where
\begin{equation}
    H^{0,1}_{\delb,G}(T^{1,0}) = \ker[G: H^{0,1}(T^{1,0}) \to H^{1,2}(M)] 
\end{equation}
in agreement with \eqref{eq:D5/O5_moduli}. Of course, one should also restrict to the moduli with definite parity under any orientifold plane sources in the problem.

\subsection{IIA Type 3}
The general moduli equations for type 3 backgrounds in IIA are given by \eqref{App:IIA_type_3_moduli}. In section \ref{sec:IIA_moduli}, we calculate the moduli associated to this system of equations assuming the $\del\delb$-lemma holds.

\begin{equation}\label{App:IIA_type_3_moduli}
\begin{aligned}
    \delb\kappa &= 0, \\ 
    \delb v + \kappa \cdot \CH &= 0 \\
    \delb r +\kappa\cdot \CF_4 &= 0,\qquad \delta r = \delb x \\
    \delb\phi_{0,3} &= 0, \qquad \delta\phi_{0,3} = \delb\nu_{0,2}\\
    \delb\phi_{1,2} + \del\phi_{0,3}-r\cdot \CF_4 &= 0,\qquad \delta\phi_{1,2} = \delb \nu_{1,1} + \del\nu_{0,2} - x\cdot \CF_4 \\
    \delb\theta_{0,2} &= 0, \qquad \delta\theta_{0,2} = \delb\lambda_{0,1} \\ 
    \delb\theta_{1,1}+\del\theta_{0,2} - r\cdot \CH - v\cdot \CF_4 &= 0, \qquad \delta\theta_{1,1} = \delb\lambda_{1,0} + \del\lambda_{0,1} - x\cdot \CH \\
    \delb\mu_5-\theta_{1,1}\wedge\CF_4-\phi_{1,2}\wedge \CH &= 0, \qquad\delta\mu_5 = \delb\nu_4 - \CF_4\wedge\lambda - \CH\wedge\nu,\\
    & \ \qquad \qquad \delta\mu_6 = \delb\rho_5 - \CF_4\wedge\nu
\end{aligned}
\end{equation}

\section{Cohomologies for sources}\label{app:Cohomologies_for_sources}

\subsection{Relative cohomology}

We review the definition of relative de Rham cohomology \cite{Bott:1982xhp}. Similar definitions give relative Dolbeault cohomology and relative $\dd_{H}$ cohomology. In fact, one can define a notion of relative homology for more general chain and cochain sequences \cite{Hatcher_AT}.

Let $i:M'\to M$ define a map between smooth manifolds (it need not be an inclusion). We define the set $\Omega^{n}(M,M') = \Omega^{n}(M)\oplus \Omega^{n-1}(M')$ and a map $\dd:\Omega^{n}(M,M')\to \Omega^{n+1}(M,M')$ by
\begin{equation}
    \dd(\alpha,\beta) = (\dd\alpha, i^{*}\alpha-\dd\beta)
\end{equation}
where on the right hand side $\dd$ denotes the usual de Rham differential. It is easy to check that $\dd^{2} = 0$ on $\Omega^{*}(M,M')$, and the relative cohomology $H^{*}(M,M')$ is define to be the cohomology of $(\Omega^{*}(M,M'),\dd)$. Note that it consists of objects that are closed on $M$ which are exact when pulled back to $M'$.

We can also define a short exact sequence
\begin{equation}
    0\to \Omega^{n-1}(M')\xrightarrow{\eta} \Omega^{n}(M,M') \xrightarrow{\zeta} \Omega^{n}(M) \to 0
\end{equation}
where $\eta(\beta) = (0,\beta)$ and $\zeta(\alpha,\beta) = \alpha$. Both $\eta,\zeta$ define chain maps\footnote{In fact $\eta$ satisfies $\eta\dd= -\dd\eta$ but that is enough for the long exact sequence to exist.} and hence we get the long exact sequence of maps \eqref{eq:long-exact}.
\begin{equation}
\label{eq:long-exact2}
    \begin{tikzcd}
        \dots\arrow[r,"\zeta^{*}"]& H^{n}(M)\arrow[r, "i^*"]&H^{n}(M')\arrow[r,"\eta^{*}"]& H^{n+1}(M,M')\arrow[r,"\zeta^{*}"]&H^{n+1}(M)\arrow[r, "i^*"]&\dots
    \end{tikzcd}
\end{equation}

\subsection{Sources and deformations}

Supergravity theories contain higher-form field strengths $F_{n} \in \Omega^{n}(M)$ which must satisfy Bianchi identities such as \eqref{eq:bianchi}. For simplicity we will consider a  theory without Chern--Simons-like terms, in which the Bianchi identity takes the form
\begin{equation}\label{eq:simplified_bianchi}
    \dd F_{n} = j_{n+1}
\end{equation}
where $j_{n+1}$ is a current with either $\delta$-function support on the magnetic sources $\Sigma$, or with some smoothed out distribution in some small neighbourhood of $\Sigma$. In terms of the RR Bianchi identity \eqref{eq:bianchi}, one can imagine that we have set $H=0$ in the original theory, or alternatively can consider a variation of this discussion with relative $\dd_{H}$ cohomology.

As in section \ref{sec:sources_note}, we define $M'=M-N$, where $N \supset \mathrm{Supp}(j_{n+1})$ is some small neighbourhood of the sources, and $i:M'\to M$ is the inclusion map. Conservation of the current tells us that $j_{n+1}$ must be closed, while \eqref{eq:simplified_bianchi} tells us it is trivial in cohomology on $M$. Moreover, in relative cohomology, we have
\begin{equation}
    (j_{n+1},i^{*}F_{n}) = \dd(F_{n},0)
\end{equation}
and hence is trivial. The element $(j_{n+1},0)\sim (0,-i^{*}F_{n})$ is also closed in $\Omega^{n+1}(M,M')$ precisely because $i^{*}j_{n+1} = 0$ and hence defines a cohomology element which is in general not trivial.
\begin{equation}
    (j_{n+1},0)\sim(0,-i^{*}F_{n}) \in H^{n+1}(M,M')
\end{equation}

Suppose now we want to define a small deformation of the sources. This is given by
\begin{equation}
    \delta j_{n+1} = \dd\gamma \, , \quad \mathrm{Supp}(\delta j_{n+1}) \subset N \ .
\end{equation}
By the above argument, we must also have
\begin{equation}
    (\delta j_{n+1},0) \sim (0,-i^{*}\gamma) \in H^{n+1}(M,M') \ .
\end{equation}
Since we have assumed that the support of the deformation is contained within the excised manifold $N$, we must have that $i^{*}\dd\gamma = 0$ and hence $i^{*}\gamma \in H^{n}(M')$. From the definition of the long exact sequence of cohomologies \eqref{eq:long-exact2}, we find that the deformations of the sources are not generic elements of the relative cohomology, but instead must lie in the image of $\eta^{*}$.
\begin{equation}
    (\delta j_{n+1},0) \sim (0,-i^{*}\gamma) \in \im(\eta^{*}: H^{n}(M') \to H^{n+1}(M,M')) \simeq \frac{H^{n}(M')}{i^{*}H^{n}(M)}
\end{equation}
where in the last equality, we have used the fact that \eqref{eq:long-exact2} is exact. Note that this final equality reflects our intuition about the deformations of the sources. Indeed, if $\delta j_{n+1} = \dd \gamma$, then the deformations are defined by some $\gamma$ which is not closed on $M$ but is closed when pulled back to $M'$. It is defined up to addition of some closed form on $M$. The quotient given in the final equality is precisely the set of such objects.

\bibliographystyle{JHEP}
\bibliography{references.bib}

\providecommand{\href}[2]{#2}\begingroup\raggedright\begin{thebibliography}{100}

\bibitem{Grana:2005jc}
M.~Grana, \emph{{Flux compactifications in string theory: A Comprehensive
  review}}, \href{https://doi.org/10.1016/j.physrep.2005.10.008}{\emph{Phys.
  Rept.} {\bfseries 423} (2006) 91}
  [\href{https://arxiv.org/abs/hep-th/0509003}{{\ttfamily hep-th/0509003}}].

\bibitem{Blumenhagen:2006ci}
R.~Blumenhagen, B.~Kors, D.~Lust and S.~Stieberger, \emph{{Four-dimensional
  String Compactifications with D-Branes, Orientifolds and Fluxes}},
  \href{https://doi.org/10.1016/j.physrep.2007.04.003}{\emph{Phys. Rept.}
  {\bfseries 445} (2007) 1}
  [\href{https://arxiv.org/abs/hep-th/0610327}{{\ttfamily hep-th/0610327}}].

\bibitem{McAllister:2023vgy}
L.~McAllister and F.~Quevedo, \emph{{Moduli Stabilization in String Theory}},
  \href{https://arxiv.org/abs/2310.20559}{{\ttfamily 2310.20559}}.

\bibitem{Taylor:1999ii}
T.~R. Taylor and C.~Vafa, \emph{{RR flux on Calabi-Yau and partial
  supersymmetry breaking}},
  \href{https://doi.org/10.1016/S0370-2693(00)00005-8}{\emph{Phys. Lett. B}
  {\bfseries 474} (2000) 130}
  [\href{https://arxiv.org/abs/hep-th/9912152}{{\ttfamily hep-th/9912152}}].

\bibitem{Dasgupta:1999ss}
K.~Dasgupta, G.~Rajesh and S.~Sethi, \emph{{M theory, orientifolds and G -
  flux}}, \href{https://doi.org/10.1088/1126-6708/1999/08/023}{\emph{JHEP}
  {\bfseries 08} (1999) 023}
  [\href{https://arxiv.org/abs/hep-th/9908088}{{\ttfamily hep-th/9908088}}].

\bibitem{Giddings:2001yu}
S.~B. Giddings, S.~Kachru and J.~Polchinski, \emph{{Hierarchies from fluxes in
  string compactifications}},
  \href{https://doi.org/10.1103/PhysRevD.66.106006}{\emph{Phys. Rev. D}
  {\bfseries 66} (2002) 106006}
  [\href{https://arxiv.org/abs/hep-th/0105097}{{\ttfamily hep-th/0105097}}].

\bibitem{Gukov:1999ya}
S.~Gukov, C.~Vafa and E.~Witten, \emph{{CFT's from Calabi-Yau four folds}},
  \href{https://doi.org/10.1016/S0550-3213(00)00373-4}{\emph{Nucl. Phys. B}
  {\bfseries 584} (2000) 69}
  [\href{https://arxiv.org/abs/hep-th/9906070}{{\ttfamily hep-th/9906070}}].

\bibitem{Grana:2000jj}
M.~Grana and J.~Polchinski, \emph{{Supersymmetric three form flux perturbations
  on AdS(5)}}, \href{https://doi.org/10.1103/PhysRevD.63.026001}{\emph{Phys.
  Rev. D} {\bfseries 63} (2001) 026001}
  [\href{https://arxiv.org/abs/hep-th/0009211}{{\ttfamily hep-th/0009211}}].

\bibitem{Grana:2001xn}
M.~Grana and J.~Polchinski, \emph{{Gauge / gravity duals with holomorphic
  dilaton}}, \href{https://doi.org/10.1103/PhysRevD.65.126005}{\emph{Phys. Rev.
  D} {\bfseries 65} (2002) 126005}
  [\href{https://arxiv.org/abs/hep-th/0106014}{{\ttfamily hep-th/0106014}}].

\bibitem{Gubser:2000vg}
S.~S. Gubser, \emph{{Supersymmetry and F theory realization of the deformed
  conifold with three form flux}},
  \href{https://arxiv.org/abs/hep-th/0010010}{{\ttfamily hep-th/0010010}}.

\bibitem{Bena:2020xrh}
I.~Bena, J.~Bl\r{a}b\"ack, M.~Gra\~na and S.~L\"ust, \emph{{The tadpole
  problem}}, \href{https://doi.org/10.1007/JHEP11(2021)223}{\emph{JHEP}
  {\bfseries 11} (2021) 223}
  [\href{https://arxiv.org/abs/2010.10519}{{\ttfamily 2010.10519}}].

\bibitem{DeWolfe:2002nn}
O.~DeWolfe and S.~B. Giddings, \emph{{Scales and hierarchies in warped
  compactifications and brane worlds}},
  \href{https://doi.org/10.1103/PhysRevD.67.066008}{\emph{Phys. Rev. D}
  {\bfseries 67} (2003) 066008}
  [\href{https://arxiv.org/abs/hep-th/0208123}{{\ttfamily hep-th/0208123}}].

\bibitem{Giddings:2005ff}
S.~B. Giddings and A.~Maharana, \emph{{Dynamics of warped compactifications and
  the shape of the warped landscape}},
  \href{https://doi.org/10.1103/PhysRevD.73.126003}{\emph{Phys. Rev. D}
  {\bfseries 73} (2006) 126003}
  [\href{https://arxiv.org/abs/hep-th/0507158}{{\ttfamily hep-th/0507158}}].

\bibitem{Shiu:2008ry}
G.~Shiu, G.~Torroba, B.~Underwood and M.~R. Douglas, \emph{{Dynamics of Warped
  Flux Compactifications}},
  \href{https://doi.org/10.1088/1126-6708/2008/06/024}{\emph{JHEP} {\bfseries
  06} (2008) 024} [\href{https://arxiv.org/abs/0803.3068}{{\ttfamily
  0803.3068}}].

\bibitem{Douglas:2008jx}
M.~R. Douglas and G.~Torroba, \emph{{Kinetic terms in warped
  compactifications}},
  \href{https://doi.org/10.1088/1126-6708/2009/05/013}{\emph{JHEP} {\bfseries
  05} (2009) 013} [\href{https://arxiv.org/abs/0805.3700}{{\ttfamily
  0805.3700}}].

\bibitem{Frey:2008xw}
A.~R. Frey, G.~Torroba, B.~Underwood and M.~R. Douglas, \emph{{The Universal
  Kahler Modulus in Warped Compactifications}},
  \href{https://doi.org/10.1088/1126-6708/2009/01/036}{\emph{JHEP} {\bfseries
  01} (2009) 036} [\href{https://arxiv.org/abs/0810.5768}{{\ttfamily
  0810.5768}}].

\bibitem{Burgess:2005jx}
C.~P. Burgess, C.~Escoda and F.~Quevedo, \emph{{Nonrenormalization of flux
  superpotentials in string theory}},
  \href{https://doi.org/10.1088/1126-6708/2006/06/044}{\emph{JHEP} {\bfseries
  06} (2006) 044} [\href{https://arxiv.org/abs/hep-th/0510213}{{\ttfamily
  hep-th/0510213}}].

\bibitem{Ashmore:2019qii}
A.~Ashmore, C.~Strickland-Constable, D.~Tennyson and D.~Waldram,
  \emph{{Generalising G$_\text{2}$ geometry: involutivity, moment maps and
  moduli}}, \href{https://doi.org/10.1007/JHEP01(2021)158}{\emph{JHEP}
  {\bfseries 01} (2021) 158}
  [\href{https://arxiv.org/abs/1910.04795}{{\ttfamily 1910.04795}}].

\bibitem{Grana:2004bg}
M.~Grana, R.~Minasian, M.~Petrini and A.~Tomasiello, \emph{{Supersymmetric
  backgrounds from generalized Calabi-Yau manifolds}},
  \href{https://doi.org/10.1088/1126-6708/2004/08/046}{\emph{JHEP} {\bfseries
  08} (2004) 046} [\href{https://arxiv.org/abs/hep-th/0406137}{{\ttfamily
  hep-th/0406137}}].

\bibitem{Coimbra:2011ky}
A.~Coimbra, C.~Strickland-Constable and D.~Waldram, \emph{{$E_{d(d)} \times
  \mathbb{R}^+$ generalised geometry, connections and M theory}},
  \href{https://doi.org/10.1007/JHEP02(2014)054}{\emph{JHEP} {\bfseries 02}
  (2014) 054} [\href{https://arxiv.org/abs/1112.3989}{{\ttfamily 1112.3989}}].

\bibitem{Coimbra:2012af}
A.~Coimbra, C.~Strickland-Constable and D.~Waldram, \emph{{Supergravity as
  Generalised Geometry II: $E_{d(d)} \times \mathbb{R}^+$ and M theory}},
  \href{https://doi.org/10.1007/JHEP03(2014)019}{\emph{JHEP} {\bfseries 03}
  (2014) 019} [\href{https://arxiv.org/abs/1212.1586}{{\ttfamily 1212.1586}}].

\bibitem{Coimbra:2014uxa}
A.~Coimbra, C.~Strickland-Constable and D.~Waldram, \emph{{Supersymmetric
  Backgrounds and Generalised Special Holonomy}},
  \href{https://doi.org/10.1088/0264-9381/33/12/125026}{\emph{Class. Quant.
  Grav.} {\bfseries 33} (2016) 125026}
  [\href{https://arxiv.org/abs/1411.5721}{{\ttfamily 1411.5721}}].

\bibitem{Ashmore:2015joa}
A.~Ashmore and D.~Waldram, \emph{{Exceptional Calabi-Yau spaces: the geometry
  of $\mathcal{N}=2$ backgrounds with flux}},
  \href{https://doi.org/10.1002/prop.201600109}{\emph{Fortsch. Phys.}
  {\bfseries 65} (2017) 1600109}
  [\href{https://arxiv.org/abs/1510.00022}{{\ttfamily 1510.00022}}].

\bibitem{Gauntlett:2002sc}
J.~P. Gauntlett, D.~Martelli, S.~Pakis and D.~Waldram, \emph{{G structures and
  wrapped NS5-branes}},
  \href{https://doi.org/10.1007/s00220-004-1066-y}{\emph{Commun. Math. Phys.}
  {\bfseries 247} (2004) 421}
  [\href{https://arxiv.org/abs/hep-th/0205050}{{\ttfamily hep-th/0205050}}].

\bibitem{Gauntlett:2002fz}
J.~P. Gauntlett and S.~Pakis, \emph{{The Geometry of D = 11 killing spinors}},
  \href{https://doi.org/10.1088/1126-6708/2003/04/039}{\emph{JHEP} {\bfseries
  04} (2003) 039} [\href{https://arxiv.org/abs/hep-th/0212008}{{\ttfamily
  hep-th/0212008}}].

\bibitem{Gauntlett:2003wb}
J.~P. Gauntlett, J.~B. Gutowski and S.~Pakis, \emph{{The Geometry of D = 11
  null Killing spinors}},
  \href{https://doi.org/10.1088/1126-6708/2003/12/049}{\emph{JHEP} {\bfseries
  12} (2003) 049} [\href{https://arxiv.org/abs/hep-th/0311112}{{\ttfamily
  hep-th/0311112}}].

\bibitem{Grana:2005sn}
M.~Grana, R.~Minasian, M.~Petrini and A.~Tomasiello, \emph{{Generalized
  structures of N=1 vacua}},
  \href{https://doi.org/10.1088/1126-6708/2005/11/020}{\emph{JHEP} {\bfseries
  11} (2005) 020} [\href{https://arxiv.org/abs/hep-th/0505212}{{\ttfamily
  hep-th/0505212}}].

\bibitem{Martucci:2006ij}
L.~Martucci, \emph{{D-branes on general N=1 backgrounds: Superpotentials and
  D-terms}}, \href{https://doi.org/10.1088/1126-6708/2006/06/033}{\emph{JHEP}
  {\bfseries 06} (2006) 033}
  [\href{https://arxiv.org/abs/hep-th/0602129}{{\ttfamily hep-th/0602129}}].

\bibitem{Koerber:2007xk}
P.~Koerber and L.~Martucci, \emph{{From ten to four and back again: How to
  generalize the geometry}},
  \href{https://doi.org/10.1088/1126-6708/2007/08/059}{\emph{JHEP} {\bfseries
  08} (2007) 059} [\href{https://arxiv.org/abs/0707.1038}{{\ttfamily
  0707.1038}}].

\bibitem{Koerber:2006hh}
P.~Koerber and L.~Martucci, \emph{{Deformations of calibrated D-branes in flux
  generalized complex manifolds}},
  \href{https://doi.org/10.1088/1126-6708/2006/12/062}{\emph{JHEP} {\bfseries
  12} (2006) 062} [\href{https://arxiv.org/abs/hep-th/0610044}{{\ttfamily
  hep-th/0610044}}].

\bibitem{Tomasiello:2007zq}
A.~Tomasiello, \emph{{Reformulating supersymmetry with a generalized Dolbeault
  operator}}, \href{https://doi.org/10.1088/1126-6708/2008/02/010}{\emph{JHEP}
  {\bfseries 02} (2008) 010} [\href{https://arxiv.org/abs/0704.2613}{{\ttfamily
  0704.2613}}].

\bibitem{Martucci:2009sf}
L.~Martucci, \emph{{On moduli and effective theory of N=1 warped flux
  compactifications}},
  \href{https://doi.org/10.1088/1126-6708/2009/05/027}{\emph{JHEP} {\bfseries
  05} (2009) 027} [\href{https://arxiv.org/abs/0902.4031}{{\ttfamily
  0902.4031}}].

\bibitem{Tseng:2009gr}
L.-S. Tseng and S.-T. Yau, \emph{{Cohomology and Hodge Theory on Symplectic
  Manifolds. I.}}, {\emph{J. Diff. Geom.} {\bfseries 91} (2012) 383}
  [\href{https://arxiv.org/abs/0909.5418}{{\ttfamily 0909.5418}}].

\bibitem{Tseng:2010kt}
L.-S. Tseng and S.-T. Yau, \emph{{Cohomology and Hodge Theory on Symplectic
  Manifolds. II}}, {\emph{J. Diff. Geom.} {\bfseries 91} (2012) 417}
  [\href{https://arxiv.org/abs/1011.1250}{{\ttfamily 1011.1250}}].

\bibitem{Tseng:2011gv}
L.-S. Tseng and S.-T. Yau, \emph{{Generalized Cohomologies and Supersymmetry}},
  \href{https://doi.org/10.1007/s00220-014-1895-2}{\emph{Commun. Math. Phys.}
  {\bfseries 326} (2014) 875}
  [\href{https://arxiv.org/abs/1111.6968}{{\ttfamily 1111.6968}}].

\bibitem{Smith:2022baw}
G.~R. Smith and D.~Waldram, \emph{{M-theory Moduli from Exceptional Complex
  Structures}},  \href{https://arxiv.org/abs/2211.09517}{{\ttfamily
  2211.09517}}.

\bibitem{Gualtieri:2003dx}
M.~Gualtieri, \emph{{Generalized complex geometry}}, Ph.D. thesis, University
  of Oxford, 2003.
\newblock \href{https://arxiv.org/abs/math/0401221}{{\ttfamily math/0401221}}.

\bibitem{Cavalcanti:2005hq}
G.~R. Cavalcanti, \emph{{New aspects of the $dd^c$-lemma}}, Ph.D. thesis,
  University of Oxford, 2004.
\newblock \href{https://arxiv.org/abs/math/0501406}{{\ttfamily math/0501406}}.

\bibitem{Rohm:1985jv}
R.~Rohm and E.~Witten, \emph{{The Antisymmetric Tensor Field in Superstring
  Theory}}, \href{https://doi.org/10.1016/0003-4916(86)90099-0}{\emph{Annals
  Phys.} {\bfseries 170} (1986) 454}.

\bibitem{Anderson:2010mh}
L.~B. Anderson, J.~Gray, A.~Lukas and B.~Ovrut, \emph{{Stabilizing the Complex
  Structure in Heterotic Calabi-Yau Vacua}},
  \href{https://doi.org/10.1007/JHEP02(2011)088}{\emph{JHEP} {\bfseries 02}
  (2011) 088} [\href{https://arxiv.org/abs/1010.0255}{{\ttfamily 1010.0255}}].

\bibitem{Anderson:2011ty}
L.~B. Anderson, J.~Gray, A.~Lukas and B.~Ovrut, \emph{{The Atiyah Class and
  Complex Structure Stabilization in Heterotic Calabi-Yau Compactifications}},
  \href{https://doi.org/10.1007/JHEP10(2011)032}{\emph{JHEP} {\bfseries 10}
  (2011) 032} [\href{https://arxiv.org/abs/1107.5076}{{\ttfamily 1107.5076}}].

\bibitem{Gray:2018kss}
J.~Gray and H.~Parsian, \emph{{Moduli identification methods in Type II
  compactifications}},
  \href{https://doi.org/10.1007/JHEP07(2018)158}{\emph{JHEP} {\bfseries 07}
  (2018) 158} [\href{https://arxiv.org/abs/1803.08176}{{\ttfamily
  1803.08176}}].

\bibitem{Coimbra:2011nw}
A.~Coimbra, C.~Strickland-Constable and D.~Waldram, \emph{{Supergravity as
  Generalised Geometry I: Type II Theories}},
  \href{https://doi.org/10.1007/JHEP11(2011)091}{\emph{JHEP} {\bfseries 11}
  (2011) 091} [\href{https://arxiv.org/abs/1107.1733}{{\ttfamily 1107.1733}}].

\bibitem{Bergshoeff:2001pv}
E.~Bergshoeff, R.~Kallosh, T.~Ortin, D.~Roest and A.~Van~Proeyen, \emph{{New
  formulations of D = 10 supersymmetry and D8 - O8 domain walls}},
  \href{https://doi.org/10.1088/0264-9381/18/17/303}{\emph{Class. Quant. Grav.}
  {\bfseries 18} (2001) 3359}
  [\href{https://arxiv.org/abs/hep-th/0103233}{{\ttfamily hep-th/0103233}}].

\bibitem{Chiossi:2002tw}
S.~Chiossi and S.~Salamon, \emph{{The Intrinsic torsion of SU(3) and G(2)
  structures}},  in \emph{{International Conference on Differential Geometry
  held in honor of the 60th Birthday of A.M. Naveira}}, 2, 2002,
  \href{https://arxiv.org/abs/math/0202282}{{\ttfamily math/0202282}}.

\bibitem{Collinucci:2008pf}
A.~Collinucci, F.~Denef and M.~Esole, \emph{{D-brane Deconstructions in IIB
  Orientifolds}},
  \href{https://doi.org/10.1088/1126-6708/2009/02/005}{\emph{JHEP} {\bfseries
  02} (2009) 005} [\href{https://arxiv.org/abs/0805.1573}{{\ttfamily
  0805.1573}}].

\bibitem{Maharana:2012tu}
A.~Maharana and E.~Palti, \emph{{Models of Particle Physics from Type IIB
  String Theory and F-theory: A Review}},
  \href{https://doi.org/10.1142/S0217751X13300056}{\emph{Int. J. Mod. Phys. A}
  {\bfseries 28} (2013) 1330005}
  [\href{https://arxiv.org/abs/1212.0555}{{\ttfamily 1212.0555}}].

\bibitem{Hull:1986kz}
C.~M. Hull, \emph{{Compactifications of the Heterotic Superstring}},
  \href{https://doi.org/10.1016/0370-2693(86)91393-6}{\emph{Phys. Lett. B}
  {\bfseries 178} (1986) 357}.

\bibitem{Strominger:1986uh}
A.~Strominger, \emph{{Superstrings with Torsion}},
  \href{https://doi.org/10.1016/0550-3213(86)90286-5}{\emph{Nucl. Phys. B}
  {\bfseries 274} (1986) 253}.

\bibitem{Giryavets:2003vd}
A.~Giryavets, S.~Kachru, P.~K. Tripathy and S.~P. Trivedi, \emph{{Flux
  compactifications on Calabi-Yau threefolds}},
  \href{https://doi.org/10.1088/1126-6708/2004/04/003}{\emph{JHEP} {\bfseries
  04} (2004) 003} [\href{https://arxiv.org/abs/hep-th/0312104}{{\ttfamily
  hep-th/0312104}}].

\bibitem{Giryavets:2004zr}
A.~Giryavets, S.~Kachru and P.~K. Tripathy, \emph{{On the taxonomy of flux
  vacua}}, \href{https://doi.org/10.1088/1126-6708/2004/08/002}{\emph{JHEP}
  {\bfseries 08} (2004) 002}
  [\href{https://arxiv.org/abs/hep-th/0404243}{{\ttfamily hep-th/0404243}}].

\bibitem{DeWolfe:2005gy}
O.~DeWolfe, \emph{{Enhanced symmetries in multiparameter flux vacua}},
  \href{https://doi.org/10.1088/1126-6708/2005/10/066}{\emph{JHEP} {\bfseries
  10} (2005) 066} [\href{https://arxiv.org/abs/hep-th/0506245}{{\ttfamily
  hep-th/0506245}}].

\bibitem{Conlon:2004ds}
J.~P. Conlon and F.~Quevedo, \emph{{On the explicit construction and statistics
  of Calabi-Yau flux vacua}},
  \href{https://doi.org/10.1088/1126-6708/2004/10/039}{\emph{JHEP} {\bfseries
  10} (2004) 039} [\href{https://arxiv.org/abs/hep-th/0409215}{{\ttfamily
  hep-th/0409215}}].

\bibitem{Balasubramanian:2005zx}
V.~Balasubramanian, P.~Berglund, J.~P. Conlon and F.~Quevedo,
  \emph{{Systematics of moduli stabilisation in Calabi-Yau flux
  compactifications}},
  \href{https://doi.org/10.1088/1126-6708/2005/03/007}{\emph{JHEP} {\bfseries
  03} (2005) 007} [\href{https://arxiv.org/abs/hep-th/0502058}{{\ttfamily
  hep-th/0502058}}].

\bibitem{Conlon:2005ki}
J.~P. Conlon, F.~Quevedo and K.~Suruliz, \emph{{Large-volume flux
  compactifications: Moduli spectrum and D3/D7 soft supersymmetry breaking}},
  \href{https://doi.org/10.1088/1126-6708/2005/08/007}{\emph{JHEP} {\bfseries
  08} (2005) 007} [\href{https://arxiv.org/abs/hep-th/0505076}{{\ttfamily
  hep-th/0505076}}].

\bibitem{Candelas:1984yd}
P.~Candelas and D.~J. Raine, \emph{{Spontaneous Compactification and
  Supersymmetry in $d=11$ Supergravity}},
  \href{https://doi.org/10.1016/0550-3213(84)90604-7}{\emph{Nucl. Phys. B}
  {\bfseries 248} (1984) 415}.

\bibitem{Candelas:1985ux}
P.~Candelas, \emph{{Compactification and Supersymmetry of Chiral $N=2 D=10$
  Supergravity}},
  \href{https://doi.org/10.1016/0550-3213(85)90400-6}{\emph{Nucl. Phys. B}
  {\bfseries 256} (1985) 385}.

\bibitem{deWit:1986mwo}
B.~de~Wit, D.~J. Smit and N.~D. Hari~Dass, \emph{{Residual Supersymmetry of
  Compactified D=10 Supergravity}},
  \href{https://doi.org/10.1016/0550-3213(87)90267-7}{\emph{Nucl. Phys. B}
  {\bfseries 283} (1987) 165}.

\bibitem{Maldacena:2000mw}
J.~M. Maldacena and C.~Nunez, \emph{{Supergravity description of field theories
  on curved manifolds and a no go theorem}},
  \href{https://doi.org/10.1142/S0217751X01003937}{\emph{Int. J. Mod. Phys. A}
  {\bfseries 16} (2001) 822}
  [\href{https://arxiv.org/abs/hep-th/0007018}{{\ttfamily hep-th/0007018}}].

\bibitem{Gauntlett:2003cy}
J.~P. Gauntlett, D.~Martelli and D.~Waldram, \emph{{Superstrings with intrinsic
  torsion}}, \href{https://doi.org/10.1103/PhysRevD.69.086002}{\emph{Phys. Rev.
  D} {\bfseries 69} (2004) 086002}
  [\href{https://arxiv.org/abs/hep-th/0302158}{{\ttfamily hep-th/0302158}}].

\bibitem{Figueroa-OFarrill:2000lcd}
J.~M. Figueroa-O'Farrill and S.~Stanciu, \emph{{D-brane charge, flux
  quantization and relative (co)homology}},
  \href{https://doi.org/10.1088/1126-6708/2001/01/006}{\emph{JHEP} {\bfseries
  01} (2001) 006} [\href{https://arxiv.org/abs/hep-th/0008038}{{\ttfamily
  hep-th/0008038}}].

\bibitem{Alvarez:2003fw}
M.~Alvarez and D.~I. Olive, \emph{{Charges and fluxes in Maxwell theory on
  compact manifolds with boundary}},
  \href{https://doi.org/10.1007/s00220-006-0065-6}{\emph{Commun. Math. Phys.}
  {\bfseries 267} (2006) 279}
  [\href{https://arxiv.org/abs/hep-th/0303229}{{\ttfamily hep-th/0303229}}].

\bibitem{Moore-minicourse}
G.~W. Moore, ``A minicourse on generalized abelian gauge theory, self-dual
  theories, and differential cohomology.''
  \url{https://www.physics.rutgers.edu/~gmoore/SCGP-Minicourse.pdf}.

\bibitem{PiresPacheco:2008qik}
P.~Pires~Pacheco and D.~Waldram, \emph{{M-theory, exceptional generalised
  geometry and superpotentials}},
  \href{https://doi.org/10.1088/1126-6708/2008/09/123}{\emph{JHEP} {\bfseries
  09} (2008) 123} [\href{https://arxiv.org/abs/0804.1362}{{\ttfamily
  0804.1362}}].

\bibitem{Coimbra:2016ydd}
A.~Coimbra and C.~Strickland-Constable, \emph{{Supersymmetric Backgrounds, the
  Killing Superalgebra, and Generalised Special Holonomy}},
  \href{https://doi.org/10.1007/JHEP11(2016)063}{\emph{JHEP} {\bfseries 11}
  (2016) 063} [\href{https://arxiv.org/abs/1606.09304}{{\ttfamily
  1606.09304}}].

\bibitem{deWit:1986mz}
B.~de~Wit and H.~Nicolai, \emph{{$d=11$ Supergravity With Local SU(8)
  Invariance}}, \href{https://doi.org/10.1016/0550-3213(86)90290-7}{\emph{Nucl.
  Phys. B} {\bfseries 274} (1986) 363}.

\bibitem{Hohm:2013pua}
O.~Hohm and H.~Samtleben, \emph{{Exceptional Form of D=11 Supergravity}},
  \href{https://doi.org/10.1103/PhysRevLett.111.231601}{\emph{Phys. Rev. Lett.}
  {\bfseries 111} (2013) 231601}
  [\href{https://arxiv.org/abs/1308.1673}{{\ttfamily 1308.1673}}].

\bibitem{Hohm:2013vpa}
O.~Hohm and H.~Samtleben, \emph{{Exceptional Field Theory I: $E_{6(6)}$
  covariant Form of M-Theory and Type IIB}},
  \href{https://doi.org/10.1103/PhysRevD.89.066016}{\emph{Phys. Rev. D}
  {\bfseries 89} (2014) 066016}
  [\href{https://arxiv.org/abs/1312.0614}{{\ttfamily 1312.0614}}].

\bibitem{Hohm:2013uia}
O.~Hohm and H.~Samtleben, \emph{{Exceptional field theory. II. E$_{7(7)}$}},
  \href{https://doi.org/10.1103/PhysRevD.89.066017}{\emph{Phys. Rev. D}
  {\bfseries 89} (2014) 066017}
  [\href{https://arxiv.org/abs/1312.4542}{{\ttfamily 1312.4542}}].

\bibitem{Tennyson:2021qwl}
D.~Tennyson and D.~Waldram, \emph{{Exceptional complex structures and the
  hypermultiplet moduli of 5d Minkowski compactifications of M-theory}},
  \href{https://doi.org/10.1007/JHEP08(2021)088}{\emph{JHEP} {\bfseries 08}
  (2021) 088} [\href{https://arxiv.org/abs/2104.09900}{{\ttfamily
  2104.09900}}].

\bibitem{Grana:2005ny}
M.~Grana, J.~Louis and D.~Waldram, \emph{{Hitchin functionals in N=2
  supergravity}},
  \href{https://doi.org/10.1088/1126-6708/2006/01/008}{\emph{JHEP} {\bfseries
  01} (2006) 008} [\href{https://arxiv.org/abs/hep-th/0505264}{{\ttfamily
  hep-th/0505264}}].

\bibitem{Koerber:2008sx}
P.~Koerber and L.~Martucci, \emph{{Warped generalized geometry
  compactifications, effective theories and non-perturbative effects}},
  \href{https://doi.org/10.1002/prop.200810552}{\emph{Fortsch. Phys.}
  {\bfseries 56} (2008) 862} [\href{https://arxiv.org/abs/0803.3149}{{\ttfamily
  0803.3149}}].

\bibitem{Becker:2003yv}
K.~Becker, M.~Becker, K.~Dasgupta and P.~S. Green, \emph{{Compactifications of
  heterotic theory on nonKahler complex manifolds. 1.}},
  \href{https://doi.org/10.1088/1126-6708/2003/04/007}{\emph{JHEP} {\bfseries
  04} (2003) 007} [\href{https://arxiv.org/abs/hep-th/0301161}{{\ttfamily
  hep-th/0301161}}].

\bibitem{LopesCardoso:2003dvb}
G.~Lopes~Cardoso, G.~Curio, G.~Dall'Agata and D.~Lust, \emph{{BPS action and
  superpotential for heterotic string compactifications with fluxes}},
  \href{https://doi.org/10.1088/1126-6708/2003/10/004}{\emph{JHEP} {\bfseries
  10} (2003) 004} [\href{https://arxiv.org/abs/hep-th/0306088}{{\ttfamily
  hep-th/0306088}}].

\bibitem{Gurrieri:2004dt}
S.~Gurrieri, A.~Lukas and A.~Micu, \emph{{Heterotic on half-flat}},
  \href{https://doi.org/10.1103/PhysRevD.70.126009}{\emph{Phys. Rev. D}
  {\bfseries 70} (2004) 126009}
  [\href{https://arxiv.org/abs/hep-th/0408121}{{\ttfamily hep-th/0408121}}].

\bibitem{Benmachiche:2006df}
I.~Benmachiche and T.~W. Grimm, \emph{{Generalized N=1 orientifold
  compactifications and the Hitchin functionals}},
  \href{https://doi.org/10.1016/j.nuclphysb.2006.05.003}{\emph{Nucl. Phys. B}
  {\bfseries 748} (2006) 200}
  [\href{https://arxiv.org/abs/hep-th/0602241}{{\ttfamily hep-th/0602241}}].

\bibitem{delaOssa:2015maa}
X.~de~la Ossa, E.~Hardy and E.~E. Svanes, \emph{{The Heterotic Superpotential
  and Moduli}}, \href{https://doi.org/10.1007/JHEP01(2016)049}{\emph{JHEP}
  {\bfseries 01} (2016) 049}
  [\href{https://arxiv.org/abs/1509.08724}{{\ttfamily 1509.08724}}].

\bibitem{McOrist:2016cfl}
J.~McOrist, \emph{{On the Effective Field Theory of Heterotic Vacua}},
  \href{https://doi.org/10.1007/s11005-017-1025-0}{\emph{Lett. Math. Phys.}
  {\bfseries 108} (2018) 1031}
  [\href{https://arxiv.org/abs/1606.05221}{{\ttfamily 1606.05221}}].

\bibitem{Grimm:2004uq}
T.~W. Grimm and J.~Louis, \emph{{The Effective action of N = 1 Calabi-Yau
  orientifolds}},
  \href{https://doi.org/10.1016/j.nuclphysb.2004.08.005}{\emph{Nucl. Phys. B}
  {\bfseries 699} (2004) 387}
  [\href{https://arxiv.org/abs/hep-th/0403067}{{\ttfamily hep-th/0403067}}].

\bibitem{Cassani:2016ncu}
D.~Cassani, O.~de~Felice, M.~Petrini, C.~Strickland-Constable and D.~Waldram,
  \emph{{Exceptional generalised geometry for massive IIA and consistent
  reductions}}, \href{https://doi.org/10.1007/JHEP08(2016)074}{\emph{JHEP}
  {\bfseries 08} (2016) 074}
  [\href{https://arxiv.org/abs/1605.00563}{{\ttfamily 1605.00563}}].

\bibitem{Tsai:2014ela}
C.-J. Tsai, L.-S. Tseng and S.-T. Yau, \emph{{Cohomology and Hodge Theory on
  Symplectic Manifolds: III}}, {\emph{J. Diff. Geom.} {\bfseries 103} (2016)
  83} [\href{https://arxiv.org/abs/1402.0427}{{\ttfamily 1402.0427}}].

\bibitem{Tanaka_2018}
H.~L. Tanaka and L.-S. Tseng, \emph{Odd sphere bundles, symplectic manifolds,
  and their intersection theory},
  \href{https://doi.org/10.4310/cjm.2018.v6.n3.a1}{\emph{Cambridge Journal of
  Mathematics} {\bfseries 6} (2018) 213}.

\bibitem{Woodward}
C.~T. {Woodward}, \emph{{Moment maps and geometric invariant theory}},
  \href{https://arxiv.org/abs/0912.1132}{{\ttfamily 0912.1132}}.

\bibitem{Thomas}
R.~P. Thomas, \emph{{Notes on GIT and symplectic reduction for bundles and
  varieties}},  \href{https://arxiv.org/abs/math/0512411}{{\ttfamily
  math/0512411}}.

\bibitem{Klebanov:2000hb}
I.~R. Klebanov and M.~J. Strassler, \emph{{Supergravity and a confining gauge
  theory: Duality cascades and chi SB resolution of naked singularities}},
  \href{https://doi.org/10.1088/1126-6708/2000/08/052}{\emph{JHEP} {\bfseries
  08} (2000) 052} [\href{https://arxiv.org/abs/hep-th/0007191}{{\ttfamily
  hep-th/0007191}}].

\bibitem{Derendinger:1991hq}
J.~P. Derendinger, S.~Ferrara, C.~Kounnas and F.~Zwirner, \emph{{On loop
  corrections to string effective field theories: Field dependent gauge
  couplings and sigma model anomalies}},
  \href{https://doi.org/10.1016/0550-3213(92)90315-3}{\emph{Nucl. Phys. B}
  {\bfseries 372} (1992) 145}.

\bibitem{Antoniadis:1997eg}
I.~Antoniadis, S.~Ferrara, R.~Minasian and K.~S. Narain, \emph{{R**4 couplings
  in M and type II theories on Calabi-Yau spaces}},
  \href{https://doi.org/10.1016/S0550-3213(97)00572-5}{\emph{Nucl. Phys. B}
  {\bfseries 507} (1997) 571}
  [\href{https://arxiv.org/abs/hep-th/9707013}{{\ttfamily hep-th/9707013}}].

\bibitem{Antoniadis:2003sw}
I.~Antoniadis, R.~Minasian, S.~Theisen and P.~Vanhove, \emph{{String loop
  corrections to the universal hypermultiplet}},
  \href{https://doi.org/10.1088/0264-9381/20/23/009}{\emph{Class. Quant. Grav.}
  {\bfseries 20} (2003) 5079}
  [\href{https://arxiv.org/abs/hep-th/0307268}{{\ttfamily hep-th/0307268}}].

\bibitem{Berg:2005ja}
M.~Berg, M.~Haack and B.~Kors, \emph{{String loop corrections to Kahler
  potentials in orientifolds}},
  \href{https://doi.org/10.1088/1126-6708/2005/11/030}{\emph{JHEP} {\bfseries
  11} (2005) 030} [\href{https://arxiv.org/abs/hep-th/0508043}{{\ttfamily
  hep-th/0508043}}].

\bibitem{Haack:2018ufg}
M.~Haack and J.~U. Kang, \emph{{Field redefinitions and K\"ahler potential in
  string theory at 1-loop}},
  \href{https://doi.org/10.1007/JHEP08(2018)019}{\emph{JHEP} {\bfseries 08}
  (2018) 019} [\href{https://arxiv.org/abs/1805.00817}{{\ttfamily
  1805.00817}}].

\bibitem{1987mast.conf..629T}
G.~{Tian}, \emph{{Smoothness of the Universal Deformation Space of Compact
  Calabi-Yau Manifolds and Its Peterson-Weil Metric}},  in \emph{Mathematical
  Aspects of String Theory}, pp.~629--646, Jan., 1987,
  \href{https://doi.org/10.1142/9789812798411_0029}{DOI}.

\bibitem{1989CMaPh.126..325T}
A.~N. {Todorov}, \emph{{The Weil-Petersson geometry of the moduli space of SU(
  n{\ensuremath{\geqq}}3) (Calabi-Yau) manifolds I}},
  \href{https://doi.org/10.1007/BF02125128}{\emph{Communications in
  Mathematical Physics} {\bfseries 126} (1989) 325}.

\bibitem{Ashmore:2019rkx}
A.~Ashmore, C.~Strickland-Constable, D.~Tennyson and D.~Waldram,
  \emph{{Heterotic backgrounds via generalised geometry: moment maps and
  moduli}}, \href{https://doi.org/10.1007/JHEP11(2020)071}{\emph{JHEP}
  {\bfseries 11} (2020) 071}
  [\href{https://arxiv.org/abs/1912.09981}{{\ttfamily 1912.09981}}].

\bibitem{Kachru:2000ih}
S.~Kachru, S.~H. Katz, A.~E. Lawrence and J.~McGreevy, \emph{{Open string
  instantons and superpotentials}},
  \href{https://doi.org/10.1103/PhysRevD.62.026001}{\emph{Phys. Rev. D}
  {\bfseries 62} (2000) 026001}
  [\href{https://arxiv.org/abs/hep-th/9912151}{{\ttfamily hep-th/9912151}}].

\bibitem{Plauschinn:2021hkp}
E.~Plauschinn, \emph{{The tadpole conjecture at large complex-structure}},
  \href{https://doi.org/10.1007/JHEP02(2022)206}{\emph{JHEP} {\bfseries 02}
  (2022) 206} [\href{https://arxiv.org/abs/2109.00029}{{\ttfamily
  2109.00029}}].

\bibitem{Lust:2021xds}
S.~L\"ust, \emph{{Large complex structure flux vacua of IIB and the Tadpole
  Conjecture}},  \href{https://arxiv.org/abs/2109.05033}{{\ttfamily
  2109.05033}}.

\bibitem{Bena:2021qty}
I.~Bena, C.~Brodie and M.~Gra\~na, \emph{{D7 moduli stabilization: the tadpole
  menace}}, \href{https://doi.org/10.1007/JHEP01(2022)138}{\emph{JHEP}
  {\bfseries 01} (2022) 138}
  [\href{https://arxiv.org/abs/2112.00013}{{\ttfamily 2112.00013}}].

\bibitem{Grimm:2021ckh}
T.~W. Grimm, E.~Plauschinn and D.~van~de Heisteeg, \emph{{Moduli stabilization
  in asymptotic flux compactifications}},
  \href{https://doi.org/10.1007/JHEP03(2022)117}{\emph{JHEP} {\bfseries 03}
  (2022) 117} [\href{https://arxiv.org/abs/2110.05511}{{\ttfamily
  2110.05511}}].

\bibitem{Grana:2022dfw}
M.~Gra\~na, T.~W. Grimm, D.~van~de Heisteeg, A.~Herraez and E.~Plauschinn,
  \emph{{The tadpole conjecture in asymptotic limits}},
  \href{https://doi.org/10.1007/JHEP08(2022)237}{\emph{JHEP} {\bfseries 08}
  (2022) 237} [\href{https://arxiv.org/abs/2204.05331}{{\ttfamily
  2204.05331}}].

\bibitem{Fidanza:2003zi}
S.~Fidanza, R.~Minasian and A.~Tomasiello, \emph{{Mirror symmetric SU(3)
  structure manifolds with NS fluxes}},
  \href{https://doi.org/10.1007/s00220-004-1221-5}{\emph{Commun. Math. Phys.}
  {\bfseries 254} (2005) 401}
  [\href{https://arxiv.org/abs/hep-th/0311122}{{\ttfamily hep-th/0311122}}].

\bibitem{MGraña_2007}
M.~Graña, J.~Louis and D.~Waldram, \emph{Su(3) × su(3) compactification and
  mirror duals of magnetic fluxes},
  \href{https://doi.org/10.1088/1126-6708/2007/04/101}{\emph{Journal of High
  Energy Physics} {\bfseries 2007} (2007) 101}.

\bibitem{BENBASSAT2006533}
O.~Ben-Bassat, \emph{Mirror symmetry and generalized complex manifolds: Part i.
  the transform on vector bundles, spinors, and branes},
  \href{https://doi.org/https://doi.org/10.1016/j.geomphys.2005.03.004}{\emph{Journal
  of Geometry and Physics} {\bfseries 56} (2006) 533}.

\bibitem{BENBASSAT20061096}
O.~Ben-Bassat, \emph{Mirror symmetry and generalized complex manifolds: Part
  ii. integrability and the transform for torus bundles},
  \href{https://doi.org/https://doi.org/10.1016/j.geomphys.2005.04.020}{\emph{Journal
  of Geometry and Physics} {\bfseries 56} (2006) 1096}.

\bibitem{Bouwknegt:2003vb}
P.~Bouwknegt, J.~Evslin and V.~Mathai, \emph{{T duality: Topology change from H
  flux}}, \href{https://doi.org/10.1007/s00220-004-1115-6}{\emph{Commun. Math.
  Phys.} {\bfseries 249} (2004) 383}
  [\href{https://arxiv.org/abs/hep-th/0306062}{{\ttfamily hep-th/0306062}}].

\bibitem{Bouwknegt:2003zg}
P.~Bouwknegt, K.~Hannabuss and V.~Mathai, \emph{{T duality for principal torus
  bundles}}, \href{https://doi.org/10.1088/1126-6708/2004/03/018}{\emph{JHEP}
  {\bfseries 03} (2004) 018}
  [\href{https://arxiv.org/abs/hep-th/0312284}{{\ttfamily hep-th/0312284}}].

\bibitem{Bunke:2005sn}
U.~Bunke and T.~Schick, \emph{{On the topology of T-duality}},
  \href{https://doi.org/10.1142/S0129055X05002315}{\emph{Rev. Math. Phys.}
  {\bfseries 17} (2005) 77}
  [\href{https://arxiv.org/abs/math/0405132}{{\ttfamily math/0405132}}].

\bibitem{Tomasiello:2005bp}
A.~Tomasiello, \emph{{Topological mirror symmetry with fluxes}},
  \href{https://doi.org/10.1088/1126-6708/2005/06/067}{\emph{JHEP} {\bfseries
  06} (2005) 067} [\href{https://arxiv.org/abs/hep-th/0502148}{{\ttfamily
  hep-th/0502148}}].

\bibitem{Cavalcanti:2011wu}
G.~R. Cavalcanti and M.~Gualtieri, \emph{{Generalized complex geometry and
  T-duality}},  6, 2011, \href{https://arxiv.org/abs/1106.1747}{{\ttfamily
  1106.1747}}.

\bibitem{Lau:2014fia}
S.-C. Lau, L.-S. Tseng and S.-T. Yau, \emph{{Non-K\"ahler SYZ Mirror
  Symmetry}}, \href{https://doi.org/10.1007/s00220-015-2454-1}{\emph{Commun.
  Math. Phys.} {\bfseries 340} (2015) 145}
  [\href{https://arxiv.org/abs/1409.2765}{{\ttfamily 1409.2765}}].

\bibitem{Kapustin:2004gv}
A.~Kapustin and Y.~Li, \emph{{Topological sigma-models with H-flux and twisted
  generalized complex manifolds}},
  \href{https://doi.org/10.4310/ATMP.2007.v11.n2.a3}{\emph{Adv. Theor. Math.
  Phys.} {\bfseries 11} (2007) 269}
  [\href{https://arxiv.org/abs/hep-th/0407249}{{\ttfamily hep-th/0407249}}].

\bibitem{Pestun:2005rp}
V.~Pestun and E.~Witten, \emph{{The Hitchin functionals and the topological
  B-model at one loop}},
  \href{https://doi.org/10.1007/s11005-005-0007-9}{\emph{Lett. Math. Phys.}
  {\bfseries 74} (2005) 21}
  [\href{https://arxiv.org/abs/hep-th/0503083}{{\ttfamily hep-th/0503083}}].

\bibitem{Pestun:2006rj}
V.~Pestun, \emph{{Topological strings in generalized complex space}},
  \href{https://doi.org/10.4310/ATMP.2007.v11.n3.a3}{\emph{Adv. Theor. Math.
  Phys.} {\bfseries 11} (2007) 399}
  [\href{https://arxiv.org/abs/hep-th/0603145}{{\ttfamily hep-th/0603145}}].

\bibitem{Ashmore:2023vji}
A.~Ashmore, J.~J.~M. Ibarra, D.~D. McNutt, C.~Strickland-Constable, E.~E.
  Svanes, D.~Tennyson et~al., \emph{{A heterotic Kodaira-Spencer theory at
  one-loop}}, \href{https://doi.org/10.1007/JHEP10(2023)130}{\emph{JHEP}
  {\bfseries 10} (2023) 130}
  [\href{https://arxiv.org/abs/2306.10106}{{\ttfamily 2306.10106}}].

\bibitem{Ashmore:2021pdm}
A.~Ashmore, A.~Coimbra, C.~Strickland-Constable, E.~E. Svanes and D.~Tennyson,
  \emph{{Topological G$_{2}$ and Spin(7) strings at 1-loop from double
  complexes}}, \href{https://doi.org/10.1007/JHEP02(2022)089}{\emph{JHEP}
  {\bfseries 02} (2022) 089}
  [\href{https://arxiv.org/abs/2108.09310}{{\ttfamily 2108.09310}}].

\bibitem{Kupka:2024rvl}
J.~Kupka, C.~Strickland-Constable, E.~E. Svanes, D.~Tennyson and F.~Valach,
  \emph{{BPS complexes and Chern--Simons theories from $G$-structures in gauge
  theory and gravity}},  \href{https://arxiv.org/abs/2406.03550}{{\ttfamily
  2406.03550}}.

\bibitem{Hitchin:1999fh}
N.~J. Hitchin, \emph{{Lectures on special Lagrangian submanifolds}},
  {\emph{AMS/IP Stud. Adv. Math.} {\bfseries 23} (2001) 151}
  [\href{https://arxiv.org/abs/math/9907034}{{\ttfamily math/9907034}}].

\bibitem{Bott:1982xhp}
R.~Bott and L.~W. Tu, \emph{{Differential Forms in Algebraic Topology}}.
  Springer, 1982,
  \href{https://doi.org/10.1007/978-1-4757-3951-0}{10.1007/978-1-4757-3951-0}.

\bibitem{Hatcher_AT}
A.~Hatcher, \emph{Algebraic topology}. Cambridge University Press, Cambridge,
  2002.

\end{thebibliography}\endgroup
\end{document}